\documentclass[12pt,a4paper]{article}
\usepackage[numbers,sort&compress]{natbib}
\usepackage{amsmath,amssymb,tabularx}
\usepackage[title]{appendix}
\usepackage{verbatim}
\usepackage{titlesec}
\titlelabel{\thetitle.\quad}
\usepackage{bm}
\usepackage{subcaption}
\usepackage{dsfont}
\usepackage{color}
\usepackage{graphicx}
\usepackage{physics}
\usepackage{slashed}
\usepackage[svgnames]{xcolor}
\usepackage{hyperref}
\usepackage{hyperref}
\hypersetup{
	colorlinks,
	citecolor=blue,
	filecolor=blue,
	linkcolor=blue,
	urlcolor=blue
}
\usepackage{caption}	
\usepackage{microtype}
\usepackage{ctable}

\newcommand\varpm{\mathbin{\vcenter{\hbox{%
				\oalign{\hfil$\scriptstyle+$\hfil\cr
					\noalign{\kern-.3ex}
					$\scriptscriptstyle({-})$\cr}%
}}}}
\newcommand\varmp{\mathbin{\vcenter{\hbox{%
				\oalign{$\scriptstyle({+})$\cr
					\noalign{\kern-.3ex}
					\hfil$\scriptscriptstyle-$\hfil\cr}%
}}}}
\newcommand{\nn}{\nonumber}
\allowdisplaybreaks
\newcommand{\bea}{\begin{eqnarray}}
\newcommand{\eea}{\end{eqnarray}}
\newcommand{\ba}{\begin{align}}
\newcommand{\ea}{\end{align}}
\begin{document}
\title{
\begin{flushright}
\ \\*[-80pt] 
\begin{minipage}{0.2\linewidth}
\normalsize

%arXiv:YYMM.NNNN \\
HUPD-2405 \\*[5pt]
\end{minipage}
\end{flushright}
{\Large 
The third family quark mass hierarchy and FCNC \\in the universal seesaw model\\*[5pt]} }
\author{\normalsize
\centerline{
Takuya Morozumi$^{1,2}$\footnote{morozumi@hiroshima-u.ac.jp},
Albertus Hariwangsa Panuluh$^{1,3}$\footnote{panuluh@usd.ac.id}
} \\ \normalsize
\centerline{
%Author 3 $^{1}$\footnote{E-mail address},
%Author 4$^{1}$\footnote{E-mail address},
}
%%%%%%%%%%%%%%%%%%%%%%
\\*[10pt]
\centerline{
\begin{minipage}{\linewidth}
\begin{center}
$^1${\it \small
Physics Program, Graduate School of Advanced Science and Engineering, Hiroshima University,\\ Higashi-Hiroshima 739-8526, Japan} \\*[5pt]
$^2${\it \small
Core of Research for the Energetic Universe, Hiroshima University, Higashi-Hiroshima 739-8526,\\ Hiroshima, Japan} \\*[5pt]
$^3${\it \small
Department of Physics Education, Sanata Dharma University, Paingan, Maguwohardjo,\\
Sleman, Yogyakarta 55282, Indonesia} \\*[5pt]
\end{center}
\end{minipage}}
\\*[50pt]}
\date{
\centerline{\small \bf Abstract}
\begin{minipage}{0.9\linewidth}
\medskip
\medskip
\small
We present the study of the quark sector of the universal seesaw model with $\mathrm{SU(2)_L \times SU(2)_R \times U(1)_{Y'}}$ gauge symmetry in the massless case of the two lightest quark families. This model aims to explain the mass hierarchy of the third family quark by introducing a vector-like quark (VLQ) partner for each quark. In this model, we introduce $\mathrm{SU(2)_L}$ and $\mathrm{SU(2)_R}$ Higgs doublets. We derive explicitly the Lagrangian for the quark sector, Higgs sector, and kinetic terms of the gauge fields, starting from the Lagrangian, which is invariant under $\mathrm{SU(2)_L \times SU(2)_R \times U(1)_{Y'}}$ gauge symmetry. At each stage of the symmetry breaking, we present the Lagrangian with the remaining gauge symmetry. Additionally, we investigate the flavor-changing neutral currents (FCNCs) of Higgs ($h$) and $Z$-boson in the interaction with the top, heavy top, bottom, and heavy bottom quarks. 
\end{minipage}
}
\begin{titlepage}
\maketitle
\thispagestyle{empty}
\end{titlepage}
\counterwithin*{equation}{section}
\renewcommand\theequation{\thesection.\arabic{equation}}
\section{Introduction}\label{sec:introduction}
The seesaw mechanism is a well-known approach to explain the smallness of neutrino masses \citep{neutrino minkowski,neutrino yanagida kek,gell mann,neutrino yanagida,neutrino,neutrino 2,neutrino 3}. It introduces heavy right-handed neutrinos that mix with the light left-handed neutrinos, giving them a small mass. This inspired the construction of a similar model, which can be applied to other cases. One problem that the Standard Model (SM) cannot explain is the fermion mass hierarchy. In this paper, we study the quark sector of the universal seesaw model \citep{ber0,ber,raj,chang,dav,ber2,ber3,koide,satou,kiyo,umeeda,universal seesaw 3,universal seesaw, universal seesaw 2, babu, dcruz}, an extension of the SM that applies a seesaw-like mechanism to the quark sector to solve the mass hierarchy problem. In the quark sector, an interesting aspect is the large mass of the top quark compared to the other quarks. Our focus is on the third family of quarks, and within our framework, the two lightest quark families are massless.

Introducing vector-like quarks (VLQs) into this model is essential. VLQs have left- and right-handed components that transform identically under some gauge group. Using this property, they can mix with SM quarks, resulting in modified mass matrices that can be diagonalized and generate a tiny seesaw-like mass. Various studies about adding VLQs to the SM have been explored, for example, introducing one down-type isosinglet VLQ \citep{downvlq}, one up-type isosinglet VLQ \citep{upvlq}, and both one up-type and down-type isosinglet VLQ \citep{updownvlq}. The presence of VLQs also has implications for flavor physics, as they can introduce flavor-changing neutral currents (FCNCs) \citep{vlq fcnc}, and weak-basis invariants have been analyzed to understand the flavor structures \citep{weakbasis 1,weakbasis 2}. Effective field theory approaches to VLQs have been studied to understand their contributions to low-energy observables \citep{eft,eftproc}. A review of the theory and phenomenology of isosinglet VLQs can be found in Ref.\citep{review vlq}.

This paper aims to study the quark sector within the universal seesaw model in the massless case of the two lightest quark families. We derive the Lagrangian, including the quark and Higgs sectors, and gauge kinetic terms. We also demonstrate how the model can naturally explain the observed quark mass hierarchies in the third family, particularly the significant mass of the top quark. We also explore the phenomenological implications of this model by investigating the interaction between the Higgs and $Z$-boson with quarks, which includes FCNC processes.

The outline of this paper is as follows. In section \ref{sec:the model}, we introduce the model with the particle contents and the Lagrangian. Section \ref{sec:quark doublet and yukawa interaction} focuses on the quark sector and Yukawa interactions. We explain the derivation of the Lagrangian of the kinetic terms and Yukawa interactions. Starting with the Lagrangian, which is invariant under $\mathrm{SU(2)_L \times SU(2)_R \times U(1)_{Y'}}$, in each stage of the symmetry breaking, we present the Lagrangian with the remaining gauge symmetry.  The quark mass eigenvalues and the identification of FCNC within the massive third family quarks and their VLQ partners are discussed.

Section \ref{sec: Higgs sector} discusses the Higgs sector of this model. The kinetic terms and Higgs potential are also derived step by step. In the end, we classify the terms based on the number of the fields in the term as linear, quadratic, cubic, and quartic, ensuring a clear understanding of the interactions of the gauge sector. In addition, we also provide the exact diagonal mass of $Z-Z'$ bosons and $h-H$ bosons. 

The kinetic terms of gauge fields are discussed in Section \ref{sec:gauge field}. In the final derivation, we show the difference between our model and SM. Finally, in Section \ref{sec:discussion}, we present some phenomenological implications of our model. We start the discussion with the hierarchy of VLQ's mass parameters, the non-zero vacuum expectation value of $\mathrm{SU(2)_L}$ Higgs doublet $(v_L)$, and the non-zero vacuum expectation value of $\mathrm{SU(2)_R}$ Higgs doublet $(v_R)$. Then, we analyze the interaction between Higgs $(h)$ and $Z$-boson with the quarks. This leads to a discussion about FCNCs in this model.
\section{The model}\label{sec:the model}
%\subsection{Particle contents and Lagrangian}
We consider an extension of SM with $\mathrm{SU(3)_C}\times\mathrm{SU(2)_L \times SU(2)_R \times U(1)_{Y'}}$ gauge symmetry in the massless case of the two lightest quark families. In addition to $\mathrm{SU(2)_L}$ SM Higgs doublet ($\phi_L$), we have a $\mathrm{SU(2)_R}$ Higgs doublet ($\phi_R$).  We also introduce one up-type and one down-type isosinglet vector-like quarks (VLQs), denoted by $T$ and $B$, respectively. The charge convention we use in this model is,
\begin{equation}
	Q = I^3_{L} + I^3_{R} + Y' \label{charge convention},
\end{equation}
where $Q, I^3_{L(R)}$ and $Y'$ are electromagnetic charge, left(right) weak-isospin, and $\mathrm{U(1)_{Y'}}$ hypercharge respectively.
The particle contents and their charge assignments under the model's gauge group are given in Table \ref{tab:model}.

The Lagrangian of this model (excluding the QCD part) is as follows, 
\begin{align}
	\mathcal{L} &={\cal L}_{q}+ {\cal L}_{H}+{\cal L}_{\text{gauge}}, \label{full Lagrangian} \\
{\cal L}_{q}&= 
 \overline{q^i_{L}} i \gamma^\mu D_{L\mu} q^i_{L} + \overline{q^i_{R}} i \gamma^\mu D_{R\mu} q^i_{R} + \overline{T} i \gamma^{\mu} D_{T\mu} T + \overline{B} i \gamma^{\mu} D_{B\mu} B \nn \\
 &\hspace{10pt}- Y_{u_L}^3\overline{q_L^3}\tilde{\phi}_L T_R -Y_{u_R}^3 \overline{T_L} \tilde{\phi}_R^\dagger q_R^3 - \overline{q^i_{L}} y^i_{d_L} \phi_L B_R - \overline{B_L} y^{i\ast}_{d_R} \phi_R^\dagger q_R^i -h.c.\nn\\ &\hspace{10pt} - \overline{T_L}M_T T_R - \overline{B_L} M_B B_R -h.c., \label{quark Lagrangian}\\
{\cal L}_{H}&= (D_L^\mu\phi_L)^\dagger(D_{L\mu}\phi_L)+(D_R^\mu\phi_R)^\dagger(D_{R\mu}\phi_R) - V(\phi_L ,\phi_R) \label{higgs Lagrangian} , \\
{\cal L}_{\text{gauge}}&= -\frac{1}{4}F^a_{L\mu\nu}F_L^{a\mu\nu}-\frac{1}{4}F^a_{R\mu\nu}F_R^{a\mu\nu} -\frac{1}{4} B'_{\mu\nu}B^{\prime\mu\nu}, \label{gauge Lagrangian}
\end{align}
where,
\begin{align}
	V(\phi_L,\phi_R) & = \mu_L^2 \phi_L^\dagger \phi_L +\mu_R^2 \phi_R^\dagger \phi_R + \lambda_L(\phi_L^\dagger \phi_L)^2 +\lambda_R(\phi_R^\dagger \phi_R)^2+2\lambda_{LR}(\phi_L^\dagger \phi_L)(\phi_R^\dagger \phi_R), \label{higgs potential} \\
		D_{L(R)\mu}q^i_{L(R)} &= \left( \partial_\mu + i g_{L(R)} \frac{\tau^a}{2} W^a_{L(R)\mu}  + i g'_{1} Y'_q B'_{\mu}\right) q^i_{L(R)}, \label{cov der quark doublet}\\
		D_{L(R)\mu}\phi_{L(R)} &= \left( \partial_\mu + i g_{L(R)} \frac{\tau^a}{2} W^a_{L(R)\mu}  + i g'_{1} Y'_\phi B'_{\mu}\right) \phi_{L(R)}, \label{cov der higgs doublet}\\
		D_{T\mu} T &= \left( \partial_\mu + i g'_{1} Y'_{T} B'_{\mu}\right) T, \label{cov der top VLQ} \\
		D_{B\mu} B &= \left( \partial_\mu + i g'_{1} Y'_{B} B'_{\mu}\right) B, \label{cov der bottom VLQ} \\
		F^a_{L\mu\nu} &=\partial_\mu W^a_{L\nu} - \partial_\nu W^a_{L\mu} -g_L \epsilon^{abc}W^b_{L\mu}W^c_{L\nu}, \\
		F^a_{R\mu\nu} &=\partial_\mu W^a_{R\nu} - \partial_\nu W^a_{R\mu} -g_R \epsilon^{abc}W^b_{R\mu}W^c_{R\nu},\\
		B'_{\mu\nu} &= \partial_\mu B'_\nu - \partial_\nu B'_\mu. 
\end{align}
\begin{table}[t]
	\captionsetup{font=small}
	\caption{\label{tab:model}Quark and Higgs fields with their quantum numbers under the $\mathrm{SU(3)_C}\times\mathrm{SU(2)_L \times SU(2)_R \times U(1)_{Y'}}$ gauge groups, where $i\in \{1,2,3\}$ is the family index. } 
	
	\begin{center}
		\begin{tabular}{|c|cccc|}
			\hline
			Quark and Higgs Fields	& $\mathrm{SU(3)_C}$ & $\mathrm{SU(2)_L}$ &$\mathrm{SU(2)_R}$  & $\mathrm{U(1)_{Y'}}$ \\
			\hline\hline 
			$q_L^i=\left( \begin{array}{c}
				u_L^i	\\d_L^i
			\end{array}\right)$ 	& \textbf{3} & \textbf{2} & \textbf{1} & 1/6 \\ [0.8\normalbaselineskip]
			
			$q_R^i=\left( \begin{array}{c}
				u_R^i	\\d_R^i
			\end{array}\right)$	& \textbf{3} &  \textbf{1}& \textbf{2} & 1/6 \\[0.8\normalbaselineskip]
			
			$T_{L,R}$	& \textbf{3} & \textbf{1} &  \textbf{1}& 2/3 \\ [0.8\normalbaselineskip]
			
			$B_{L,R}$	& \textbf{3} &\textbf{1}  & \textbf{1} & $-1/3$ \\ [0.8\normalbaselineskip]
			
			$\phi_L=\left( \begin{array}{c}
				\chi_L^+	\\\chi_L^0
			\end{array}\right)$ 	& \textbf{1} &\textbf{2}  &\textbf{1}  &1/2  \\ [0.8\normalbaselineskip]
			
			$\phi_R=\left( \begin{array}{c}
				\chi_R^+	\\\chi_R^0
			\end{array}\right)$	&\textbf{1}  & \textbf{1} &\textbf{2}  &  1/2\\
			\hline
		\end{tabular}
	\end{center}
\end{table}
The Lagrangian in Eq.(\ref{full Lagrangian}) is divided into three parts. The first part is the kinetic terms of quark doublet and isosinglet VLQs, Yukawa interactions, and mass terms of isosinglet VLQs, which are contained in Eq.(\ref{quark Lagrangian}). The second part is the kinetic terms and potential of Higgs doublet, which are contained in Eq.(\ref{higgs Lagrangian}). The third part is the kinetic terms of the gauge fields, which are written in Eq.(\ref{gauge Lagrangian}).

The first line of  Eq.(\ref{quark Lagrangian}) is the kinetic terms of quark doublet and isosinglet VLQs where the definition of the covariant derivatives are written in Eqs.(\ref{cov der quark doublet}), (\ref{cov der top VLQ}) and (\ref{cov der bottom VLQ}) respectively, where $g_{L(R)}$ is $\mathrm{SU(2)_{L(R)}}$ gauge coupling, $\tau^a$ is the Pauli matrix, $g_1'$ is $\mathrm{U(1)_{Y'}}$ gauge coupling and $Y'$ is the corresponding $\mathrm{U(1)_{Y'}}$ hypercharge. For the Yukawa interaction part, one can choose in a weak-basis where the Yukawa couplings of up-type quark doublet ($Y^3_{u_L}$ and $Y^3_{u_R}$) are real positive numbers. In contrast, the Yukawa couplings of down-type quark are general complex vectors as shown in the second line of  Eq.(\ref{quark Lagrangian}). The derivation of this weak-basis is briefly explained in Appendix \ref{sec:appendix weak basis} The family index for SM quarks is denoted as $i \in \{1,2,3\}$, the charge conjugation of Higgs fields is defined as $\tilde{\phi}_{L(R)}=i\tau^2 \phi^\ast_{L(R)}$. In the third line of Eq.(\ref{quark Lagrangian}), $M_T$ and $M_B$ are isosinglet VLQs mass parameters that we take as real numbers.

The first two terms of Eq.(\ref{higgs Lagrangian}) are the kinetic terms of Higgs doublet where the definition of the covariant derivatives are written in Eq.(\ref{cov der higgs doublet}). The third term is the Higgs potential, which is shown in Eq.(\ref{higgs potential}), containing the mass terms and quartic interactions of Higgs doublet, including the interaction between $\phi_L$ and $\phi_R$. Later $\phi_R$ and $\phi_L$ acquire non-zero vacuum expectation values (vevs) denoted as $v_R$ and $v_L$ that break $\mathrm{SU(2)_R}$ and $\mathrm{SU(2)_L}$  respectively and satisfy $v_R \gg v_L$. 

\section{Quark sector and Yukawa interactions}\label{sec:quark doublet and yukawa interaction} 
In this section, we derived the kinetic terms of quark doublet and isosinglet VLQs, Yukawa interactions, and mass terms of isosinglet VLQs, which are contained in Eq.(\ref{quark Lagrangian}) with $\mathrm{SU(2)_L\times SU(2)_R \times U(1)_{Y'}}$ symmetric Lagrangian. After $\mathrm{SU(2)_R}$ Higgs doublet acquires non-zero vev, we obtain the Lagrangian, which is invariant under SM gauge symmetry. Furthermore,  the SM gauge group is broken into $\mathrm{U(1)_{\text{em}}}$ after $\mathrm{SU(2)_L}$ Higgs doublet acquires non-zero vev. Finally, we obtain the masses of top and bottom quark, their heavy partners', $Z, Z',h$, and $H$. FCNC and the CKM matrix are also generated.
\subsection{ $ \mathbf{SU(2)_R \times U(1)_{Y'}} \rightarrow\mathbf{U(1)_Y}$}
In this stage, the neutral scalar component of $\mathrm{SU(2)_R}$ Higgs doublet acquires non-zero vev and is expanded around the vev as follows,
\begin{equation}
	\phi_R =\left( \begin{array}{c}
		\chi_R^+	\\\chi_R^0
	\end{array}\right)= \frac{1}{\sqrt{2}}\left(\begin{array}{c}
		\sqrt{2}\chi_R^+ \\
		v_R+h_R+i \chi_R^3
	\end{array} \right),  \label{right higgs parameterization}
\end{equation}
where $v_R$ is the non-zero vev. $h_R$ is the neutral CP-even state and $\chi_R^3$ is the neutral CP-odd state. The charged component is denoted as, $\chi_R^+ = \frac{1}{\sqrt{2}} (\chi_R^1 + i \chi_R^2)$. In addition, we rotate the gauge fields with the following transformation,
\begin{equation}
	\left( \begin{array}{c}
		B'_\mu \\
		W^3_{R\mu}
	\end{array}\right) = \left( \begin{array}{cc}
	 \cos\theta_R & -\sin\theta_R \\
	 \sin\theta_R & \cos\theta_R
	\end{array}\right)\left( \begin{array}{c}
	B_\mu \\
	Z_{R\mu}
	\end{array}\right), \label{B WR mixing}
\end{equation} 
where the mixing angle,
\begin{align}
	\sin\theta_R = \frac{g'_1}{\sqrt{g_R^2+g_1^{\prime 2}}}, \qquad \cos\theta_R = \frac{g_R}{\sqrt{g_R^2+g_1^{\prime 2}}}. 
\end{align}
We also define the SM $\mathrm{U(1)_Y}$ gauge coupling as,
\begin{equation}
	g'=g'_1 \cos\theta_R = g_R \sin\theta_R.\label{relation thetaR}
\end{equation} 
After this spontaneously symmetry breaking, the Lagrangian in Eq.(\ref{quark Lagrangian}) becomes,
\begin{align}
	{\cal L}_{q}&= 
	\overline{q^i_{L}} i \gamma^\mu D_{\text{SM}\mu} q^i_{L} + \overline{T_L} i \gamma^{\mu} D_{\text{SM}\mu} T_L + \overline{B_L} i \gamma^{\mu} D_{\text{SM}\mu} B_L \nn \\
	&\hspace{10pt}+ \overline{u^i_{R}} i \gamma^\mu D_{\text{SM}\mu} u^i_{R}+ \overline{d^i_{R}} i \gamma^\mu D_{\text{SM}\mu} d^i_{R} + \overline{T_R} i \gamma^{\mu} D_{\text{SM}\mu} T_R + \overline{B_R} i \gamma^{\mu} D_{\text{SM}\mu} B_R\nn\\&\hspace{10pt}-\frac{g_R}{\sqrt{2}}\overline{u_{R}^i}\gamma^\mu d_R^i W^+_{R\mu} -h.c. \nn\\&\hspace{10pt}+g'\tan\theta_R\left(\overline{q^i_{L}}  \gamma^\mu Y_{q} q^i_{L}+\frac{2}{3}\overline{T_L} \gamma^\mu T_L-\frac{1}{3}\overline{B_L} \gamma^\mu B_L  \right) Z_{R\mu}\nn\\&\hspace{10pt} -\left\lbrace\frac{g_R}{2\cos\theta_R}(\overline{u^i_{R}} \gamma^\mu u^i_{R}-\overline{d^i_{R}} \gamma^\mu d^i_{R}) - g'\tan\theta_R\left(\frac{2}{3}(\overline{u^i_{R}} \gamma^\mu u^i_{R}+\overline{T_R}  \gamma^{\mu}  T_R) \right.\right. \nn\\  &\hspace{10pt}\left. \left. - \frac{1}{3}(\overline{d^i_{R}} \gamma^\mu d^i_{R}+\overline{B_R}  \gamma^{\mu}  B_R)  \right)   \right\rbrace  Z_{R\mu} \nn\\&\hspace{10pt}- Y_{u_L}^3\overline{q_L^3}\tilde{\phi}_L T_R -Y_{u_R}^3 \frac{v_R}{\sqrt{2}} \overline{T_L}  u_R^3 -\overline{T_L}M_T T_R -h.c.\nn\\&\hspace{10pt} -Y_{u_R}^3\overline{T_L}\left(\frac{1}{\sqrt{2}} u_R^3(h_R+i \chi_R^3)  -  d_R^3 \chi_R^+  \right) -h.c. \nn\\&\hspace{10pt}- \overline{q^i_{L}} y^i_{d_L} \phi_L B_R - \overline{B_L} y^{i\ast}_{d_R} \frac{v_R}{\sqrt{2}} d_R^i  - \overline{B_L} M_B B_R -h.c.\nn\\ &\hspace{10pt} -\overline{B_L}y^{i\ast}_{d_R}\left( \frac{1}{\sqrt{2}} d_R^i(h_R-i\chi_R^3) + u_R^i \chi_R^-\right) -h.c., \label{quark Lagrangian2}
\end{align}
where $i\in\{1,2,3\}$ is the family index, and the SM covariant derivatives have the following expressions,
\begin{align}
	D_{\text{SM}\mu} q^i_{L} &= \left( \partial_\mu + i g_{L} \frac{\tau^a}{2} W^a_{L\mu}  + i g' Y_{q_{L}} B_{\mu}\right)  q^i_{L}, \label{cov der SM quark doublet}\\
	D_{\text{SM}\mu} f_u  &= \left( \partial_\mu  +\frac{2}{3} i g' B_{\mu}\right) f_u, \label{cov der SM up type} \\
	D_{\text{SM}\mu} f_d &=\left( \partial_\mu  -\frac{1}{3} i g' B_{\mu}\right) f_d, \label{cov der SM down type}
\end{align}
where $f_u \in \left\lbrace u_R^i, T_{L,R}\right\rbrace$ and $f_d \in  \left\lbrace d_R^i, B_{L,R}\right\rbrace$. At this stage, $\mathrm{U(1)_Y}$ hypercharge can be obtained as following Eq.(\ref{charge convention}), $Y=I^3_R + Y'$. In Eqs.(\ref{cov der SM up type}) and (\ref{cov der SM down type}), we write the $\mathrm{U(1)_Y}$ hypercharge of the corresponding fields explicitly. Next, we follow several steps to reach the Lagrangian invariant under $\mathrm{SU(2)_L\times U(1)_Y}$ gauge symmetry.

\begin{itemize}
	\item \textbf{Step 1}: Rotate $d_R^i$ by the following transformation,
\begin{equation}
	d_R^i = (V_{d_R})^{ij} (d'_R)^j, \label{step 1}
\end{equation}
where $V_{d_R}$ is $3\times 3$ unitary matrix, which is related to Yukawa coupling parameterization as shown in Eq.(\ref{down yukawa general}),
	\begin{align}
		y_{d_{R}}&=\left(\begin{array}{c}
			\sin\theta^d_{R} \sin \phi^d_{R} e^{i \alpha^1_{d_{R}}} \\
			\sin\theta^d_{R} \cos \phi^d_{R} e^{i \alpha^2_{d_{R}}}	\\
			\cos\theta^d_{R} e^{i \alpha^3_{d_{R}}}
		\end{array} \right) Y^3_{d_{R}} = \mathbf{e}^d_{R_3} Y^3_{d_{R}}, \label{right handed down yukawa} \\
		V_{d_R}&=\left(\begin{array}{ccc}
		\mathbf{e}^d_{R_1}	& \mathbf{e}^d_{R_2} & \mathbf{e}^d_{R_3}
		\end{array} \right)\label{Vdr}. 
	\end{align}
	
	If we multiply Eq.(\ref{Vdr}) by the Hermitian conjugate of Eq.(\ref{right handed down yukawa}) from the left, it can be shown that the terms in Eq.(\ref{quark Lagrangian2}) which proportional to complex vector $y^\ast_{d_R}$ are replaced by a real positive number $Y^3_{d_R}$ multiply with $\delta^{3j}$. Then, we can extract the mass terms from the Lagrangian as follows,
\begin{align}
	\mathcal{L}_q \supset \mathcal{L}_{\text{mass}} &=  -\overline{T_L}\left( \begin{array}{cc}
		Y_{u_R}^3 \frac{v_R}{\sqrt{2}} 	& M_T
	\end{array}\right)  \left( \begin{array}{c}
		u_R^3\\
		T_R
	\end{array}\right) -h.c.\nn\\&\hspace{10pt} -\overline{B_L}\left( \begin{array}{cc}
		Y_{d_R}^3 \frac{v_R}{\sqrt{2}} 	& M_B
	\end{array}\right)  \left( \begin{array}{c}
		(d'_R)^3\\
		B_R
	\end{array}\right) -h.c.. \label{mass right handed}
\end{align}	

After doing transformation in Eq.(\ref{step 1}), $V_{d_R}$ appears as a CKM-like matrix in the right-handed charged current term,
\begin{equation}
	\mathcal{L}_q \supset	\mathcal{L}_{\text{RCC}} =-\frac{g_R}{\sqrt{2}}\sum_{i,j=1}^{3}\overline{u_{R}^i}\gamma^\mu (V_{d_R})^{ij} (d'_R)^j W^+_{R\mu} -h.c.. \label{right charge current}
\end{equation}
Eq.(\ref{mass right handed}) shows that the first and second families are decoupled from the Yukawa coupling. This leads to the fact that we have the freedom to do another $\mathrm{U(2)}$ transformation for the right-handed quark fields. This rotation should keep the third family unchanged. 
	
	\item \textbf{Step 2:} Rotate $u_R^i$ and $(d'_R)^i$ by the following transformations,
	\begin{align}
		u_R^ i &= \sum_{j=1}^{3}(\widetilde{U}_{u_R})^{ij} (\tilde{u}_R)^j, \label{Uur}\\ 
		(d'_R)^i &=\sum_{j=1}^{3}(\widetilde{W}_{d_R})^{ij} (\tilde{d}'_R)^j, \label{Udr}
	\end{align}
	where $\widetilde{U}_{u_R}$ and $\widetilde{W}_{d_R}$ are $3\times 3$ unitary matrices and written in matrix form as follows,
	\begin{align}
		\widetilde{U}_{u_R} & = \left( \begin{array}{ccc}
			&  &  0\\
			\multicolumn{2}{c}{\smash{\raisebox{.5\normalbaselineskip}{$U_{u_R}$}}}
			&  0\\
			0& 0 &1 
		\end{array}\right),\label{Uur 2} \\ \widetilde{W}_{d_R} & = \left( \begin{array}{ccc}
		&  &  0\\
		\multicolumn{2}{c}{\smash{\raisebox{.5\normalbaselineskip}{$W_{d_R}$}}}
		&  0\\
		0& 0 &1 
		\end{array}\right), \label{Wdr 2}
	\end{align}
	 with $U_{u_R}$ and $W_{d_R}$ are $2\times 2$ unitary matrices that rotate $(u_R^1, u_R^2)$ and $((d'_R)^1,(d'_R)^2)$, respectively. By applying the transformations in Eqs.(\ref{Uur}) and (\ref{Udr}) to the charged current in Eq.(\ref{right charge current}), we further define, 
	\begin{equation}
		\widetilde{V}_{d_R} = \widetilde{U}^\dagger_{u_R} V_{d_R} \widetilde{W}_{d_R}. \label{removing phase Vdr}
	\end{equation}
	As shown in Eq.(\ref{B9}),  by choosing $\widetilde{U}_{u_R}$ and $\widetilde{W}_{d_R}$ properly, the unphysical phases and angles in $V_{d_R}$ are removed and $\widetilde{V}_{d_R}$ has the following matrix form, 
	\begin{equation}
		\widetilde{V}_{d_R} = \left( \begin{array}{ccc}
			1 & 0 & 0 \\
			0 & \cos\theta^d_{R} & \sin\theta^d_{R} e^{i \frac{\alpha^3_{d_R}}{2} }  \\
			0 & -\sin\theta^d_{R} e^{i \frac{\alpha^3_{d_R}}{2} }  & \cos\theta^d_{R} e^{i\alpha^3_{d_R}}
		\end{array}\right). \label{Vdr removed}
	\end{equation}
	The details of the parameterization and the procedure for the removal of unphysical phases and angles of $V_{d_R}$ are shown in Appendix \ref{sec:appendix removing phase}
	
	\item \textbf{Step 3}: Rotate $(\tilde{u}_R)^\alpha$ and $(\tilde{d}'_R)^\alpha$ by the following transformations,
	\begin{align}
		(\tilde{u}_R)^\alpha &= \sum_{\beta=1}^{4} (\widetilde{W}_{T_R})^{\alpha \beta} (\tilde{u}'_R)^\beta, \\
		 (\tilde{d}'_R)^\alpha &=\sum_{\beta=1}^{4} (\widetilde{W}_{B_R})^{\alpha \beta}(\tilde{d}^{\prime\prime}_R)^\beta, 
	\end{align}
	where $\alpha=\{1,2,3,4\}$, $(\tilde{u}_R)^4=T_R$, and $(\tilde{d}'_R)^4 = B_R$. The $4\times 4$ unitary matrices $\widetilde{W}_{T_R}$ and $\widetilde{W}_{B_R}$ are expressed as follows,
	\begin{align}
		\widetilde{W}_{T_R} = \left(\begin{array}{cc}
			I_2 & 0_2 \\
			0_2 & {W}_{T_R}
		\end{array} \right), \\
		\widetilde{W}_{B_R} = \left(\begin{array}{cc}
			I_2 & 0_2 \\
			0_2 & {W}_{B_R}
		\end{array} \right),
	\end{align}
		where $I_2$ and $0_2$ are the $2\times 2$ identity matrix and zero matrix, respectively. The $2 \times 2$ submatrices $W_{T_R}$ and $W_{B_R}$ rotate $((\tilde{u}_R)^3,(\tilde{u}_R)^4)$ and $((\tilde{d}'_R)^3 ,(\tilde{d}'_R)^4 )$, respectively by the following expressions,
			\begin{align}
			(\tilde{u}_R)^i &= \sum_{j=3}^{4} (W_{T_R})^{ij} (\tilde{u}'_R)^j, \label{right handed up mixing}\\
			(\tilde{d}'_R)^i &=\sum_{j=3}^{4} (W_{B_R})^{ij}(\tilde{d}^{\prime\prime}_R)^j, \label{right handed down mixing}
		\end{align}
		where $i\in\{3,4\}$. The explicit matrix form of $W_{T_R}$ and $W_{B_R}$ are as follows,
			\begin{align}
		{W}_{T_R} &= \left(\begin{array}{cc}
			\cos \theta_{T_R} & \sin\theta_{T_R} \\
			-\sin\theta_{T_R} & \cos \theta_{T_R}
		\end{array} \right), \label{WTR} \\
			{W}_{B_R} &= \left(\begin{array}{cc}
							\cos \theta_{B_R} & \sin\theta_{B_R} \\
		-\sin\theta_{B_R} & \cos \theta_{B_R}
		\end{array} \right), \label{WBR}
		\end{align}
	where the mixing angles have the following expressions,
	\begin{align}
		\cos \theta_{T_R} &= \frac{M_T}{m_{u_4}},\quad \sin\theta_{T_R} = \frac{Y^3_{u_R} }{m_{u_4}}\frac{v_R}{\sqrt{2}},\quad \cos \theta_{B_R} = \frac{M_B}{m_{d_4}},\quad \sin\theta_{B_R} = \frac{Y^3_{d_R} }{m_{d_4}}\frac{v_R}{\sqrt{2}}, \nn\\
		&m_{u_4} = \sqrt{\frac{(Y^3_{u_R})^2 v_R^2}{2} + M_T^2}, \qquad m_{d_4} = \sqrt{\frac{(Y^3_{d_R})^2 v_R^2}{2} + M_B^2} \label{WTR WBR mixing angles}.
		\end{align}
	 By using Eqs.(\ref{right handed up mixing}) and (\ref{right handed down mixing}), the mass terms in Eq.(\ref{mass right handed}) transform into,
	\begin{align}
	\mathcal{L}_q \supset	\mathcal{L}_{\text{mass}} =  - m_{u_4}\overline{T_L}(\tilde{u}'_R)^4  -m_{d_4}\overline{B_L}(\tilde{d}^{\prime\prime}_R)^4 -h.c.. \label{mass right handed 2}
	\end{align}	
	
	The right-handed charged current in Eq.(\ref{right charge current}) becomes,
	\begin{equation}
	\mathcal{L}_q \supset \mathcal{L}_{\text{RCC}} =-\frac{g_R}{\sqrt{2}}\sum_{\alpha,\beta=1}^{4}\overline{(\tilde{u}'_R)^\alpha}\gamma^\mu (V^{\text{CKM}}_R)^{\alpha\beta} (\tilde{d}^{\prime\prime}_R)^\beta W^+_{R\mu} -h.c., \label{right charge current 2}
	\end{equation}
	where
	\begin{equation}
		(V^{\text{CKM}}_R)^{\alpha\beta} =\sum_{i,j=1}^{3} ({\widetilde{W}}^\dagger_{T_R})^{\alpha i} (\tilde{V}_{d_R})^{ij} (\widetilde{W}_{B_R})^{j\beta};\quad \alpha,\beta\in \{1,2,3,4\} \label{right CKM intermediate}
	\end{equation}
	is $4\times 4$ ``intermediate" right-handed CKM-like matrix. We call this matrix intermediate because it is not the final expression of the right-handed CKM-like matrix. The explicit matrix form of $V_R^{\text{CKM}}$ is shown in Eq.(\ref{VRCKM}).
	%\begin{equation}
	%	V_R^{\text{CKM}} = \left(\begin{array}{cccc}
	%		1& 0 &0  & 0 \\
	%		0	& \cos\theta^d_{R} & \sin\theta^d_{R}\cos\theta_{B_R} e^{i \frac{\alpha^3_{d_R}}{2}}& \sin\theta^d_{R}\sin\theta_{B_R} e^{i \frac{\alpha^3_{d_R}}{2}} \\
	%		0	& -\cos\theta_{T_{R}}\sin\theta^d_{R} e^{i \frac{\alpha^3_{d_R}}{2}} & \cos\theta_{T_{R}}\cos\theta^d_{R}\cos\theta_{B_R} e^{i \alpha^3_{d_R}} &\cos\theta_{T_R}\cos\theta^d_{R}\sin\theta_{B_R} e^{i \alpha^3_{d_R}} \\
	%		0&-\sin\theta_{T_R}\sin\theta^d_{R}e^{i \frac{\alpha^3_{d_R}}{2}}&\sin\theta_{T_R}\cos\theta^d_{R}\cos\theta_{B_R}e^{i \alpha^3_{d_R}}&\sin\theta_{T_R}\cos\theta^d_{R}\sin\theta_{B_R}e^{i \alpha^3_{d_R}}
	%	\end{array} \right) 
	%\end{equation}
	
	In addition, we define the right-handed weak isospin current in Eq.(\ref{quark Lagrangian2}) as
\begin{align}
	j^\mu_{3R}\equiv \overline{u^i_{R}} \gamma^\mu u^i_{R}-\overline{d^i_{R}} \gamma^\mu d^i_{R}. \label{right weak isospin}
\end{align}
Following steps 1 to 3, Eq.(\ref{right weak isospin}) transforms into,
\begin{align}
	j^\mu_{3R}&=\sum_{i=1}^{2}\overline{(\tilde{u}'_R)^i} \gamma^\mu (\tilde{u}'_R)^i+\sum_{j,k=3}^{4}\overline{(\tilde{u}'_R)^j} \gamma^\mu (Z_{T_R})^{jk} (\tilde{u}'_R)^k\nn\\ &-\sum_{i=1}^{2}\overline{(\tilde{d}^{\prime\prime}_R)^i} \gamma^\mu (\tilde{d}^{\prime\prime}_R)^i-\sum_{j,k=3}^{4}\overline{(\tilde{d}^{\prime\prime}_R)^j} \gamma^\mu (Z_{B_R})^{jk}(\tilde{d}^{\prime\prime}_R)^k\label{right weak isospin 2},
\end{align}
where the tree-level FCNC couplings are generated with the following definitions,
	\begin{align}
		(Z_{T_R})^{jk}&\equiv ({W}^\dagger_{T_R})^{j3}({W}_{T_R})^{3k} \label{def ZTR}, \\
		(Z_{B_R})^{jk}&\equiv ({W}^\dagger_{B_R})^{j3}({W}_{B_R})^{3k}\label{def ZBR},
	\end{align}
	with $j, k \in \{3, 4\}$.
	Furthermore, Eqs.(\ref{def ZTR}) and (\ref{def ZBR}) can be expressed explicitly in $2\times 2$ matrix form as follows,
	\begin{align}
		Z_{T_R}&= \left(\begin{array}{cc}
			\cos^2 \theta_{T_R} & \sin\theta_{T_R}\cos\theta_{T_R} \\
			\sin\theta_{T_R}\cos\theta_{T_R} & \sin^2\theta_{T_R}
		\end{array} \right), \label{ZTR 1} \\
		Z_{B_R}&= \left(\begin{array}{cc}
			\cos^2 \theta_{B_R} & \sin\theta_{B_R}\cos\theta_{B_R} \\
			\sin\theta_{B_R}\cos\theta_{B_R} & \sin^2\theta_{B_R}
		\end{array} \right). \label{ZBR 1}
	\end{align}
	These tree-level FCNC couplings are generated due to mixing between the third flavor of up and down quark with their corresponding isosinglet right-handed VLQ. 
	\end{itemize}
\vspace{5pt}
After following steps 1 to 3, the Lagrangian in Eq.(\ref{quark Lagrangian2}) becomes,
\begin{align}
	{\cal L}_{q}&= 
	\overline{q^i_{L}} i \gamma^\mu D_{\text{SM}\mu} q^i_{L} + \overline{T_L} i \gamma^{\mu} D_{\text{SM}\mu} T_L + \overline{B_L} i \gamma^{\mu} D_{\text{SM}\mu} B_L \nn \\
	&\hspace{10pt}+ \overline{(\tilde{u}'_R)^\alpha} i \gamma^\mu D_{\text{SM}\mu} (\tilde{u}'_R)^\alpha+ \overline{(\tilde{d}^{\prime\prime}_R)^\alpha} i \gamma^\mu D_{\text{SM}\mu} (\tilde{d}^{\prime\prime}_R)^\alpha \nn\\&\hspace{10pt}-\frac{g_R}{\sqrt{2}}\sum_{\alpha,\beta=1}^{4}\overline{(\tilde{u}'_R)^\alpha}\gamma^\mu (V^{\text{CKM}}_R)^{\alpha\beta} (\tilde{d}^{\prime\prime}_R)^\beta W^+_{R\mu} -h.c \nn\\&\hspace{10pt}+g'\tan\theta_R\left(\overline{q^i_{L}}  \gamma^\mu Y_{q} q^i_{L}+\frac{2}{3}\overline{T_L} \gamma^\mu T_L-\frac{1}{3}\overline{B_L} \gamma^\mu B_L  \right) Z_{R\mu}\nn\\&\hspace{10pt} -\left\lbrace\frac{g_R}{2\cos\theta_R}(j^\mu_{3_R}) - g'\tan\theta_R\left(\frac{2}{3}\overline{(\tilde{u}'_R)^\alpha} \gamma^\mu (\tilde{u}'_R)^\alpha  - \frac{1}{3}\overline{(\tilde{d}^{\prime\prime}_R)^\alpha} \gamma^\mu (\tilde{d}^{\prime\prime}_R)^\alpha \right)   \right\rbrace  Z_{R\mu} \nn\\&\hspace{10pt}- Y_{u_L}^3\overline{q_L^3}\tilde{\phi}_L \left(\sum_{j=3}^{4}(W_{T_R})^{4j} (\tilde{u}'_R)^j \right) -m_{u_4} \overline{T_L}(\tilde{u}'_R)^4  -h.c.\nn\\&\hspace{10pt} -\frac{m_{u_4}}{v_R}\overline{T_L}\left[ \left(\sum_{j=3}^{4}(Z_{T_R})^{4j}(\tilde{u}'_R)^j \right) (h_R+i \chi_R^3)  - \sqrt{2} \left(\sum_{\beta=2}^{4}(V_R^{\text{CKM}})^{4\beta}(\tilde{d}^{\prime\prime}_R)^\beta  \right)  \chi_R^+  \right] -h.c. \nn\\&\hspace{10pt}- \overline{q^i_{L}} y^i_{d_L} \phi_L \left(\sum_{j=3}^{4}(W_{B_R})^{4j} (\tilde{d}^{\prime\prime}_R)^j \right) -m_{d_4} \overline{B_L}(\tilde{d}^{\prime\prime}_R)^4 -h.c.\nn\\ &\hspace{10pt} - \frac{m_{d_4}}{v_R}\overline{B_L}\left[ \left(\sum_{j=3}^{4}(Z_{B_R})^{4j}(\tilde{d}^{\prime\prime}_R)^j \right) (h_R-i\chi_R^3) +\sqrt{2}\left(\sum_{\beta=2}^{4}(V_R^{\text{CKM}\dagger})^{4\beta}(\tilde{u}'_R)^\beta  \right)  \chi_R^-\right] -h.c., \label{quark Lagrangian3}
\end{align}
where $i=\{1,2,3\}$, $\alpha=\{1,2,3,4\}$ and the definition of $W_{T_R},W_{B_R},m_{u_4},m_{d_4},V_R^{\text{CKM}},Z_{T_R}$, and $Z_{B_R}$ are written in Eqs.(\ref{WTR}),(\ref{WBR}),(\ref{WTR WBR mixing angles}),(\ref{VRCKM}),(\ref{ZTR 1}), and (\ref{ZBR 1}) respectively.
One can show that Lagrangian in Eq.(\ref{quark Lagrangian3}) is invariant under $\mathrm{SU(2)_L \times U(1)_Y}$ gauge symmetry.
\subsection{$\mathbf{SU(2)_L \times U(1)_Y}\rightarrow \mathbf{U(1)_\text{em}}$ }
In this stage, the neutral scalar component of $\mathrm{SU(2)_L}$ Higgs doublet acquires non-zero vev and is expanded around vev's as follows,
\begin{equation}
	\phi_L =\left( \begin{array}{c}
		\chi_L^+	\\\chi_L^0
	\end{array}\right)= \frac{1}{\sqrt{2}}\left(\begin{array}{c}
		\sqrt{2}\chi_L^+ \\
		v_L+h_L+i \chi_L^3
	\end{array} \right), \label{higgs left parameterization}
\end{equation}
where $v_L$ is the non-zero vev, $h_L$ is the neutral CP-even state and $\chi_L^3$ is the neutral CP-odd state. The charged component, $\chi_L^+ = \frac{1}{\sqrt{2}} (\chi_L^1 + i \chi_L^2)$. In addition, we rotate the gauge fields with the following transformation,
\begin{equation}
	\left( \begin{array}{c}
		B_\mu \\
		W^3_{L\mu}
	\end{array}\right) = \left( \begin{array}{cc}
		\cos\theta_W & -\sin\theta_W \\
		\sin\theta_W & \cos\theta_W
	\end{array}\right)\left( \begin{array}{c}
		A_\mu \\
		Z_{L\mu}
	\end{array}\right), \label{B WL mixing}
\end{equation} 
where the mixing angles are defined as,
\begin{align}
	\cos\theta_W = \frac{g_L}{\sqrt{g_L^2+g^{\prime 2}}}, \qquad \sin\theta_W = \frac{g'}{\sqrt{g_L^2+g^{\prime 2}}}.
\end{align}
We also define the electromagnetic $\mathrm{U(1)_{\text{em}}}$ gauge coupling as,
\begin{equation}
	e=g' \cos\theta_W = g_L \sin\theta_W.
\end{equation}
After this breaking, the Lagrangian in Eq.(\ref{quark Lagrangian3}) becomes
\begin{align}
	{\cal L}_{q}&= 
	\overline{u^i_{L}} i \gamma^\mu D_{\text{em}\mu} u^i_{L} + \overline{T_L} i \gamma^{\mu} D_{\text{em}\mu} T_L+\overline{d^i_{L}} i \gamma^\mu D_{\text{em}\mu} d^i_{L} + \overline{B_L} i \gamma^{\mu} D_{\text{em}\mu} B_L \nn \\
	&\hspace{10pt}+ \overline{(\tilde{u}'_R)^\alpha} i \gamma^\mu D_{\text{em}\mu} (\tilde{u}'_R)^\alpha+ \overline{(\tilde{d}^{\prime\prime}_R)^\alpha} i \gamma^\mu D_{\text{em}\mu} (\tilde{d}^{\prime\prime}_R)^\alpha \nn\\&\hspace{10pt}-\frac{g_L}{\sqrt{2}}\overline{u^i_{L}} \gamma^\mu d_L^i W^+_{L\mu}-h.c. \nn\\&\hspace{10pt} -\left( \frac{g_L}{2\cos\theta_W}(j^\mu_{3L})-e \tan\theta_W (j^\mu_{\text{em}}) \right)Z_{L\mu}   \nn\\&\hspace{10pt}-\frac{g_R}{\sqrt{2}}\sum_{\alpha,\beta=1}^{4}\overline{(\tilde{u}'_R)^\alpha}\gamma^\mu (V^{\text{CKM}}_R)^{\alpha\beta} (\tilde{d}^{\prime\prime}_R)^\beta W^+_{R\mu} -h.c. \nn\\&\hspace{10pt} -\left\lbrace\frac{g_R}{2\cos\theta_R}(j^\mu_{3_R}) - g'\tan\theta_R\left((j^\mu_{\text{em}})-\frac{1}{2}(j^\mu_{3_L})\right)   \right\rbrace  Z_{R\mu} \nn\\&\hspace{10pt}- Y_{u_L}^3\frac{v_L}{\sqrt{2}}\overline{u_L^3} \left(\sum_{j=3}^{4}(W_{T_R})^{4j} (\tilde{u}'_R)^j \right) -m_{u_4} \overline{T_L}(\tilde{u}'_R)^4  -h.c.\nn\\&\hspace{10pt}- Y_{u_L}^3\left(\frac{1}{\sqrt{2}} \overline{u_L^3}\left(\sum_{j=3}^{4}(W_{T_R})^{4j} (\tilde{u}'_R)^j \right)(h_L-i\chi_L^3) -\overline{d_L^3}\left(\sum_{j=3}^{4}(W_{T_R})^{4j} (\tilde{u}'_R)^j \right)\chi_L^-   \right)-h.c.  \nn\\&\hspace{10pt} -\frac{m_{u_4}}{v_R}\overline{T_L}\left[ \left(\sum_{j=3}^{4}(Z_{T_R})^{4j}(\tilde{u}'_R)^j \right) (h_R+i \chi_R^3)  - \sqrt{2} \left(\sum_{\beta=2}^{4}(V_R^{\text{CKM}})^{4\beta}(\tilde{d}^{\prime\prime}_R)^\beta  \right)  \chi_R^+  \right] -h.c. \nn\\&\hspace{10pt}- y^i_{d_L} \frac{v_L}{\sqrt{2}} \overline{d^i_{L}}  \left(\sum_{j=3}^{4}(W_{B_R})^{4j} (\tilde{d}^{\prime\prime}_R)^j \right) -m_{d_4} \overline{B_L}(\tilde{d}^{\prime\prime}_R)^4 -h.c. \nn\\&\hspace{10pt} - y^i_{d_L}\left( \frac{1}{\sqrt{2}} \overline{d^i_{L}}  \left(\sum_{j=3}^{4}(W_{B_R})^{4j} (\tilde{d}^{\prime\prime}_R)^j \right) (h_L+i \chi_L^3) + \overline{u^i_{L}}\left(\sum_{j=3}^{4}(W_{B_R})^{4j} (\tilde{d}^{\prime\prime}_R)^j \right) \chi_L^+  \right) -h.c.   \nn\\ &\hspace{10pt} - \frac{m_{d_4}}{v_R}\overline{B_L}\left[ \left(\sum_{j=3}^{4}(Z_{B_R})^{4j}(\tilde{d}^{\prime\prime}_R)^j \right) (h_R-i\chi_R^3) +\sqrt{2}\left(\sum_{\beta=2}^{4}(V_R^{\text{CKM}\dagger})^{4\beta}(\tilde{u}'_R)^\beta  \right)  \chi_R^-\right] -h.c., \label{quark Lagrangian4}
\end{align}
where the  covariant derivatives are,
\begin{align}
	D_{\text{em}\mu} f'_u &= \left( \partial_\mu + \frac{2}{3} i e A_\mu \right) f'_u  \label{em cov der up type},\\ 
	D_{\text{em}\mu} f'_d &= \left( \partial_\mu - \frac{1}{3} i e A_\mu \right) f'_d.  \label{em cov der down type}
\end{align}
The left-handed weak isospin current and electromagnetic current are
\begin{align}
j^\mu_{3L} & =\overline{u^i_{L}} \gamma^\mu u^i_{L} - \overline{d^i_{L}} \gamma^\mu d^i_{L}, \label{left handed weak isospin} \\
	j^\mu_{\text{em}} &= \frac{2}{3}\left(\overline{u^i_{L}} \gamma^\mu u^i_{L} + \overline{T_L}\gamma^\mu T_L  +\overline{(\tilde{u}'_R)^\alpha}\gamma^\mu (\tilde{u}'_R)^\alpha  \right) \nn\\ &\hspace{10pt}- \frac{1}{3}\left(\overline{d^i_{L}} \gamma^\mu d^i_{L} + \overline{B_L}\gamma^\mu B_L +(\tilde{d}^{\prime\prime}_R)^\alpha \gamma^\mu (\tilde{d}^{\prime\prime}_R)^\alpha \right), \label{electromagnetic current}
\end{align}
where $f'_u \in \left\lbrace u^i_{L},(\tilde{u}'_R)^\alpha,T_L \right\rbrace$, $f'_d \in \left\lbrace d_L^i,(\tilde{d}^{\prime\prime}_R)^\alpha,B_L \right\rbrace$, $i\in\{1,2,3\}$, $\alpha\in\{1,2,3,4\}$ and the right-handed weak isospin current $j^\mu_{3R}$ is written in Eq.(\ref{right weak isospin 2}). Our main goal is to obtain the mass eigenvalues of the top and bottom quarks and their heavy partners. The following steps outline our approach: (the number of counting steps continues from the previous subsection)
\begin{itemize}
	\item \textbf{Step 4:} Rotate $d_L^i$ by the following transformation, 
	
	\begin{equation}
		d_L^i = (V_{d_L})^{ij} (d'_L)^j \label{step 4},
	\end{equation}
	where $V_{d_L}$ is $3\times 3$ unitary matrix, associated with the parameterization of Yukawa couplings as demonstrated in Eq.(\ref{down yukawa general}),
	\begin{align}
		y_{d_{L}}&=\left(\begin{array}{c}
			\sin\theta^d_{L} \sin \phi^d_{L} e^{i \alpha^1_{d_{L}}} \\
			\sin\theta^d_{L} \cos \phi^d_{L} e^{i \alpha^2_{d_{L}}}	\\
			\cos\theta^d_{L} e^{i \alpha^3_{d_{L}}}
		\end{array} \right) Y^3_{d_{L}} = \mathbf{e}^d_{L_3} Y^3_{d_{L}} \label{left handed down yukawa}, \\
		V_{d_L}&=\left(\begin{array}{ccc}
			\mathbf{e}^d_{L_1}	& \mathbf{e}^d_{L_2} & \mathbf{e}^d_{L_3}
		\end{array} \right)\label{Vdl}. 
	\end{align}
	If we multiply Eq.(\ref{left handed down yukawa}) by the hermitian conjugate of Eq.(\ref{Vdl}) from the left, it can be shown that the terms in Eq.(\ref{quark Lagrangian4}) that proportional to the complex vector  $y_{d_L}$ are replaced by the product of a real positive number $Y^3_{d_{L}}$ and $\delta^{j3}$. The mass terms can be extracted from the Lagrangian and written as follows,
		\begin{align}
		\mathcal{L}_q \supset	\mathcal{L}_{\text{mass}}  = & - \left( \begin{array}{cc}
		\overline{u_L^3}	& \overline{T_L}
		\end{array}\right)  \left(\begin{array}{cc}
		 Y_{u_L}^3\frac{v_L}{\sqrt{2}}({W}_{T_R})^{43}	&  Y_{u_L}^3\frac{v_L}{\sqrt{2}}({W}_{T_R})^{44} \\
			0 & m_{u_4}
		\end{array} \right) \left(\begin{array}{c}
		(\tilde{u}'_R)^3\\
		(\tilde{u}'_R)^4
		\end{array} \right) -h.c. \nn\\ &-\left( \begin{array}{cc}
		\overline{(d'_L)^3}	& \overline{B_L}
		\end{array}\right)  \left(\begin{array}{cc}
		Y_{d_L}^3\frac{v_L}{\sqrt{2}}({W}_{B_R})^{43}	&  Y_{d_L}^3\frac{v_L}{\sqrt{2}}({W}_{B_R})^{44} \\
		0 & m_{d_4}
		\end{array} \right) \left(\begin{array}{c}
		(\tilde{d}^{\prime\prime}_R)^3\\
		(\tilde{d}^{\prime\prime}_R)^4
		\end{array} \right) -h.c.. \label{mass term}
	\end{align}	
	 	Additionally, an important outcome of the transformation in Eq.(\ref{step 4}) is that $V_{d_L}$ appears as CKM-like matrix in the left-handed charged current term,
	 	\begin{equation}
	 	\mathcal{L}_q \supset	\mathcal{L}_{\text{LCC}} =-\frac{g_L}{\sqrt{2}}\sum_{i,j=1}^{3}\overline{u_{L}^i}\gamma^\mu (V_{d_L})^{ij} (d'_L)^j W^+_{L\mu} -h.c.. \label{left charge current}
	 \end{equation}
	 From Eq.(\ref{mass term}), we have freedom to do another $\mathrm{U(2)}$ transformation to the left-handed quark fields with keeping the third family unchanged. 
	
	\item \textbf{Step 5:} Rotate $u_L^i$ and $(d'_L)^i$ by the following transformations
	\begin{align}
		u_L^ i &= \sum_{j=1}^{3}(\widetilde{U}_{u_L})^{ij} (\tilde{u}_L)^j \label{Uul},\\ 
		(d'_L)^i &=\sum_{j=1}^{3}(\widetilde{W}_{d_L})^{ij} (\tilde{d}'_L)^j \label{Udl},
	\end{align}
	where $\widetilde{U}_{u_L}$ and $\widetilde{W}_{d_L}$ are $3\times 3$ unitary matrices and written in the matrix form as follows,
	\begin{align}
		\widetilde{U}_{u_L} & = \left( \begin{array}{ccc}
			&  &  0\\
			\multicolumn{2}{c}{\smash{\raisebox{.5\normalbaselineskip}{$U_{u_L}$}}}
			&  0\\
			0& 0 &1 
		\end{array}\right),\label{Uul 2} \\ \widetilde{W}_{d_L} & = \left( \begin{array}{ccc}
			&  &  0\\
			\multicolumn{2}{c}{\smash{\raisebox{.5\normalbaselineskip}{$W_{d_L}$}}}
			&  0\\
			0& 0 &1 
		\end{array}\right), \label{Wdl 2}
	\end{align}
	with $U_{u_L}$ and $W_{d_L}$ are $2\times 2$ unitary matrices which rotate $(u_L^1, u_L^2)$ and $((d'_L)^1,(d'_L)^2)$, respectively. By applying the transformations in Eqs.(\ref{Uul}) and (\ref{Udl}) to the charged current in Eq.(\ref{left charge current}), we further define
	\begin{equation}
		\widetilde{V}_{d_L} = \widetilde{U}^\dagger_{u_L} V_{d_L} \widetilde{W}_{d_L} \label{removing phase Vdl}.
	\end{equation}
	By properly choosing $\widetilde{U}_{u_L}$ and $\widetilde{W}_{d_L}$, the unphysical phases and angles in $V_{d_L}$ are eliminated, resulting in $\widetilde{V}_{d_L}$, which has the same matrix form as Eq.(\ref{Vdr removed}), with the $R$ index replaced by $L$. 
	
	\item \textbf{Step 6:} Rotate $(\tilde{u}_L)^\alpha$, $(\tilde{u}'_R)^\alpha$, $(\tilde{d}'_L)^\alpha$, and $(\tilde{d}^{\prime\prime}_R)^\alpha$ into the mass basis by the following transformations,
	\begin{align}
		(\tilde{u}_L)^\alpha	
			 &= \sum_{\beta= 1}^{4}(\widetilde{K}_{T_L})^{\alpha \beta} 	({u}_L^m)^\beta	
		\label{KTL},\\
		(\tilde{u}'_R)^\alpha &=\sum_{\beta= 1}^{4} (\widetilde{K}_{T_R} )^{\alpha \beta} 
			({u}_R^m)^\beta				
		\label{KTR}, \\
			(\tilde{d}'_L)^\alpha	
			 &= \sum_{\beta= 1}^{4}(\widetilde{K}_{B_L} )^{\alpha\beta}	
			({d}_L^m)^\beta	
			\label{KBL},  \\
		(\tilde{d}^{\prime\prime}_R)^\alpha
			 &= \sum_{\beta= 1}^{4}(\widetilde{K}_{B_R})^{\alpha\beta} 	
			({d}_R^m)^\beta
			 \label{KBR},
	\end{align}
	where $\alpha\in\{1,2,3,4\}$, $(\tilde{u}_L)^4 = T_L$ and $(\tilde{d}'_L)^4 = B_L$. The $4 \times 4$ unitary matrices $\widetilde{K}_{T_L}$, $\widetilde{K}_{T_R}$, $\widetilde{K}_{B_L}$, and $\widetilde{K}_{B_R}$ are expressed as follows,
		\begin{align}
		\widetilde{K}_{T_L} = \left(\begin{array}{cc}
			I_2 & 0_2 \\
			0_2 & {K}_{T_L}
		\end{array} \right), \\
		\widetilde{K}_{T_R} = \left(\begin{array}{cc}
			I_2 & 0_2 \\
			0_2 & {K}_{T_R}
		\end{array} \right), \\
		\widetilde{K}_{B_L} = \left(\begin{array}{cc}
			I_2 & 0_2 \\
			0_2 & {K}_{B_L}
		\end{array} \right), \\
		\widetilde{K}_{B_R} = \left(\begin{array}{cc}
			I_2 & 0_2 \\
			0_2 & {K}_{B_R}
		\end{array} \right),
	\end{align}
	where $I_2$ and $0_2$ are the $2\times 2$ identity matrix and zero matrix, respectively. The $2\times 2$ unitary submatrices  $K_{T_L},K_{T_R}, K_{B_L}$, and $K_{B_R}$ rotate  ($(\tilde{u}_L)^3, (\tilde{u}_L)^4$),$((\tilde{u}'_R)^3,(\tilde{u}'_R)^4)$, $((\tilde{d}'_L)^3,(\tilde{d}'_L)^4)$ and $((\tilde{d}^{\prime\prime}_R)^3,(\tilde{d}^{\prime\prime}_R)^4)$ pairs, respectively where the explicit forms are written in Eqs.(\ref{KTL explicit}), (\ref{KTR explicit}), (\ref{KBL explicit}), and (\ref{KBR explicit}). 	 
		
		We denote the top and bottom quarks as the third component of the fields in the mass basis, while the heavy top and bottom quarks are the fourth component. We can diagonalize the mass matrices in Eq.(\ref{mass term}), which are defined as
	\begin{align}
		\mathbb{M}_t &\equiv \left(\begin{array}{cc}
			Y_{u_L}^3\frac{v_L}{\sqrt{2}}({W}_{T_R})^{43}	&  Y_{u_L}^3\frac{v_L}{\sqrt{2}}({W}_{T_R})^{44} \\
			0 & m_{u_4}
		\end{array} \right), \label{top mass matrix} \\
		\mathbb{M}_b &\equiv  \left(\begin{array}{cc}
			Y_{d_L}^3\frac{v_L}{\sqrt{2}}({W}_{B_R})^{43}	&  Y_{d_L}^3\frac{v_L}{\sqrt{2}}({W}_{B_R})^{44} \\
			0 & m_{d_4}
		\end{array} \right) \label{bottom mass matrix},
	\end{align}
    by using the appropriate submatrices in Eqs.(\ref{KTL}) - (\ref{KBR}) resulting in:
    \begin{align}
    	K^\dagger_{T_L} \mathbb{M}_t K_{T_R} &= (m^{\text{diag}}_t) = \text{diag}(m_t,m_{t'}) \label{diagonalizing top mass matrix}, \\
    	K^\dagger_{B_L} \mathbb{M}_b K_{B_R} &= (m^{\text{diag}}_b) = \text{diag}(m_b,m_{b'}) \label{diagonalizing bottom mass matrix}. 
    \end{align}
    From this diagonalization process, we obtain,
    \begin{align}
    	m_{t(b)} &= - \frac{\sqrt{M_{T(B)}^2 + (m_{u(d)_R} - m_{u(d)_L})^2}}{2} +  \frac{\sqrt{M_{T(B)}^2 + (m_{u(d)_R} + m_{u(d)_L})^2}}{2} \label{top and bottom exact mass},\\
    	m_{t'(b')} &=  \frac{\sqrt{M_{T(B)}^2 + (m_{u(d)_R} - m_{u(d)_L})^2}}{2} +  \frac{\sqrt{M_{T(B)}^2 + (m_{u(d)_R} + m_{u(d)_L})^2}}{2} \label{top and bottom partner exact mass},
    \end{align}
    where $m_{t(b)}$ and $m_{t'(b')}$ are the exact mass eigenvalues for the top(bottom) and heavy-top(bottom), respectively. The definitions of $m_{u_L}, m_{u_R}, m_{d_L}$, and $m_{d_R}$ are shown in Eqs.(\ref{mur and mul}) and (\ref{mdr and mdl}). The diagonalization procedure is explained in Appendix \ref{appendix quark mass} The mass eigenvalues for $t$ and $t'$ in Eqs.(\ref{top and bottom exact mass}) and (\ref{top and bottom partner exact mass}) agree with Eq.(10) of Ref.\cite{eftproc}.
    
    Moreover, the left-handed and right-handed charged currents in Eqs.(\ref{left charge current}) and (\ref{right charge current 2}), now become
	\begin{align}
		\mathcal{L}_q \supset	\mathcal{L}_{\text{CC}} &= \mathcal{L}_{\text{LCC}} + \mathcal{L}_{\text{RCC}} \nn\\ &=-\frac{g_L}{\sqrt{2}}\sum_{\alpha,\beta=1}^{4}\overline{({u}_L^m)^\alpha}\gamma^\mu (\mathcal{V}^{\text{CKM}}_{L})^{\alpha\beta} ({d}_L^m)^\beta W^+_{L\mu} -h.c. \nn\\ &\hspace{10pt}-\frac{g_R}{\sqrt{2}}\sum_{\alpha,\beta=1}^{4}\overline{	({u}_R^m)^{\alpha}}\gamma^\mu (\mathcal{V}^{\text{CKM}}_R)^{\alpha\beta} ({d}_R^m)^{\beta} W^+_{R\mu} -h.c., \label{charge current final}
		\end{align}
	where
	\begin{align}
		(\mathcal{V}^{\text{CKM}}_{L})^{\alpha\beta} &=\sum_{i,j=1}^{3} (\widetilde{K}^\dagger_{T_L})^{\alpha i} (\widetilde{V}_{d_L})^{ij} (\widetilde{K}_{B_L})^{j\beta} \label{left CKM final}, \\
		(\mathcal{V}^{\text{CKM}}_R)^{\alpha\beta} &=\sum_{\rho,\eta=1}^{4} (\widetilde{K}^\dagger_{T_R})^{\alpha \rho}(V^{\text{CKM}}_R)^{\rho\eta} (\widetilde{K}_{B_R})^{\eta\beta} \label{right CKM final}
	\end{align}
	are the left-handed and right-handed CKM-like matrices. The matrix forms are shown in Eqs.(\ref{VL CKM final}) and (\ref{VR CKM final}), respectively. However, there are some unphysical phases which can be eliminated from the left-handed and right-handed CKM-like matrices. We have the freedom to rephase the quark fields with the following transformations,
	\begin{align}
		(u_{L(R)}^m)^\alpha &= (\theta_{u_{L(R)}})^\alpha \delta^{\alpha \beta} (\hat{u}_{L(R)}^m)^\beta, \label{rephase up type 1}\\
		(d_{L(R)}^m)^\alpha &= (\theta_{d_{L(R)}})^\alpha \delta^{\alpha \beta} (\hat{d}_{L(R)}^m)^\beta \label{rephase down typec 1},
	\end{align}
	where,
	\begin{align}
		\theta_{u_{L(R)}} &= \mathrm{diag}(e^{i\theta_{u_{L(R)1}}},e^{i\theta_{u_{L(R)2}}},e^{i\theta_{u_3}},e^{i\theta_{u_4}}), \\ \theta_{d_{L(R)}}&=\mathrm{diag}(e^{i\theta_{d_{L(R)1}}},e^{i\theta_{d_{L(R)2}}},e^{i\theta_{d_3}},e^{i\theta_{d_4}}).
	\end{align}
	  After rephasing the quark fields, the left-handed and right-handed CKM-like matrices become the final versions denoted as $\mathcal{\hat{V}}^{\text{CKM}}_{L}$ and $\mathcal{\hat{V}}^{\text{CKM}}_{R}$, whose matrix forms are as follows,
	  \begin{align}
	  	\mathcal{\hat{V}}_L^{\text{CKM}}& =  \left(\begin{array}{cccc}
	  		1& 0 &0  & 0 \\
	  		0	& c_{\theta^d_{L}} & s_{\theta^d_{L}} c_{\phi_{B_L}} & -s_{\theta^d_{L}}s_{\phi_{B_L}}  \\
	  		0	& -c_{\phi_{T_{L}}}s_{\theta^d_{L}}  & c_{\phi_{T_{L}}}c_{\theta^d_{L}}c_{\phi_{B_L}}  &-c_{\phi_{T_L}}c_{\theta^d_{L}}s_{\phi_{B_L}} \\
	  		0&s_{\phi_{T_L}}s_{\theta^d_{L}}&-s_{\phi_{T_L}}c_{\theta^d_{L}}c_{\phi_{B_L}}&s_{\phi_{T_L}}c_{\theta^d_{L}}s_{\phi_{B_L}}
	  	\end{array} \right)  \label{VL CKM rephased}, \\ \nn\\
	  	\mathcal{\hat{V}}_R^{\text{CKM}}& =  \left(\begin{array}{cccc}
	  		1& 0 &0  & 0 \\
	  		0	& c_{\theta^d_{R}} & -s_{\theta^d_{R}} c_{\beta_{B_R}} e^{i \frac{\delta}{2}}& s_{\theta^d_{R}}s_{\beta_{B_R}} e^{i \frac{\delta}{2}} \\
	  		0	& c_{\beta_{T_{R}}}s_{\theta^d_{R}} e^{i \frac{\delta}{2}} & c_{\beta_{T_{R}}}c_{\theta^d_{R}}c_{\beta_{B_R}} e^{i \delta} &-c_{\beta_{T_R}}c_{\theta^d_{R}}s_{\beta_{B_R}} e^{i \delta} \\
	  		0&-s_{\beta_{T_R}}s_{\theta^d_{R}}e^{i \frac{\delta}{2}}&-s_{\beta_{T_R}}c_{\theta^d_{R}}c_{\beta_{B_R}}e^{i \delta}&s_{\beta_{T_R}}c_{\theta^d_{R}}s_{\beta_{B_R}}e^{i \delta}
	  	\end{array} \right)  \label{VR CKM rephased},
	  \end{align}
	  where
	   \begin{align}
	  	c_{\theta^d_{L}}&=\cos\theta^d_{L},\quad s_{\theta^d_{L}}=\sin\theta^d_{L},\quad c_{\phi_{T_{L}}}=\cos\phi_{T_{L}},\nn\\ s_{\phi_{T_{L}}}&=\sin\phi_{T_{L}},\quad c_{\phi_{B_{L}}}=\cos\phi_{B_{L}},\quad s_{\phi_{B_{L}}}=\sin\phi_{B_{L}},\nn\\
	  	c_{\theta^d_{R}}&=\cos\theta^d_{R},\quad s_{\theta^d_{R}}=\sin\theta^d_{R},\quad c_{\beta_{T_{R}}}=\cos\beta_{T_{R}},\nn\\ s_{\beta_{T_{R}}}&=\sin\beta_{T_{R}},\quad c_{\beta_{B_{R}}}=\cos\beta_{B_{R}},\quad s_{\beta_{B_{R}}}=\sin\beta_{B_{R}}, \nn\\  \beta_{T_R} & = \theta_{T_R} -\phi_{T_R}, \quad \beta_{B_R} = \theta_{B_R} -\phi_{B_R}, \quad \delta = \alpha^3_{d_R}-\alpha^3_{d_L}.
	  \end{align}
	   The number of $CP$ violating phase in this model is one. This agrees with the result in Ref.\cite{umeeda} for the $N=1$ case. The details of the rephasing process is explained in Appendix \ref{appendix CKM matrix}
	
	In addition, the final expression of the left-handed FCNC couplings, which appears in the left-handed weak isospin current in Eq.(\ref{left handed weak isospin}), are defined as follows,
	\begin{align}
		(\mathcal{Z}_{T_L})^{ij} &\equiv (K^\dagger_{T_L})^{i3}(K_{T_L})^{3j} \label{ZTL final}, \\
		(\mathcal{Z}_{B_L})^{ij}& \equiv (K^\dagger_{B_L})^{i3}(K_{B_L})^{3j} \label{ZBL final},
	\end{align}
	where $i,j \in \{3,4\}$. These have explicit matrix form as follows,
	\begin{align}
		\mathcal{Z}_{T_L}&= \left(\begin{array}{cc}
			\cos^2 \phi_{T_L} & -\sin\phi_{T_L}\cos\phi_{T_L} \\
			 -\sin\phi_{T_L}\cos\phi_{T_L}  & \sin^2\phi_{T_L}
		\end{array} \right) \label{ZTL}, \\
		\mathcal{Z}_{B_L}&= \left(\begin{array}{cc}
		\cos^2 \phi_{B_L} & -\sin\phi_{B_L}\cos\phi_{B_L} \\
		-\sin\phi_{B_L}\cos\phi_{B_L}  & \sin^2\phi_{B_L}
		\end{array} \right). \label{ZBL}
	\end{align}
	Similarly, for the right-handed weak isospin current from Eq.(\ref{right weak isospin 2}), the intermediate right-handed FCNC couplings transforms into the final expressions as,
	\begin{align}
		(\mathcal{Z}_{T_R})^{ij} &\equiv \sum_{k,l=3}^{4} (K^\dagger_{T_R})^{ik}(Z_{T_R})^{kl}(K_{T_R})^{lj} \label{ZTR final}, \\
		(\mathcal{Z}_{B_R})^{ij} &\equiv \sum_{k,l=3}^{4}(K^\dagger_{B_R})^{ik}(Z_{B_R})^{kl}(K_{B_R})^{lj}, \label{ZBR final}
	\end{align}
	where $i,j \in \{3,4\}$. These can be expressed in matrix form as follows,
	\begin{align}
		\mathcal{Z}_{T_R}&= \left(\begin{array}{cc}
			\cos^2 \beta_{T_R} & -\sin\beta_{T_R}\cos\beta_{T_R} \\
			-\sin\beta_{T_R}\cos\beta_{T_R}  & \sin^2\beta_{T_R}
		\end{array} \right) \label{ZTR}, \\
		\mathcal{Z}_{B_R}&= \left(\begin{array}{cc}
			\cos^2 \beta_{B_R} & -\sin\beta_{B_R}\cos\beta_{B_R} \\
			-\sin\beta_{B_R}\cos\beta_{B_R}  & \sin^2\beta_{B_R}
		\end{array} \right), \label{ZBR}
	\end{align}
	with $\beta_{T_R} = \theta_{T_{R}} - \phi_{T_R}$ and $\beta_{B_R} = \theta_{B_{R}} - \phi_{B_R}$.
	\end{itemize}

Finally, we obtain the expression of the Lagrangian for the quark and Yukawa interaction after following all steps as follows,
\begin{align}
	{\cal L}_{q}&= 
	\sum_{\alpha=1}^{4}\overline{(\hat{u}^m)^\alpha }i \gamma^\mu D_{\text{em}\mu} (\hat{u}^m)^\alpha +\sum_{\alpha=1}^{4} \overline{(\hat{d}^m)^\alpha }i \gamma^\mu D_{\text{em}\mu} (\hat{d}^m)^\alpha\nn \\
&\hspace{10pt}-\frac{g_L}{\sqrt{2}}\left(\sum_{\alpha,\beta=1}^{4}\overline{({\hat{u}}_L^m)^\alpha}\gamma^\mu (\mathcal{\hat{V}}^{\text{CKM}}_{L})^{\alpha\beta} ({\hat{d}}_L^m)^\beta W^+_{L\mu} +h.c.\right) \nn\\&\hspace{10pt} -\left( \frac{g_L}{2\cos\theta_W}(j^\mu_{3L})-e \tan\theta_W (j^\mu_{\text{em}}) \right)Z_{L\mu}   \nn\\&\hspace{10pt}-\frac{g_R}{\sqrt{2}}\left(\sum_{\alpha,\beta=1}^{4}\overline{	({\hat{u}}_R^m)^{\alpha}}\gamma^\mu (\mathcal{\hat{V}}^{\text{CKM}}_R)^{\alpha\beta} ({\hat{d}}_R^m)^{\beta} W^+_{R\mu} +h.c. \right)\nn\\&\hspace{10pt} -\left\lbrace\frac{g_R}{2\cos\theta_R}(j^\mu_{3_R}) - g'\tan\theta_R\left((j^\mu_{\text{em}})-\frac{1}{2}(j^\mu_{3_L})\right)   \right\rbrace  Z_{R\mu} \nn\\&\hspace{10pt}- \sum_{j=3}^{4}(m_t^{\text{diag}})^{jj} \overline{(\hat{u}^m)^j} (\hat{u}^m)^j- \sum_{j=3}^{4}(m_b^{\text{diag}})^{jj} \overline{(\hat{d}^m)^j} (\hat{d}^m)^j\nn\\&\hspace{10pt}-\frac{1}{v_L} \sum_{k,i=3}^{4}\left(  (\mathcal{Z}_{T_L} m_t^{\text{diag}}  )^{ki} \overline{(\hat{u}_L^m)^k }  (\hat{u}_R^m)^i + (m_t^{\text{diag}} \mathcal{Z}_{T_L} )^{ki} \overline{(\hat{u}_R^m)^k }  (\hat{u}_L^m)^i\right. \nn\\ &\hspace{80pt} \left. +  (\mathcal{Z}_{B_L} m_b^{\text{diag}}  )^{ki} \overline{(\hat{d}_L^m)^k }  (\hat{d}_R^m)^i + ( m_b^{\text{diag}} \mathcal{Z}_{B_L} )^{ki} \overline{(\hat{d}_R^m)^k }  (\hat{d}_L^m)^i   \right) h_L \nn\\ &\hspace{10pt} - \frac{\sqrt{2}}{v_L}\left[\sum_{k=3}^{4}\sum_{\alpha=2}^{4}\left( \overline{(\hat{u}_L^m)^\alpha} (\mathcal{\hat{V}}_L^{\text{CKM}}m_b^{\text{diag}} )^{\alpha k} (\hat{d}_R^m)^k  -\overline{(\hat{u}_R^m)^k} (m_t^{\text{diag}}\mathcal{\hat{V}}_L^{\text{CKM}} )^{k \alpha} (\hat{d}_L^m)^\alpha     \right) \chi_L^+ +h.c. \right]\nn\\ &\hspace{10pt}     
+ \frac{1}{v_L}\sum_{k,i=3}^{4} \left(  (\mathcal{Z}_{T_L} m_t^{\text{diag}}  )^{ki} \overline{(\hat{u}_L^m)^k }  (\hat{u}_R^m)^i - ( m_t^{\text{diag}} \mathcal{Z}_{T_L} )^{ki} \overline{(\hat{u}_R^m)^k }  (\hat{u}_L^m)^i \right. \nn\\ &\hspace{80pt} \left.  -  (\mathcal{Z}_{B_L} m_b^{\text{diag}}  )^{ki} \overline{(\hat{d}_L^m)^k } (\hat{d}_R^m)^i  +(m_b^{\text{diag}} \mathcal{Z}_{B_L}  )^{ki} \overline{(\hat{d}_R^m)^k } (\hat{d}_L^m)^i  \right) i \chi_L^3 \nn\\ &\hspace{10pt}-\frac{1}{v_R}\sum_{k,i=3}^{4} \left(  ((1-\mathcal{Z}_{T_L}) m_t^{\text{diag}} \mathcal{Z}_{T_R}  )^{ki} \overline{(\hat{u}_L^m)^k }  (\hat{u}_R^m)^i + (\mathcal{Z}_{T_R}  m_t^{\text{diag}} (1-\mathcal{Z}_{T_L}) )^{ki} \overline{(\hat{u}_R^m)^k }  (\hat{u}_L^m)^i  \right. \nn\\ & \hspace{60pt}\left. +  ( ( 1-  \mathcal{Z}_{B_L}) m_b^{\text{diag}} \mathcal{Z}_{B_R} )^{ki} \overline{(\hat{d}_L^m)^k }  (\hat{d}_R^m)^i +(\mathcal{Z}_{B_R}  m_b^{\text{diag}} ( 1-  \mathcal{Z}_{B_L}) )^{ki} \overline{(\hat{d}_R^m)^k }  (\hat{d}_L^m)^i   \right) h_R \nn\\ &\hspace{10pt} - \frac{\sqrt{2}}{v_R} \left[  \sum_{k=3}^{4}\sum_{\alpha=2}^{4}\left( \overline{(\hat{u}_R^m)^\alpha} (\mathcal{\hat{V}}_R^{\text{CKM}}m_b^{\text{diag}}(1-\mathcal{Z}_{B_L}) )^{\alpha k} (\hat{d}_L^m)^k \right.\right. \nn\\ &\hspace{80pt} \left.\left.   - \overline{(\hat{u}_L^m)^k} ((1-\mathcal{Z}_{T_L})  m_t^{\text{diag}}\mathcal{\hat{V}}_R^{\text{CKM}} )^{k \alpha} (\hat{d}_R^m)^\alpha     \right) \chi_R^+ +h.c. \right] \nn\\ &\hspace{10pt}     
+ \frac{1}{v_R}\sum_{k,i=3}^{4} \left(  ( ( 1-  \mathcal{Z}_{B_L}) m_b^{\text{diag}} \mathcal{Z}_{B_R} )^{ki} \overline{(\hat{d}_L^m)^k }  (\hat{d}_R^m)^i - (\mathcal{Z}_{B_R}  m_b^{\text{diag}} ( 1-  \mathcal{Z}_{B_L}) )^{ki} \overline{(\hat{d}_R^m)^k }  (\hat{d}_L^m)^i  \right. \nn\\ & \hspace{10pt}\left. - ((1-\mathcal{Z}_{T_L}) m_t^{\text{diag}} \mathcal{Z}_{T_R}  )^{ki} \overline{(\hat{u}_L^m)^k } (\hat{u}_R^m)^i + (\mathcal{Z}_{T_R}  m_t^{\text{diag}}(1-\mathcal{Z}_{T_L})  )^{ki} \overline{(\hat{u}_R^m)^k } (\hat{u}_L^m)^i \right) i \chi_R^3  \label{quark Lagrangian5},
\end{align}
where we define $\hat{u}^m = \hat{u}_L^m + \hat{u}_R^m $ and $\hat{d}^m = \hat{d}_L^m + \hat{d}_R^m$. As mentioned before, the top and bottom quarks are the third component of the fields in the mass basis, while the heavy partners are the fourth component,
\begin{align}
	(\hat{u}_{L(R)}^m)^3 = t_{L(R)},\quad (\hat{u}_{L(R)}^m)^4 = t'_{L(R)},\quad (\hat{d}_{L(R)}^m)^3 = b_{L(R)},\quad (\hat{d}_{L(R)}^m)^4 = b'_{L(R)}. \label{top and bottom definition}
\end{align}
 The left-handed, right-handed weak-isospin, and electromagnetic current in Eq.(\ref{quark Lagrangian5}) now have the following final expressions,
\begin{align}
	j^\mu_{3L} & =\sum_{i=1}^{2}\overline{(\hat{u}_L^m)^i} \gamma^\mu (\hat{u}_L^m)^i + \sum_{l,j=3}^{4} \overline{(\hat{u}_L^m)^l} \gamma^\mu (\mathcal{Z}_{T_L})^{lj} (\hat{u}_L^m)^j \nn\\ &- \sum_{i=1}^{2}\overline{(\hat{d}_L^m)^i} \gamma^\mu (\hat{d}_L^m)^i -\sum_{l,j=3}^{4} \overline{(\hat{d}_L^m)^l} \gamma^\mu (\mathcal{Z}_{B_L})^{lj} (\hat{d}_L^m)^j\label{left handed weak isospin final}, \\
	j^\mu_{3R}&=\sum_{i=1}^{2}\overline{(\hat{u}^m_R)^i} \gamma^\mu (\hat{u}^m_R)^i+\sum_{l,j=3}^{4}\overline{(\hat{u}^m_R)^l} \gamma^\mu (\mathcal{Z}_{T_R})^{lj} (\hat{u}^m_R)^j\nn\\ &-\sum_{i=1}^{2}\overline{(\hat{d}^m_R)^i} \gamma^\mu (\hat{d}^m_R)^i+\sum_{l,j=3}^{4}\overline{(\hat{d}^m_R)^l} \gamma^\mu (\mathcal{Z}_{B_R})^{lj} (\hat{d}^m_R)^j\label{right weak isospin final}, \\
	j^\mu_{\text{em}} &= \frac{2}{3}\sum_{\alpha=1}^{4}\overline{(\hat{u}^m)^\alpha } \gamma^\mu (\hat{u}^m)^\alpha - \frac{1}{3}\sum_{\alpha=1}^{4}\overline{(\hat{d}^m)^\alpha } \gamma^\mu(\hat{d}^m)^\alpha \label{electromagnetic current final},
\end{align}
where the definition and matrix forms of FCNC couplings are shown in Eqs.(\ref{ZTL final})-(\ref{ZBR}). It should be noted that the Lagrangian, written in Eq.(\ref{quark Lagrangian5}), can be expressed in the mass eigenstate of Higgs and $Z$ boson. We will discuss about this in Section \ref{sec: Higgs sector}. 

\section{Higgs Sector} \label{sec: Higgs sector}
In this section, we derived the kinetic terms and potential of Higgs, which are contained in Eq.(\ref{higgs Lagrangian}). In the same way as in Section \ref{sec:quark doublet and yukawa interaction}, we derive it step by step from the $\mathrm{SU(2)_R \times U(1)_{Y'}}$ breaking into $\mathrm{U(1)_Y}$ and finally $\mathrm{SU(2)_L \times U(1)_Y}$ breaking into $\mathrm{U(1)_{\text{em}}}$.
\subsection{ $ \mathbf{SU(2)_R \times U(1)_{Y'}}\rightarrow \mathbf{U(1)_{Y}}$ }
This stage occurs after $\mathrm{SU(2)_R}$ Higgs doublet acquires non-zero vev and takes the parameterization in Eq.(\ref{right higgs parameterization}). In addition, there is a mixing between $B'_\mu$ and $W^3_{R\mu}$ into $B_\mu$ and $Z_{R\mu}$ following the transformation shown in Eq.(\ref{B WR mixing}). We will analyze the kinetic terms and potential separately. Furthermore, we classify the terms based on the number of the fields in the term as linear, quadratic, cubic, and quartic. The gauge fields inside the covariant derivatives are not counted as fields.
\subsubsection{Kinetic terms}
 The kinetic terms in Eq.(\ref{higgs Lagrangian}) becomes,
\begin{align}
	\mathcal{L}_H\supset \mathcal{L}_{\text{kin}} &= (D^\mu_{\text{SM}}\phi_L )^\dagger (D_{\text{SM}\mu}\phi_L )\nn\\ &\hspace{10pt}-i g' Y_{\phi} \tan \theta_R Z_{R\mu}\{(D_{\text{SM}\mu}\phi_L )^\dagger \phi_L - \phi_L^\dagger (D^\mu_{\text{SM}}\phi_L )  \}\nn\\&\hspace{10pt} + g^{\prime 2} Y_{\phi}^2 \tan^2 \theta_R Z_R^\mu Z_{R\mu} \phi_L^\dagger \phi_L \nn\\ &\hspace{10pt} 
	+(D^\mu_{\text{SM}}\chi_R^-)(D_{\text{SM}\mu}\chi_R^+ )+ i\frac{g_R v_R}{2}\{W_R^{+\mu}(D_{\text{SM}\mu}\chi_R^- ) - W_R^{-\mu}(D_{\text{SM}\mu}\chi_R^+ ) \} +\frac{g_R^2 v_R^2}{4} W_R^{-\mu}W_{R\mu}^+ \nn\\ &\hspace{10pt}
	+\frac{1}{2} (\partial_\mu h_R)^2 + \frac{1}{2}\left(\partial_\mu \chi_R^3 - \frac{g_R v_R}{2 \cos \theta_R}Z_{R\mu} \right)^2 \nn\\ &\hspace{10pt} 
	-\frac{g_R}{2} \chi_R^3 \{(W_R^{+\mu}D_{\text{SM}\mu}\chi_R^- )  +W_R^{-\mu} (D_{\text{SM}\mu}\chi_R^+ ) \} \nn\\ &\hspace{10pt} 
	+ i\frac{g_R}{2} \{W_R^{+\mu}  (D_{\text{SM}\mu}\chi_R^- ) - W_R^{-\mu} (D_{\text{SM}\mu}\chi_R^+ )  \}h_R +\frac{g_R^2 v_R}{2}h_R W_R^{-\mu}W_{R\mu}^+ \nn\\&\hspace{10pt}  + i \frac{g_R}{2} \frac{\cos 2\theta_R}{\cos\theta_R} Z_R^\mu \{\chi_R^+(D_{\text{SM}\mu}\chi_R^- ) - \chi_R^-(D_{\text{SM}\mu}\chi_R^+ ) \} \nn\\  &\hspace{10pt} + \frac{g_R^2 v_R}{4} \left( \frac{\cos 2\theta_R -1}{\cos \theta_R}\right)  ( W^+_{R\mu} \chi_R^-+W^-_{R\mu} \chi_R^+ ) Z_R^\mu \nn\\ &\hspace{10pt}
		+\frac{g_R}{2} (W^+_{R\mu} \chi_R^-+W^-_{R\mu} \chi_R^+  ) \partial^\mu \chi_R^3 -i \frac{g_R}{2}  (W^+_{R\mu} \chi_R^--W^-_{R\mu} \chi_R^+ ) \partial^\mu h_R\nn\\ &\hspace{10pt}
		+\frac{g_R}{2\cos \theta_R} \{\chi_R^3 (\partial^\mu h_R) - (\partial^\mu \chi_R^3) h_R\}Z_{R\mu} + \left( \frac{g_R}{2 \cos\theta_R}\right)^2 v_R h_R Z_R^\mu Z_{R\mu} \nn\\ &\hspace{10pt} + \frac{g_R^2 (\cos 2\theta_R)-1 }{4\cos\theta_R} \{  (W^+_{R\mu}\chi_R^--W^-_{R\mu}\chi_R^+ ) i\chi_R^3 + (W^+_{R\mu}\chi_R^-+W^-_{R\mu}\chi_R^+ )h_R  \}Z_R^\mu \nn\\&\hspace{10pt} 
		+\frac{g_R^2}{4} \left( \frac{1}{2 \cos^2 \theta_R} Z_{R\mu}Z_R^\mu + W_R^{+\mu}W^-_{R \mu} \right)( (\chi_R^3)^2 +h_R^2  )\nn\\&\hspace{10pt}  +\frac{g_R^2}{2}\left(W_R^{+\mu}W^-_{R \mu} + \frac{\cos^2 2\theta_R}{2 \cos^2 \theta_R} Z_R^\mu Z_{R\mu}    \right) (\chi_R^- \chi_R^+) \label{higgs kinetic term 1},
		\end{align}
where,
\begin{align}
	D_{\text{SM} \mu} \phi_L &= \left( \partial_\mu +i g_L W^a_{L\mu} \frac{\tau_L^a}{2} +i g' Y_{\phi} B_\mu\right)\phi_L,  \\
	D_{\text{SM} \mu} \chi_R^+& = (\partial_\mu + i g' B_\mu)\chi_R^+,
\end{align}
are the definition of SM covariant derivatives for $\phi_L$ and $\chi_R^+$ respectively.
\subsubsection{Higgs Potential}
The Higgs potential which is written in Eq.(\ref{higgs potential}) now becomes,
\begin{align}
	V(\phi_L,\phi_R) &= (\mu_L^2+\lambda_{LR} v_R^2) \phi_L^\dagger \phi_L + \lambda_L (\phi_L^\dagger \phi_L)^2 \nn\\&\hspace{10pt} + 2 \lambda_{LR} v_R (\phi_L^\dagger \phi_L) h_R + 2\lambda_{LR}(\phi_L^\dagger \phi_L) \left(\chi_R^- \chi_R^+ +\frac{1}{2} (h_R^2 + (\chi_R^3)^2 ) \right) \nn\\&\hspace{10pt} + \frac{\mu_R^2}{2}v_R^2 + \frac{\lambda_R}{4} v_R^4 \nn\\ &\hspace{10pt} + h_R ( \mu_R^2 v_R + \lambda_R v_R^3 ) \nn\\ &\hspace{10pt}+
	\frac{h_R^2}{2} (\mu_R^2 + 3 \lambda_R v_R^2) + (\mu_R^2 + \lambda_R v_R^2) \left( \chi_R^- \chi_R^+ + \frac{1}{2} (\chi_R^3)^2 \right) \nn\\ &\hspace{10pt}+
	2 v_R \lambda_R  h_R \left( \chi_R^- \chi_R^+ + \frac{1}{2} (h_R^2 + (\chi_R^3)^2) \right)\nn\\ &\hspace{10pt}+ \lambda_R \left( \chi_R^- \chi_R^+ + \frac{1}{2} (h_R^2 + (\chi_R^3)^2) \right)^2 \label{higgs potential 1}.
\end{align}
\subsection{$\mathbf{SU(2)_L \times U(1)_Y}\rightarrow \mathbf{U(1)_\text{em}}$} 
This stage occurs after $\mathrm{SU(2)_L}$ Higgs doublet acquires non-zero vev and takes the parameterization, which is written in Eq.(\ref{higgs left parameterization}). As happens in SM, there is a mixing between $B_\mu$ and $W^3_{L\mu}$ into $A_\mu$ and $Z_{L\mu}$ following the transformation shown in Eq.(\ref{B WL mixing}).
	
\subsubsection{Kinetic terms}
At this stage, one can show that the first line of Eq.(\ref{higgs kinetic term 1}) has similar result with $\mathrm{SU(2)_R \times U(1)_{Y'}}$ breaking by replacing $R \rightarrow L$, $\theta_R \rightarrow \theta_W$, and $D_{\text{SM}} \rightarrow D_{\text{em}}$. After computing all terms, the kinetic terms of Higgs in Eq.(\ref{higgs kinetic term 1}) becomes,
\begin{align}
	\mathcal{L}_H\supset \mathcal{L}_{\text{kin}} &=(D^\mu_{\text{em}}\chi_L^-)(D_{\text{em}\mu}\chi_L^+ )+(D^\mu_{\text{em}}\chi_R^-)(D_{\text{em}\mu}\chi_R^+ ) \nn\\&\hspace{10pt}+ i\frac{g_L v_L}{2}\{W_L^{+\mu}(D_{\text{em}\mu}\chi_L^- ) - W_L^{-\mu}(D_{\text{em}\mu}\chi_L^+ ) \}+\frac{g_L^2 v_L^2}{4} W_L^{-\mu}W_{L\mu}^+   \nn\\&\hspace{10pt} +i\frac{g_R v_R}{2}\{W_R^{+\mu}(D_{\text{em}\mu}\chi_R^- ) - W_R^{-\mu}(D_{\text{em}\mu}\chi_R^+ ) \}+\frac{g_R^2 v_R^2}{4} W_R^{-\mu}W_{R\mu}^+ \nn\\ &\hspace{10pt}
	+\frac{1}{2} (\partial_\mu h_L)^2 + \frac{1}{2}\left(\partial_\mu \chi_L^3 - \frac{g_L v_L}{2 \cos \theta_W}Z_{L\mu} \right)^2 +\frac{1}{2} (\partial_\mu h_R)^2 + \frac{1}{2}\left(\partial_\mu \chi_R^3 - \frac{g_R v_R}{2 \cos \theta_R}Z_{R\mu} \right)^2 \nn\\ &\hspace{10pt} +\frac{1}{2}  g'  \tan \theta_R Z_{R\mu} \left\{ - v_L (\partial^{\mu} \chi_L^3) + \frac{g_L v_L^2}{2 \cos \theta_W} Z_L^{\mu} \right\} + \frac{1}{8} v_L^2 g'^2 \tan^2 \theta_R Z_R^{\mu} Z_{R \mu}  \nn\\ &\quad
	-\frac{g_L}{2} \chi_L^3 \{W_L^{+\mu}(D_{\text{em}\mu}\chi_L^- ) +  W_L^{-\mu} (D_{\text{em}\mu}\chi_L^+ )    \}  \nn\\ &\hspace{10pt} -\frac{g_R}{2} \chi_R^3 \{W_R^{+\mu} (D_{\text{em}\mu}\chi_R^- ) +W_R^{-\mu} (D_{\text{em}\mu}\chi_R^+ )   \}\nn\\ &\hspace{10pt}
	+ i\frac{g_L}{2} \{W_L^{+\mu} (D_{\text{em}\mu}\chi_L^- )  - W_L^{-\mu} (D_{\text{em}\mu}\chi_L^+ )  \}h_L +\frac{g_L^2 v_L}{2}h_L W_L^{-\mu}W_{L\mu}^+ \nn\\ &\quad +i \frac{g_R}{2} \left\{W_R^{+\mu} (D_{\text{em} \mu} \chi_R^-)  - W_R^{-\mu}(D_{\text{em} \mu} \chi_R^+)  \right\} h_R +\frac{g_R^2 v_R}{2}h_R W_R^{-\mu}W_{R\mu}^+ \nn\\&\hspace{10pt}  + i \frac{g_L}{2} \frac{\cos 2\theta_W}{\cos\theta_W}  \{\chi_L^+(D_{\text{em}\mu}\chi_L^- ) -\chi_L^- (D_{\text{em}\mu}\chi_L^+ ) \}Z_L^\mu \nn\\&\quad + i \frac{g_R}{2} \frac{\cos 2\theta_R}{\cos\theta_R} \{\chi_R^+(D_{\text{em}\mu}\chi_R^- ) - \chi_R^-(D_{\text{em}\mu}\chi_R^+ \} ) Z_R^\mu  \nn\\  &\hspace{10pt} + \frac{g_L^2 v_L}{4} \left( \frac{\cos 2\theta_W -1}{\cos \theta_W}\right)  ( W^+_{L\mu} \chi_L^- + W^-_{L\mu} \chi_L^+ ) Z_L^\mu\nn\\&\quad + \frac{g_R^2 v_R}{4} \left( \frac{\cos 2\theta_R -1}{\cos \theta_R}\right)  ( W^+_{R\mu} \chi_R^- + W^-_{R\mu} \chi_R^+ ) Z_R^\mu \nn\\ &\hspace{10pt}
	+\frac{g_L}{2} (W^+_{L\mu} \chi_L^- + W^-_{L\mu} \chi_L^+  ) \partial^\mu \chi_L^3 -i \frac{g_L}{2}  (W^+_{L\mu} \chi_L^- - W^-_{L\mu} \chi_L^+ ) \partial^\mu h_L\nn\\ &\hspace{10pt} +\frac{g_R}{2} ( W^+_{R\mu} \chi_R^-+W^-_{R\mu} \chi_R^+ ) \partial^\mu \chi_R^3 -i \frac{g_R}{2}  (W^+_{R\mu} \chi_R^- - W^-_{R\mu} \chi_R^+ ) \partial^\mu h_R\nn\\ &\hspace{10pt}
	+\frac{g_L}{2\cos \theta_W} \{\chi_L^3 (\partial_\mu h_L) - (\partial_\mu \chi_L^3) h_L\} Z^\mu_{L} + \left( \frac{g_L}{2 \cos\theta_W}\right)^2 v_L h_L Z_L^\mu Z_{L\mu} \nn\\ &\hspace{10pt} +\frac{g_R}{2\cos \theta_R} \{\chi_R^3 (\partial_\mu h_R) - (\partial_\mu \chi_R^3) h_R\}Z^\mu_{R} + \left( \frac{g_R}{2 \cos\theta_R}\right)^2 v_R h_R Z_R^\mu Z_{R\mu} \nn\\ &\hspace{10pt}-i e \tan \theta_W \{\chi_R^+(D_{\text{em}\mu} \chi_R^-)   -\chi_R^- (D_{\text{em}\mu} \chi_R^+)  \} Z^\mu_{L} \nn\\ &\quad -i \frac{1}{2} g'  \tan \theta_R \{\chi_L^+(D_{\text{em}\mu} \chi_L^-)   -\chi_L^- (D_{\text{em}\mu} \chi_L^+)  \} Z^\mu_{R} \nn\\ &\hspace{10pt}-  \frac{g_R}{2}v_R e \tan \theta_W  \left(W_{R\mu}^{+} \chi_R^- +W_{R\mu}^{-}  \chi_R^+ \right) Z^\mu_{L}- \frac{g_L}{2} v_L g' \tan \theta_R ( W_{L\mu}^{+} \chi_L^- + W_{L\mu}^{-} \chi_L^+)Z_R^{\mu}\nn\\&\quad +g' \frac{1}{2} \tan \theta_R \left\{ (\partial_{\mu} h_L) \chi_L^3 - (\partial_{\mu} \chi_L^3) h_L \right\}  Z_R^\mu   \nn\\&\quad +  g' \frac{1}{2} \tan \theta_R \frac{g_L}{\cos \theta_W} v_L h_LZ_{R\mu} Z_L^{\mu} + v_L g'^2 \frac{1}{4} \tan^2 \theta_R  h_L  Z_{R \mu} Z_R^{\mu} \nn\\&\quad   + \frac{g_L^2 (\cos 2\theta_W- 1 )}{4\cos\theta_W} \{  (W^+_{L\mu}\chi_L^- - W^-_{L\mu}\chi_L^+ ) i\chi_L^3 + ( W^+_{L\mu}\chi_L^- +  W^-_{L\mu}\chi_L^+ )h_L  \}Z_L^\mu \nn\\&\hspace{10pt} + \frac{g_R^2 (\cos 2\theta_R -1)}{4\cos\theta_R} \{  (W^+_{R\mu}\chi_R^- -  W^-_{R\mu}\chi_R^+ ) i\chi_R^3 + (W^+_{R\mu}\chi_R^- +  W^-_{R\mu}\chi_R^+ )h_R  \}Z_R^\mu \nn\\&\hspace{10pt}  -i \frac{g_R}{2} e \tan \theta_W \chi_R^3 \left(W_{R\mu}^{+} \chi_R^-  - W_{R\mu}^{-}\chi_R^+  \right) Z^\mu_{L}-i \frac{g_L}{2} g' \tan \theta_R \chi_L^3 \left(W_{L\mu}^{+} \chi_L^-  - W_{L\mu}^{-}\chi_L^+  \right) Z^\mu_{R} \nn\\ &\hspace{10pt} 
	- \frac{g_R}{2} e \tan \theta_W h_R \left(W_{R\mu}^{+} \chi_R^-  + W_{R\mu}^{-} \chi_R^+  \right) Z^\mu_{L} - \frac{g_L}{2} g' \tan \theta_R h_L \left(W_{L\mu}^{+} \chi_L^-  + W_{L\mu}^{-} \chi_L^+  \right) Z^\mu_{R}\nn\\ &\hspace{10pt}
	+\frac{g_L^2}{4} \left(  W_L^{+\mu}W^-_{L \mu} \right)((\chi_L^3)^2 +h_L^2) +\frac{g_R^2}{4} \left( W_R^{+\mu}W^-_{R \mu} \right)( (\chi_R^3)^2 +h_R^2  )\nn\\&\hspace{10pt}
		+\frac{g_L^2}{2}W_L^{+\mu}W^-_{L \mu} \chi_L^+ \chi_L^-  +\frac{g_R^2}{2}W_R^{+\mu}W^-_{R \mu}   \chi_R^+ \chi_R^-  \nn\\&\hspace{10pt} 
			+\frac{g_L^2}{2} \frac{\cos^2 2\theta_W}{2 \cos^2 \theta_W}     \chi_L^+ \chi_L^-  Z_{L\mu}Z_L^\mu +\frac{g_R^2}{2}\frac{\cos^2 2\theta_R}{2 \cos^2 \theta_R}    \chi_R^+ \chi_R^-  Z_{R\mu} Z_R^\mu\nn\\&\hspace{10pt} + e^2 \tan^2 \theta_W  \chi_R^+ \chi_R^- Z_{L\mu} Z_L^\mu   +  \frac{g'^2}{4} \tan^2 \theta_R \chi_L^- \chi_L^+ Z_{R \mu}Z_R^{\mu}  
			\nn\\ &\hspace{10pt}  
	- \frac{g_R}{2} e \tan \theta_W \frac{\cos 2\theta_R}{\cos \theta_R} (2\chi_R^+ \chi_R^-) Z_{L\mu} Z_R^\mu - \frac{g_L}{2}  g' \tan \theta_R  \frac{ \cos 2 \theta_W}{\cos \theta_W} \chi_L^- \chi_L^+ Z_{L\mu} Z^\mu_{R} \nn\\ &\hspace{10pt}
	 +\frac{g_L^2}{4}  \frac{1}{2 \cos^2 \theta_W} Z_{L\mu}Z_L^\mu ((\chi_L^3)^2 +h_L^2) +\frac{g_R^2}{4}  \frac{1}{2 \cos^2 \theta_R} Z_{R\mu}Z_R^\mu ( (\chi_R^3)^2 +h_R^2  )\nn\\&\hspace{10pt}
	 	 +  \frac{g'^2}{4} \frac{1}{2}\tan^2 \theta_R  Z_{R \mu} Z_R^{\mu} \left((\chi_L^3)^2 +h_L^2  \right)  \label{higgs kinetic term 2},
	 \end{align}
where,
\begin{align}
	D_{\text{em}\mu}\chi_{L(R)}^+ = (\partial_\mu + i e A_\mu)\chi_{L(R)}^+ \label{cov der em chi}. 
\end{align}
\subsubsection{Higgs potential}
At this stage, the Higgs potential in Eq.(\ref{higgs potential 1}) becomes,

	\begin{align}
		V(\phi_L, \phi_R) &= \frac{\mu_L^2}{2} v_L^2 + \frac{\mu_R^2}{2} v_R^2 + \frac{\lambda_L}{4} v_L^4 + \frac{\lambda_R}{4} v_R^4 + \frac{\lambda_{LR}}{2} v_R^2 v_L^2 \nn \\
		&\quad + h_L (\mu_L^2 v_L + \lambda_L v_L^3 + \lambda_{LR} v_R^2 v_L) + h_R (\mu_R^2 v_R + \lambda_R v_R^3 + \lambda_{LR} v_R v_L^2) \nn \\
		&\quad + h_L (2 \lambda_{LR} v_R v_L) h_R + \frac{h_L^2}{2} (\mu_L^2 + 3 \lambda_L v_L^2 + \lambda_{LR} v_R^2) + \frac{h_R^2}{2} (\mu_R^2 + 3 \lambda_R v_R^2 + \lambda_{LR} v_L^2) \nn \\
		&\quad + (\mu_L^2 + \lambda_L v_L^2 + \lambda_{LR} v_R^2) \left( \chi_L^- \chi_L^+ + \frac{1}{2} (\chi_L^3)^2 \right) \nn\\ &\quad+ (\mu_R^2 + \lambda_R v_R^2 + \lambda_{LR} v_L^2) \left( \chi_R^- \chi_R^+ + \frac{1}{2}  (\chi_R^3)^2 \right) \nn \\
		&\quad + 2 v_L \left\{ \lambda_L \left( \chi_L^- \chi_L^+ + \frac{1}{2} (h_L^2 + (\chi_L^3)^2) \right) + \lambda_{LR} \left( \chi_R^- \chi_R^+ + \frac{1}{2} (h_R^2 + (\chi_R^3)^2) \right) \right\} h_L \nn \\
		&\quad + 2 v_R \left\{ \lambda_R \left( \chi_R^- \chi_R^+ + \frac{1}{2} (h_R^2 + (\chi_R^3)^2) \right) + \lambda_{LR} \left( \chi_L^- \chi_L^+ + \frac{1}{2} (h_L^2 + (\chi_L^3)^2) \right) \right\} h_R \nn \\
		&\quad + \lambda_L \left( \chi_L^- \chi_L^+ + \frac{1}{2} (h_L^2 + (\chi_L^3)^2) \right)^2 + \lambda_R \left( \chi_R^- \chi_R^+ + \frac{1}{2} (h_R^2 + (\chi_R^3)^2) \right)^2 \nn \\
		&\quad + 2 \lambda_{LR} \left( \chi_L^- \chi_L^+ + \frac{1}{2} (h_L^2 + (\chi_L^3)^2) \right) \left( \chi_R^- \chi_R^+ + \frac{1}{2} (h_R^2 + (\chi_R^3)^2) \right) \label{higgs potential 2}.
	\end{align}
where $\mu_L^2$ and $\mu_R^2$ are negative. The minimization conditions of the potential are,
\begin{align}
	v_L&(\mu_L^2 + \lambda_L v_L^2 + \lambda_{LR} v_R^2 ) = 0, \label{minimization2}\\ 
	v_R&(\mu_R^2 + \lambda_R v_R^2 + \lambda_{LR} v_L^2 ) = 0.  \label{minimization}
\end{align}
We can obtain the expressions of the non-zero vevs as follows,
\begin{align}
	v_L = \sqrt{\frac{-\mu_L^2 \lambda_R +\lambda_{LR} \mu_R^2 }{\lambda_R \lambda_L - \lambda^2_{LR}}} \quad \text{and} \quad 
	v_R = \sqrt{\frac{- \mu_R^2\lambda_L +\lambda_{LR} \mu_L^2 }{\lambda_R \lambda_L - \lambda^2_{LR}}}, \label{vev}
\end{align}
where the vevs are taken to be positive. One can show that the linear terms of Higgs fields and quadratic terms of $\chi^{\pm}_{L(R)}$ and $\chi^3_{L(R)}$ will vanish by using Eqs.(\ref{minimization2}) and (\ref{minimization}).

\subsection{Boson mass}
We collect the quadratic terms from kinetic terms Eq.(\ref{higgs kinetic term 2}) and Higgs potential Eq.(\ref{higgs potential 2}) below,
\begin{align}
	\mathcal{L}_H\supset \mathcal{L}_{\text{quad}} &=(D^\mu_{\text{em}}\chi_L^-)(D_{\text{em}\mu}\chi_L^+ )+(D^\mu_{\text{em}}\chi_R^-)(D_{\text{em}\mu}\chi_R^+ ) \nn\\ 
	&\quad + i\frac{g_L v_L}{2}\{W_L^{+\mu}(D_{\text{em}\mu}\chi_L^- ) - W_L^{-\mu}(D_{\text{em}\mu}\chi_L^+ ) \}+\frac{g_L^2 v_L^2}{4} W_L^{-\mu}W_{L\mu}^+ \nn\\
	&\quad + i\frac{g_R v_R}{2}\{W_R^{+\mu}(D_{\text{em}\mu}\chi_R^- ) -W_R^{-\mu} (D_{\text{em}\mu}\chi_R^+ ) \} +\frac{g_R^2 v_R^2}{4} W_R^{-\mu}W_{R\mu}^+ \nn\\ &\quad + \frac{1}{2} \left( \frac{g_L}{2} \frac{v_L}{\cos \theta_W}\right)^2 Z_L^\mu Z_{L\mu} + \frac{1}{2}\left\lbrace \left( \frac{g_R}{2} \frac{v_R}{\cos \theta_R}\right)^2 + \left(\frac{g'}{2} v_L \tan \theta_R \right)^2  \right\rbrace Z_R^\mu Z_{R\mu} \nn\\ &\quad + \frac{g' v_L}{2} \tan \theta_R \frac{g_L}{2} \frac{v_L}{\cos \theta_W}  Z_L^\mu Z_{R\mu} \nn\\ &\quad + \frac{1}{2}(\partial_\mu \chi_L^3 )^2 +\frac{1}{2}(\partial_\mu \chi_R^3 )^2 \nn\\ &\quad - \frac{1}{2}\frac{g_L v_L}{\cos\theta_W} Z_{L\mu} (\partial^\mu \chi_L^3) - \frac{1}{2}\frac{g_R v_R}{\cos\theta_R} Z_{R\mu} (\partial^\mu \chi_R^3)-\frac{g' v_L}{2} \tan\theta_R Z_{R\mu} (\partial^\mu \chi_L^3) \nn\\&\quad+\frac{1}{2} (\partial_\mu h_L)^2 +\frac{1}{2} (\partial_\mu h_R)^2 \nn\\&\quad - h_L (2 \lambda_{LR} v_R v_L) h_R - \frac{h_L^2}{2} (2\lambda_L v_L^2) - \frac{h_R^2}{2} (2\lambda_R v_R^2) \label{quadratic}.
\end{align}
From Eq.(\ref{quadratic}), we obtain the $W_L$ and $W_R$ mass,
\begin{align}
	M_{W_L} & = \frac{g_L}{2}v_L \label{WL mass}, \\
	M_{W_R} & = \frac{g_R}{2}v_R \label{WR mass}.
\end{align}
Since there is mixing between $Z_L$ and $Z_R$ as well as $h_L$ and $h_R$, then we need to diagonalize the mass matrices to obtain the mass eigenstate for the $Z$ bosons and the Higgs bosons. In line with that, the Nambu-Goldstone bosons $\chi_L^3$ and $\chi_R^3$ also mix.

\subsubsection{$Z$ and $Z'$ boson mass}
We define the following transformation from the $Z_L$ and $Z_R$ basis into the mass eigenstates,
\begin{equation}
	\left( \begin{array}{c}
		Z_{L\mu} \\
		Z_{R\mu}
	\end{array}\right)  = \left( \begin{array}{cc}
	\cos \theta & \sin \theta \\
	-\sin \theta & \cos \theta
	\end{array}\right)  \left( \begin{array}{c}
	Z_{\mu} \\
	Z'_{\mu}
	\end{array}\right). \label{Z mixing matrix}
\end{equation}
From Eq.(\ref{quadratic}), the mass matrix in the $Z_{L}$ and $Z_{R}$ basis are as follows,
\begin{equation}
	\mathbb{M}_Z^2 = \left( \begin{array}{cc}
	(\frac{g_L v_L}{2 \cos\theta_W})^2	& \frac{1}{2}g' v_L \tan\theta_R \frac{g_L v_L}{2 \cos\theta_W} \\
	\frac{1}{2}g' v_L \tan\theta_R \frac{g_L v_L}{2 \cos\theta_W} 	& 	(\frac{g_R v_R}{2 \cos\theta_R})^2 + \left( \frac{1}{2}g' v_L \tan\theta_R\right)^2 
	\end{array}\right).
\end{equation}
The mass matrix $\mathbb{M}_Z^2$ can be diagonalized, 
\begin{equation}
	\mathcal{O}_Z^T	\mathbb{M}_Z^2 \mathcal{O}_Z = \text{diag}(M_Z^2 , M^2_{Z'}),
\end{equation}
where $\mathcal{O}_Z$ is the mixing matrix in Eq.(\ref{Z mixing matrix}). The exact mass eigenvalues and mixing angles are as follows,
\begin{align}
	M_{Z}^2 &= \frac{M_{W_R}^2}{2 c_R^2} \left\{  1 + (c_R^2 + t_W^2) \frac{M_{W_L}^2}{M_{W_R}^2}  - \sqrt{  1 - \frac{2 M_{W_L}^2}{M_{W_R}^2} \left( \frac{c_R^2 - s_W^2 s_R^2}{c_W^2} \right)  + (c_R^2 + t_W^2)^2 \left( \frac{M_{W_L}^2}{M_{W_R}^2} \right)^2 } \right\} \label{Z mass exact}, \\
	M_{Z'}^2 &= \frac{M_{W_R}^2}{2 c_R^2} \left\{  1 + (c_R^2 + t_W^2) \frac{M_{W_L}^2}{M_{W_R}^2}  + \sqrt{  1 - \frac{2 M_{W_L}^2}{M_{W_R}^2} \left( \frac{c_R^2 - s_W^2 s_R^2}{c_W^2} \right)  + (c_R^2 + t_W^2)^2 \left( \frac{M_{W_L}^2}{M_{W_R}^2} \right)^2 } \right\} \label{Z prime mass exact},\\
	\tan 2\theta &= \frac{2 c_R s_R^3 s_W \frac{v_L^2}{v_R^2}}{s_W^2 - s_R^2 \left( s_W^2 \cos 2\theta_R + c_W^2 c_R^2 \right) \frac{v_L^2}{v_R^2}}, \quad  0\leq \theta \leq\frac{\pi}{4} \label{Z mixing angle exact},
\end{align}
where,
\begin{align}
	c_R = \cos\theta_R,\hspace{6pt} s_R=\sin\theta_R,\hspace{6pt} c_W = \cos\theta_W, \hspace{6pt} s_W = \sin\theta_W, \hspace{6pt} t_W = \tan\theta_W.
\end{align}
When $M_{W_R} \gg M_{W_L}$, the masses of the $Z$ and $Z'$ bosons are approximately given as follows,
\begin{align}
		M_{Z}^2 &\simeq \frac{M_{W_L}^2}{c_W^2}\left( 1- \frac{M^2_{W_L}}{M^2_{W_R}} s_R^2 t_W^2\right), \\
		M_{Z'}^2 &\simeq \frac{M_{W_R}^2}{c_R^2}\left( 1+ \frac{M^2_{W_L}}{M^2_{W_R}} s_R^2 t_W^2\right).
\end{align}
\subsubsection{Higgs boson mass}
We define the following transformation from $h_L$ and $h_R$ basis into the mass eigenstate,
\begin{equation}
	\left( \begin{array}{c}
		h_{L} \\
		h_{R}
	\end{array}\right)  = \left( \begin{array}{cc}
		\cos \phi & \sin \phi \\
		-\sin \phi & \cos \phi
	\end{array}\right)  \left( \begin{array}{c}
		h \\
		H
	\end{array}\right) \label{Higgs mixing matrix}
\end{equation}
The mass matrix of Higgs in the $h_{L}$ and $h_{R}$ basis are as follows,
\begin{equation}
	\mathbb{M}_h^2 = \left( \begin{array}{cc}
		2\lambda_L v_L^2	& 2 \lambda_{LR} v_R v_L\\
	2 \lambda_{LR} v_R v_L	& 		2\lambda_R v_R^2 
	\end{array}\right).  \label{higgs mass matrix}
\end{equation}
By defining the mixing matrix in Eq.(\ref{Higgs mixing matrix}) as $\mathcal{O}_h$, we can diagonalize $\mathbb{M}_h$ as,
\begin{equation}
	\mathcal{O}_h^T	\mathbb{M}_h^2 \mathcal{O}_h = \text{diag}(m_h^2 , m^2_{H}) \label{higgs diagonalization},
\end{equation}
which gives the exact mass eigenvalues,
\begin{align}
	m_{h}^2 & = \lambda_L v_L^2 + \lambda_R v_R^2 - \sqrt{(\lambda_L v_L^2 - \lambda_R v_R^2)^2 + 4\lambda_{LR}^2 v_L^2 v_R^2}, \\ 
	m_H^2 & = \lambda_L v_L^2 + \lambda_R v_R^2 + \sqrt{(\lambda_L v_L^2 - \lambda_R v_R^2)^2 + 4\lambda_{LR}^2 v_L^2 v_R^2}.
\end{align}
In addition, the mixing angle in Eq.(\ref{Higgs mixing matrix}) is given by,
\begin{align}
	\tan 2\phi &= \frac{2\lambda_{LR} v_R v_L}{\lambda_R v_R^2 - \lambda_L v_L^2}, \quad  0\leq |\phi| \leq \frac{\pi}{4}.
\end{align}
Furthermore, the mass eigenvalues and mixing angle can be expressed in the following approximation forms,
\begin{align}
	m_h^2 & \simeq 2\lambda_L\left(1-\frac{\lambda^2_{LR}}{\lambda_L \lambda_R} \right)  v_L^2,  \\ 
	m_H^2 & \simeq 2\lambda_R v_R^2, \\
	\tan 2\phi & \simeq \frac{2\lambda_{LR}}{\lambda_R}\frac{v_L}{v_R} \label{phi approximate}
\end{align}
if we ignore the correction of $\mathcal{O}\left(v_L^2/v_R^2\right)$.

\subsection{$\chi_L^3$ and $\chi_R^3$ mixing}
From Eq.(\ref{quadratic}), we extract the following form,
\begin{align}
	\mathcal{L}_{\text{quad}} \supset \mathcal{L}_\chi &=\frac{1}{2}(\partial_\mu \chi_L^3 )^2 +\frac{1}{2}(\partial_\mu \chi_R^3 )^2 \nn\\ &\quad - \frac{1}{2}\frac{g_L v_L}{\cos\theta_W} Z_{L\mu} (\partial^\mu \chi_L^3) - \frac{1}{2}\frac{g_R v_R}{\cos\theta_R} Z_{R\mu} (\partial^\mu \chi_R^3)-\frac{g' v_L}{2} \tan\theta_R Z_{R\mu} (\partial^\mu \chi_L^3). \label{chi}
\end{align}
By changing into the mass eigenstate using Eq.(\ref{Z mixing matrix}) and writing in terms of the diagonal mass eigenvalues $(M_Z, M_{Z'})$, Eq.(\ref{chi}) can be written as,
\begin{align}
	\mathcal{L}_{\text{quad}} \supset \mathcal{L}_\chi &=\frac{1}{2}(\partial_\mu \chi_Z )^2 +\frac{1}{2}(\partial_\mu \chi_{Z'} )^2 \nn\\ &\quad - M_Z\ (\partial^\mu \chi_Z) Z_\mu - M_{Z'} (\partial^\mu \chi_{Z'})Z'_{\mu} \label{chi 2 },
\end{align}
where,
\begin{align}
	\left( \begin{array}{c}
		\chi_L^3 \\
		\chi_R^3
	\end{array}\right)  &= \left( \begin{array}{cc}
		\cos \alpha & \sin \alpha \\
		-\sin \alpha & \cos \alpha
	\end{array}\right)  \left( \begin{array}{c}
		\chi_{Z} \\
		\chi_{Z'}
	\end{array}\right) \label{chi mixing matrix}, \\
	\cos\alpha & = \frac{M_Z \cos\theta}{\sqrt{M_Z^2 \cos^2\theta + M_{Z'}^2 \sin^2 \theta}}, \\ \sin\alpha & = \frac{M_{Z'} \sin\theta}{\sqrt{M_Z^2 \cos^2\theta + M_{Z'}^2 \sin^2 \theta}}.
\end{align}

Therefore, the quadratic terms in Eq.(\ref{quadratic}) are written in terms of the mass basis of the $Z$ bosons, Higgs bosons, and Nambu-Goldstone bosons,
\begin{align}
	\mathcal{L}_H \supset \mathcal{L}_{\text{quad}} &=\left(D^\mu_{\text{em}} \chi_L^- -i M_{W_L}W_L^{\mu -}\right) \left(D_{\text{em}\mu} \chi_L^+ +i M_{W_L}W^+_{L\mu}\right) \nn\\ &\quad +\left(D^\mu_{\text{em}} \chi_R^- -i M_{W_R}W_R^{\mu -}\right) \left(D_{\text{em}\mu} \chi_R^+ +i M_{W_R}W^+_{R\mu}\right) \nn\\ 
	&\quad + \frac{1}{2}\left( \partial_\mu \chi_Z - M_Z Z_\mu\right)^2 + \frac{1}{2}\left( \partial_\mu \chi_{Z'} - M_{Z'} Z'_\mu\right)^2 \nn\\ &\quad
	+\frac{1}{2}\left(\partial_\mu h \right)^2 - \frac{1}{2}m_h^2 h^2+\frac{1}{2}\left(\partial_\mu H \right)^2- \frac{1}{2}m_H^2 H^2 \label{gauge final quadratic},
\end{align}
where the covariant derivatives of $\chi_L$ and $\chi_R$ are in Eq.(\ref{cov der em chi}). We have shown explicitly that $\chi^3_L$ and $\chi^3_R$ are mixed in this model. From Eq.(\ref{gauge final quadratic}), it is shown clearly that the degrees of freedom $\chi_Z$ and $\chi_{Z'}$ become the longitudinal components of the massive $Z$ and $Z'$ bosons, respectively.
\section{Kinetic terms of gauge fields}\label{sec:gauge field}
In this section we derive the kinetic terms of the gauge fields starting from Lagrangian in Eq.(\ref{gauge Lagrangian}). 
\subsection{ $ \mathbf{SU(2)_R \times U(1)_{Y'}} \rightarrow \mathbf{U(1)_{Y}}$}
At this stage, the kinetic terms of the gauge fields change from $B'_\mu$ and $W_{R\mu}$ basis into $B_\mu$ and $Z_{R\mu}$ basis. Following transformation in Eq.(\ref{B WR mixing}), the Lagrangian in Eq.(\ref{gauge Lagrangian}) becomes,
\begin{align}
	\mathcal{L}_{\text{gauge}} = & - \frac{1}{4} F_{L\mu\nu}^{a} F_L^{a\mu\nu} - \frac{1}{4} B_{\mu\nu} B^{\mu\nu} \nn\\
	& - \frac{1}{2} (\partial_\mu W_{R \nu}^+ - \partial_\nu W_{R \mu}^+)(\partial^\mu W_R^{-\nu} - \partial^\nu W_R^{-\mu}) \nn\\
	& - i (\partial_\mu W_{R\nu}^{+} - \partial_\nu W_{R\mu}^{+})(g_R \cos \theta_R Z^\nu_{R } + g' B^\nu) W_{R }^{-\mu} \nn \\
	& + i (\partial^\mu W_{R}^{-\nu} - \partial^\nu W_{R}^{-\mu})(g_R \cos \theta_R Z_{R \nu} + g' B_\nu) W_{R\mu }^{+} \nn \\
	& - \left\{(g_R \cos \theta_R Z_{R \nu} + g' B_\nu) W_{R \mu}^+ (g_R \cos \theta_R Z^\nu_{R } + g' B^\nu) W_R^{-\mu} \right. \nn \\
	& \quad \left. - (g_R \cos \theta_R Z_{R \mu} + g' B_\mu) W_{R \nu}^+ (g_R \cos \theta_R Z^\nu_{R } + g' B^\nu) W_R^{-\mu}\right\}  \nn \\
	& - \frac{1}{4} F_{Z_R\mu\nu}^{0} F_{Z_R}^{0 \mu\nu} + i W_{R \mu}^- W_{R \nu}^+ (g_R \cos \theta_R F^{0 \mu\nu}_{Z_R} + g' B^{\mu\nu}) \nn \\
	& + \frac{1}{2} g_R^2 (W_{R\mu}^{-} W_{R \nu}^+ - W_{R\mu}^{+} W_{R \nu}^-)  (W_R^{-\mu} W_{R}^{+ \nu}) \label{gauge kinetic 2},
\end{align}
where,
\begin{align}
	B_{\mu\nu} &= \partial_\mu B_\nu - \partial_\nu B_\mu, \\
		F^a_{L\mu\nu} &=\partial_\mu W^a_{L\nu} - \partial_\nu W^a_{L\mu} -g_L \epsilon^{abc}W^b_{L\mu}W^c_{L\nu}, \\
		F_{Z_R\mu\nu}^{0} &= \partial_\mu Z_{R\nu} - \partial_\nu Z_{R\mu}.
\end{align}
\subsection{ $ \mathbf{SU(2)_L \times U(1)_{Y}}\rightarrow \mathbf{U(1)_\text{em}}$}
At this stage, there is a mixing between $B_\mu$ and $W^3_{L\mu}$ into $A_\mu$ and $Z_{L\mu}$ following the transformation shown in Eq.(\ref{B WL mixing}). In addition, we also express in the diagonal basis of $Z$ and $Z'$ where the transformation is shown in Eq.(\ref{Z mixing matrix}). So the Lagrangian in Eq.(\ref{gauge kinetic 2}) becomes,
\begin{align}
	\mathcal{L}_{\text{gauge}}&= -\frac{1}{4} F_{Z\mu\nu}^0 F^{0\mu\nu}_Z - \frac{1}{4} F_{Z'\mu\nu}^0 F^{0\mu\nu}_{Z'} - \frac{1}{4} F_{\mu\nu} F^{\mu\nu} \nn \\
	&\quad - \frac{1}{2} \left( \mathcal{D}_\mu W_{L\nu}^{+} - \mathcal{D}_\nu W_{L\mu}^+ \right) \left( \mathcal{D}^\mu W_L^{-\nu} - \mathcal{D}^\nu W_{L}^{-\mu} \right) \nn \\
	&\quad - \frac{1}{2} \left( \mathcal{D}_\mu W_{R\nu}^{+} - \mathcal{D}_\nu W_{R\mu}^+ \right) \left( \mathcal{D}^\mu W_R^{-\nu} - \mathcal{D}^\nu W_{R}^{-\mu} \right) \nn \\
	&\quad + \frac{g_L^2}{2} \left( (W_L^- \cdot W_L^-) (W_L^+ \cdot W_L^+ )- \left( W_L^- \cdot W_L^+ \right)^2 \right) \nn \\
	&\quad + \frac{g_R^2}{2} \left( (W_R^- \cdot W_R^-)( W_R^+ \cdot W_R^+) - \left( W_R^- \cdot W_R^+ \right)^2 \right) \nn \\
	&\quad + i \left\{ g_L \cos \theta_W \cos \theta F_{Z}^{0\mu\nu} + g_L \cos \theta_W \sin \theta F_{Z'}^{0\mu\nu} + e F^{\mu\nu} \right\} \left( W_{L\mu}^- W_{L\nu}^+ \right) \nn \\
	&\quad + i \left\{ - (g_R \cos \theta_R \sin \theta + e \tan \theta_W \cos \theta) F_{Z}^{0\mu\nu} \right. \nn\\  &\hspace{80pt}\left. + (g_R \cos \theta_R \cos \theta - e \tan \theta_W \sin \theta) F_{Z'}^{0\mu\nu} + e F^{\mu\nu} \right\} \left( W_{R\mu}^- W_{R\nu}^+ \right),
\end{align}
where,
\begin{align}
	F_{Z\mu\nu}^0 &= \partial_\mu Z_\nu - \partial_\nu Z_\mu, \nn \\
	F_{Z'\mu\nu}^0 &= \partial_\mu Z'_\nu - \partial_\nu Z'_\mu, \nn \\
	F_{\mu\nu} &= \partial_\mu A_\nu - \partial_\nu A_\mu, \nn \\
	\mathcal{D}_\mu W_{R\nu}^+ &= (D_{\text{em}\mu} W_{R\nu}^+)  - i (e \tan \theta_W Z_{L\mu} - g_R \cos \theta_R Z_{R\mu}) W_{R\nu}^+, \nn \\
	\mathcal{D}_\mu W_{L\nu}^+ &= (D_{\text{em}\mu} W_{L\nu}^+) + i g_L \cos \theta_W Z_{L\mu} W_{L\nu}^+, \nn\\
	D_{\text{em}\mu} G_\nu  &= (\partial_\mu + i e A_\mu) G_\nu,
\end{align}
with $G_\nu \in \{ W_{R\nu}^+,W_{L\nu}^+ \}$. 

\section{Discussion}\label{sec:discussion}
\subsection{Hierarchy of VLQ's mass parameters, $v_L$ and $v_R$}
In this subsection, we discuss the Hierarchy of VLQ's mass parameters, $v_L$ and $v_R$. From Eqs.(\ref{top and bottom exact mass}) and (\ref{top and bottom partner exact mass}), we have the exact mass eigenvalues of the top and bottom quarks, as well as the heavy top and bottom quarks, respectively. One of the motivations for the universal seesaw model in the quark sector is to explain the mass hierarchy of quarks, particularly the third family quark mass hierarchy in our model. Therefore, the hierarchy of VLQ's mass parameters $M_T$, $M_B$, $v_L$, and $v_R$ is essential to be studied. We give the analytical and numerical analysis.
\subsubsection{Analytical analysis}
 The top quark exact mass eigenvalue in Eq.(\ref{top and bottom exact mass}) can be written as follows,
 \begin{align}
 		m_{t} &=  \frac{\sqrt{M_{T}^2 + m^2_{u_R} + m^2_{u_L} + 2 m_{u_L}m_{u_R} }}{2} -  \frac{\sqrt{M_{T}^2 + m^2_{u_R} + m^2_{u_L} - 2 m_{u_L}m_{u_R} }}{2} \nn\\
 		&\simeq   \left(\frac{m_{u_R}}{\sqrt{M_T^2 + m^2_{u_R}}} \right) m_{u_L}.  \label{top quark exact mass} 
 \end{align}
 From the first line to the second line of Eq.(\ref{top quark exact mass}), we use $m_{u_L} < m_{u_R}$. We can express the second line of Eq.(\ref{top quark exact mass}) in terms of Yukawa couplings using Eq.(\ref{mur and mul}) as follows,
 \begin{equation}
 	m_t \simeq \left(\frac{\frac{Y^3_{u_R} v_R}{\sqrt{2}}}{\sqrt{M_T^2 + \frac{(Y^3_{u_R})^2 v_R^2}{2}}} \right) \frac{Y^3_{u_L} v_L}{\sqrt{2}}.  \label{6.2}
 \end{equation}
 If we assume $Y^3_{u_L} = Y^3_{u_R} \simeq \mathcal{O}(1)$ and the factor inside the parenthesis is $\mathcal{O}(1)$, we can obtain the top quark mass $m_t \simeq v_L$. This implies $M_T < v_R$. In order to determine the hierarchy between $M_T$ and $v_R$ for the large top quark mass, from Eq.(\ref{6.2}) one can obtain the ratio $M_T / v_R$ as follows,
 \begin{equation}
 	\frac{M_T}{v_R} = \frac{Y^3_{u_L}Y^3_{u_R}}{\sqrt{2}} \sqrt{\frac{1}{(y_t^{\text{SM}})^2} - \frac{1}{(Y^3_{u_L})^2} },
 \end{equation}
 where $y_t^{\text{SM}}$ is the SM Yukawa coupling of top quark and $Y^3_{u_L} \geq  y_t^{\text{SM}}$. If we further require that the Yukawa couplings are in the perturbative region, $ y_t^{\text{SM}} \leq Y^3_{u_L}$,$Y^3_{u_R} \leq 1$, the upper and the lower limit of the ratio $M_T / v_R$ is given by  
  \begin{equation}
 	0 \leq \frac{M_T}{v_R} \leq \frac{1}{\sqrt{2}} \sqrt{\frac{1}{(y_t^{\text{SM}})^2} - 1 }.
 \end{equation}
 If we take $y_t^{\text{SM}} = 0.9912$, we obtain the upper limit of the ratio $M_T / v_R \leq 0.0944$.  This shows how the seesaw mechanism accommodates the top quark mass and the hierarchy between $M_T$ and $v_R$.
 
 Similarly for the bottom sector, by using $m_{d_L}< m_{d_R}$ the bottom quark mass becomes,
\begin{equation}
	m_b \simeq \left(\frac{\frac{Y^3_{d_R} v_R}{\sqrt{2}}}{\sqrt{M_B^2 + \frac{(Y^3_{d_R})^2 v_R^2}{2}}} \right) \frac{Y^3_{d_L} v_L}{\sqrt{2}}.  \label{6.5}
\end{equation}
If we assume $Y^3_{d_L} = Y^3_{d_R} \simeq \mathcal{O}(1)$ and the factor inside the parenthesis is much smaller than $\mathcal{O}(1)$, we can obtain the light bottom quark mass. This implies $M_B \gg v_R$ and we can write Eq.(\ref{6.5}) as follows,
\begin{equation}
	m_b \simeq \frac{v_R Y^{3}_{d_{R}} Y^{3}_{d_{L}} v_L}{2 M_B}.  \label{6.6}
\end{equation}
In order to determine the hierarchy between $M_B$ and $v_R$ for the light bottom quark mass, from Eq.(\ref{6.6}) one can obtain the ratio $M_B / v_R$ as follows,
\begin{align}
	\frac{M_B}{v_R} = \frac{Y^3_{d_L}Y^3_{d_R}}{\sqrt{2}} \frac{1}{y_b^{\text{SM}}},
\end{align}
where ${y_b^{\text{SM}}}$ is the SM Yukawa coupling of bottom quark.  If we further require that the Yukawa couplings are in the perturbative region, $Y^3_{d_L}$,$Y^3_{d_R} \leq 1$, the upper limit of the ratio $M_B / v_R$ is given by  
\begin{equation}
	\frac{M_B}{v_R} \leq \frac{1}{\sqrt{2}} \frac{1}{y_b^{\text{SM}}}.
\end{equation}
If we take $y_b^{\text{SM}} = 2.4 \times 10^{-2}$, we obtain the upper limit of the ratio $M_B / v_R \leq 29.46$.  The equality  holds when the Yukawa couplings $Y^3_{d_L}=Y^3_{d_R}=1$.  This shows how the seesaw mechanism accommodates the bottom quark mass and the hierarchy between $M_B$ and $v_R$. Therefore, when the all the Yukawa couplings $Y^3_{d_L}$, $Y^3_{d_R}$, $Y^3_{u_L}$ and $Y^3_{u_R}$ is $\mathcal{O}(1)$, the hierarchy for the three scales  is $M_T < v_R \ll M_B$. If we include the $v_L$, the hierarchy has two possibilities depending on the numerical inputs. The hierarchy can be $v_L < M_T < v_R \ll M_B$ or $M_T< v_L < v_R \ll M_B$.
 
 To summarize, by using the hierarchy that we discussed before, from the exact mass eigenvalues in Eqs.(\ref{top and bottom exact mass}) and (\ref{top and bottom partner exact mass}) we can obtain the approximate form as follows,
\begin{align}
	m^{\text{approx}}_t &\simeq \frac{v_R Y^{3}_{u_{R}} Y^{3}_{u_{L}}  v_L }{2 \sqrt{\frac{v_R^2}{{2}}  (Y^{3}_{u_{R}})^2+M_T^2}} \label{eq:top mass}, \\
	m^{\text{approx}}_{t'}&\simeq \sqrt{\frac{v_R^2}{{2}}  (Y^{3}_{u_{R}})^2+M_T^2} \label{eq:top VLQ mass}, \\
	m^{\text{approx}}_b &\simeq \frac{v_R Y^{3}_{d_{R}} Y^{3}_{d_{L}} v_L}{2 M_B} \label{eq:bottom mass}, \\
	m^{\text{approx}}_{b'} &\simeq M_B \label{eq:bottom VLQ mass}.
\end{align}
Our results in Eqs.(\ref{eq:top mass}) and (\ref{eq:top VLQ mass}) agree with Eqs.(7) and (8) in Ref.\cite{satou}, as well as Eqs.(3.19) and (3.17) in Ref.\cite{kiyo}, respectively. While our results in Eqs.(\ref{eq:bottom mass}) and (\ref{eq:bottom VLQ mass}) agree with Eqs.(14) and (15) in Ref.\cite{satou}, as well as Eq.(3.9) in Ref.\cite{kiyo}, respectively.

\subsubsection{Numerical analysis}
\begin{figure}[t!]
	\centering
	\begin{subfigure}[b]{0.7\textwidth}
		\centering
		\includegraphics[width=\textwidth]{"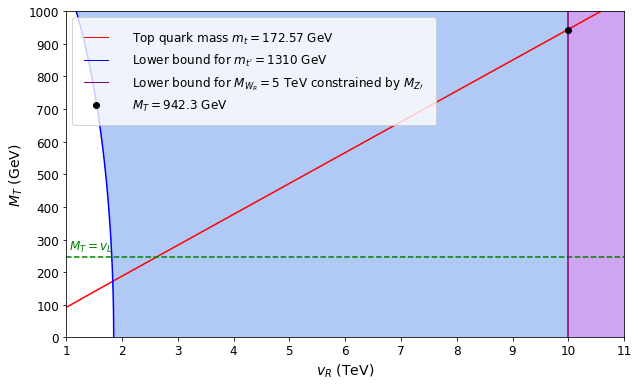"}
		\caption{}
		\label{fig:top_sector}
	\end{subfigure}
	\vfill
	\begin{subfigure}[b]{0.7\textwidth} 
		\centering
		\includegraphics[width=\textwidth]{"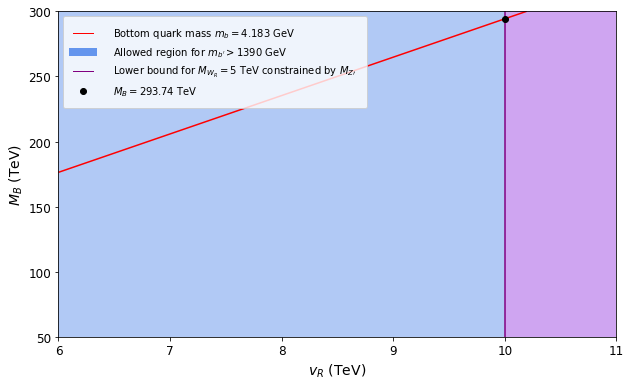"}
		\caption{}
		\label{fig:bottom_sector}
	\end{subfigure}
	\caption{Constraints on $ v_R$ and VLQ's mass parameters of different sectors. (a) Top sector. (b) Bottom sector.}
	\label{fig:combined_constraints}
\end{figure}

We start by analyzing the constraints in the top sector, as shown in Fig (\ref{fig:top_sector}). We consider an asymmetric left-right model with $g_L \ne g_R$. By assuming $g_R \simeq 1$ and using the value of $g'\simeq 0.357$, we obtain $\theta_R$ with Eq.(\ref{relation thetaR}). Additionally, we assume $Y^3_{u_R}\simeq Y^3_{u_L} \simeq 1$. Following are the constraints that we used \cite{pdg}: (1) the top quark mass obtained by the direct measurement is $m_t = 172.57$ GeV; (2) the lower bound for the heavy top quark mass is set to be $m_{t'}> 1310$ GeV; (3) the lower bound for the $Z'$ boson mass is set to be $M_{Z'} > 5150$ GeV. Using the exact mass eigenvalue for the $Z'$ boson mass in Eq.(\ref{Z prime mass exact}), we compute the lower bound for $W_R$ boson mass as $M_{W_R}\gtrsim 5$ TeV. Consequently, we find the constraint for $v_R$ using Eq. (\ref{WR mass}), yielding $v_R \gtrsim 10$ TeV. At $v_R = 10$ TeV, $M_T$ is 942.3 GeV as shown by the black dot in Fig (\ref{fig:top_sector}). Using these $v_R$ and $M_T$ values, we further calculate the heavy top quark mass with Eq.(\ref{top and bottom partner exact mass}) and obtain $m_{t'} = 7.13$ TeV.

Next, we analyze the constraints in the bottom sector, as depicted in Fig (\ref{fig:bottom_sector}). Here, we also assume $Y^3_{d_R}\simeq Y^3_{d_L} \simeq 1$. The constraints are \cite{pdg}: (1) the SM bottom quark mass we use is the running mass at bottom mass $m_b = 4.183$ GeV; (2) the lower bound for the heavy bottom quark mass is set to be $m_{b'} > 1390$ GeV; (3) the constraint for $v_R \gtrsim 10$ TeV is derived from the lower bound for the $Z'$ boson mass. For the bottom sector, at $v_R = 10$ TeV, $M_B$ is 293.74 TeV as indicated by the black dot in Fig (\ref{fig:bottom_sector}). Using these  $v_R$ and $M_B$ values, we further calculate the heavy bottom quark mass with Eq.(\ref{top and bottom partner exact mass}) and obtain $m_{b'} = 293.82$ TeV. This result indicates that $m_{b'}\simeq M_B$.

From the above facts, the mass parameter of top partner VLQ ($M_T$) is smaller than $v_R$ but it could be larger or smaller than $v_L$ depending on the other parameters. On the other hand, in the bottom sector, the mass parameter of bottom partner VLQ ($M_B$) is significantly larger compared to $v_R$. This explains the mass hierarchy problem, where the smallness of the bottom quark mass is suppressed by the large mass of the bottom VLQ through a seesaw mechanism. Mathematically, our choice of numerical input satisfies the following hierarchy: (1) for the top sector: $v_L < M_T < v_R$; (2) for the bottom sector: $v_L < v_R \ll M_B$.

One can compute the masses in the approximation form given in Eqs.(\ref{eq:top mass}), (\ref{eq:top VLQ mass}), (\ref{eq:bottom mass})
and (\ref{eq:bottom VLQ mass})
by using our choice of numerical input and obtain $m^{\text{approx}}_t = 172.58$ GeV, $m^{\text{approx}}_{t'} = 7.13$ TeV, $m^{\text{approx}}_b = 4.19$ GeV, and $m^{\text{approx}}_{b'} = 293.74$ TeV. These values are very close to the exact mass eigenvalues formula. We will use $v_R = 10$ TeV for the rest of our numerical analysis. This $v_R = 10$ TeV is also used in Ref.\cite{babu}, although unlike this paper, they considered the model with left-right symmetry where $g_L = g_R$.

\subsection{Higgs FCNC}
In this subsection, we discuss the interaction between Higgs and quarks in our model. From Eq.(\ref{quark Lagrangian5}), we extract the interactions between $h_L - h_R$ with quarks, given by
\begin{align}
	\mathcal{L}_q \supset \mathcal{L}_{hH}&=-\frac{1}{v_L} \sum_{k,i=3}^{4}\left[ (\mathcal{Z}_{T_L} m_t^{\text{diag}}  )^{ki} \overline{(\hat{u}_L^m)^k }  (\hat{u}_R^m)^i + (m_t^{\text{diag}} \mathcal{Z}_{T_L} )^{ki} \overline{(\hat{u}_R^m)^k }  (\hat{u}_L^m)^i\right. \nn\\ &\quad \left. +  (\mathcal{Z}_{B_L} m_b^{\text{diag}}  )^{ki} \overline{(\hat{d}_L^m)^k }  (\hat{d}_R^m)^i + ( m_b^{\text{diag}} \mathcal{Z}_{B_L} )^{ki} \overline{(\hat{d}_R^m)^k }  (\hat{d}_L^m)^i  \right] h_L\nn\\ &\hspace{10pt}-\frac{1}{v_R}\sum_{k,i=3}^{4} \left[ ((1-\mathcal{Z}_{T_L}) m_t^{\text{diag}} \mathcal{Z}_{T_R}  )^{ki} \overline{(\hat{u}_L^m)^k }  (\hat{u}_R^m)^i \right.\nn\\ &\quad \left.+ (\mathcal{Z}_{T_R}  m_t^{\text{diag}} (1-\mathcal{Z}_{T_L}) )^{ki} \overline{(\hat{u}_R^m)^k }  (\hat{u}_L^m)^i   +  ( ( 1-  \mathcal{Z}_{B_L}) m_b^{\text{diag}} \mathcal{Z}_{B_R} )^{ki} \overline{(\hat{d}_L^m)^k }  (\hat{d}_R^m)^i \right. \nn\\ & \quad \left. +(\mathcal{Z}_{B_R}  m_b^{\text{diag}} ( 1-  \mathcal{Z}_{B_L}) )^{ki} \overline{(\hat{d}_R^m)^k }  (\hat{d}_L^m)^i   \right] h_R \label{E1},
\end{align}
where $\mathcal{Z}_{T_L}, \mathcal{Z}_{B_L}, \mathcal{Z}_{T_R}, \mathcal{Z}_{B_R}, m_t^{\text{diag}}$, and $m_b^{\text{diag}}$ are given in Eqs.(\ref{ZTL}),(\ref{ZBL}),(\ref{ZTR}),\linebreak(\ref{ZBR}), (\ref{diagonalizing top mass matrix}), and (\ref{diagonalizing bottom mass matrix}), respectively. By transforming $h_L - h_R$ basis into $h - H$ mass eigenstate with Eq.(\ref{Higgs mixing matrix}), the Lagrangian in Eq.(\ref{E1}) transforms into,
\begin{align}
	\mathcal{L}_{hH}&=-\left\lbrace \frac{\cos\phi}{v_L} \sum_{k,i=3}^{4}\left[ (\mathcal{Z}_{T_L} m_t^{\text{diag}}  )^{ki} \overline{(\hat{u}_L^m)^k }  (\hat{u}_R^m)^i + (m_t^{\text{diag}} \mathcal{Z}_{T_L} )^{ki} \overline{(\hat{u}_R^m)^k }  (\hat{u}_L^m)^i\right.\right. \nn\\ &\hspace{80pt} \left.\left.  +  (\mathcal{Z}_{B_L} m_b^{\text{diag}}  )^{ki} \overline{(\hat{d}_L^m)^k }  (\hat{d}_R^m)^i + ( m_b^{\text{diag}} \mathcal{Z}_{B_L} )^{ki} \overline{(\hat{d}_R^m)^k }  (\hat{d}_L^m)^i \right]\right.\nn\\ &\quad\left. -\frac{\sin \phi}{v_R}\sum_{k,i=3}^{4} \left[ ((1-\mathcal{Z}_{T_L}) m_t^{\text{diag}} \mathcal{Z}_{T_R}  )^{ki} \overline{(\hat{u}_L^m)^k }  (\hat{u}_R^m)^i  + (\mathcal{Z}_{T_R}  m_t^{\text{diag}} (1-\mathcal{Z}_{T_L}) )^{ki} \overline{(\hat{u}_R^m)^k }  (\hat{u}_L^m)^i  \right. \right.\nn\\ &\quad \left.\left. +  ( ( 1-  \mathcal{Z}_{B_L}) m_b^{\text{diag}} \mathcal{Z}_{B_R} )^{ki} \overline{(\hat{d}_L^m)^k }  (\hat{d}_R^m)^i +(\mathcal{Z}_{B_R}  m_b^{\text{diag}} ( 1-  \mathcal{Z}_{B_L}) )^{ki} \overline{(\hat{d}_R^m)^k }  (\hat{d}_L^m)^i   \right] \right\rbrace h \nn\\ &-\left\lbrace 
	\frac{\sin\phi}{v_L} \sum_{k,i=3}^{4}\left[ (\mathcal{Z}_{T_L} m_t^{\text{diag}}  )^{ki} \overline{(\hat{u}_L^m)^k }  (\hat{u}_R^m)^i + (m_t^{\text{diag}} \mathcal{Z}_{T_L} )^{ki} \overline{(\hat{u}_R^m)^k }  (\hat{u}_L^m)^i\right.\right. \nn\\ &\hspace{80pt} \left.\left.  +  (\mathcal{Z}_{B_L} m_b^{\text{diag}}  )^{ki} \overline{(\hat{d}_L^m)^k }  (\hat{d}_R^m)^i + ( m_b^{\text{diag}} \mathcal{Z}_{B_L} )^{ki} \overline{(\hat{d}_R^m)^k }  (\hat{d}_L^m)^i \right]\right.\nn\\ &\quad\left. +\frac{\cos \phi}{v_R}\sum_{k,i=3}^{4} \left[ ((1-\mathcal{Z}_{T_L}) m_t^{\text{diag}} \mathcal{Z}_{T_R}  )^{ki} \overline{(\hat{u}_L^m)^k }  (\hat{u}_R^m)^i  + (\mathcal{Z}_{T_R}  m_t^{\text{diag}} (1-\mathcal{Z}_{T_L}) )^{ki} \overline{(\hat{u}_R^m)^k }  (\hat{u}_L^m)^i  \right. \right.\nn\\ &\quad \left.\left. +  ( ( 1-  \mathcal{Z}_{B_L}) m_b^{\text{diag}} \mathcal{Z}_{B_R} )^{ki} \overline{(\hat{d}_L^m)^k }  (\hat{d}_R^m)^i +(\mathcal{Z}_{B_R}  m_b^{\text{diag}} ( 1-  \mathcal{Z}_{B_L}) )^{ki} \overline{(\hat{d}_R^m)^k }  (\hat{d}_L^m)^i   \right] \right\rbrace H \label{E2},
\end{align}
where $h$ and $H$ denote as the Higgs and the heavy Higgs bosons, respectively. In this discussion, we will focus on the interaction of the Higgs boson with the quarks in our model.
\subsubsection{Top sector}
 We collect the interaction terms between the Higgs boson and the top quark ($t$) and heavy top quark ($t'$)  from Eq.(\ref{E2})
\begin{align}
	\mathcal{L}_{hH} \supset \mathcal{L}_{ht}&=-\left[\frac{\cos\phi}{v_L}\cos^2\phi_{T_L} m_t - \frac{\sin\phi}{v_R}\left( \sin^2\phi_{T_L} \cos^2\beta_{T_R} m_t \right. \right. \nn\\&\hspace{120pt} \left.\left. -\sin\phi_{T_L} \cos\phi_{T_L} \sin\beta_{T_R} \cos\beta_{T_R}m_{t'} \right) \right]\bar{t}th \nn\\&\quad + \left[\frac{\cos\phi}{v_L}\sin\phi_{T_L}\cos\phi_{T_L} m_{t'} + \frac{\sin\phi}{v_R}\left( \sin\phi_{T_L}\cos\phi_{T_L} \sin^2\beta_{T_R} m_{t'} \right. \right. \nn\\&\hspace{120pt} \left.\left. -\sin^2\phi_{T_L} \sin\beta_{T_R} \cos\beta_{T_R}m_{t} \right) \right](\bar{t}_Lt'_R + \bar{t}'_R t_L) h \nn\\&\quad + \left[\frac{\cos\phi}{v_L}\sin\phi_{T_L}\cos\phi_{T_L} m_{t} + \frac{\sin\phi}{v_R}\left( \sin\phi_{T_L}\cos\phi_{T_L} \cos^2\beta_{T_R} m_{t} \right. \right. \nn\\&\hspace{120pt} \left.\left. -\cos^2\phi_{T_L} \sin\beta_{T_R} \cos\beta_{T_R}m_{t'} \right) \right](\bar{t}^\prime_Lt_R + \bar{t}_R t'_L) h \nn\\&\quad  -\left[\frac{\cos\phi}{v_L}\sin^2\phi_{T_L} m_{t'} - \frac{\sin\phi}{v_R}\left( \cos^2\phi_{T_L} \sin^2\beta_{T_R} m_{t'} \right. \right. \nn\\&\hspace{120pt} \left.\left. -\sin\phi_{T_L} \cos\phi_{T_L} \sin\beta_{T_R} \cos\beta_{T_R}m_{t} \right) \right]\bar{t}'t'h, \label{higgs top sector}
\end{align}
where we substitute the elements of $\mathcal{Z}_{T_L}$ and $\mathcal{Z}_{T_R}$ in Eqs.(\ref{ZTL}) and (\ref{ZTR}), respectively.
Then, we take the approximation for the mixing angles with Eqs.(\ref{approximate mixing angle}) and (\ref{approximate mixing angle beta}). In addition, using the hierarchy in the top sector that is $v_L < M_T < v_R$, and approximation of mixing angle $\phi$ in Eq.(\ref{phi approximate}), we obtain the interaction between Higgs and top-sector quarks as follows
\begin{align}
	\mathcal{L}_{ht} &\simeq - \cos\phi \frac{m_t}{v_L} \left( 1 -\frac{\lambda_{LR}}{\lambda_R} \frac{M_T^2}{m^2_{u_R}}\frac{v_L^2}{v_R^2}\right)\bar{t}th - \cos\phi \frac{M_T}{m_{u_R}} \left(1+\frac{\lambda_{LR}}{\lambda_R} \frac{v_L^2}{v_R^2}\right) (\bar{t}_Lt'_R + \bar{t}'_R t_L) h \nn\\ &\quad -\cos\phi \frac{M_T}{m_{u_R}} \frac{v_L}{v_R} \left(1 + \frac{\lambda_{LR}}{\lambda_R}  \right) (\bar{t}^\prime_Lt_R + \bar{t}_R t'_L) h - \cos\phi \frac{m_{t'}}{v_R} \frac{v_L}{v_R}\left(\frac{M_T^2}{m^2_{u_R}} - \frac{\lambda_{LR}}{\lambda_R} \right)\bar{t}'t'h. \label{top sector higgs FCNC} 
\end{align}
In this expression,  we also assume that $Y^3_{u_L} \simeq Y^3_{u_R} \simeq 1$. From Eq.(\ref{top sector higgs FCNC}) we extract some useful informations regarding our model. Higgs and top quark pairs coupling receives a small correction. While Higgs and heavy top quark pairs coupling receives an overall suppression of $\mathcal{O}\left(v_L /v_R \right). $   Another important point is the tree-level FCNC interaction is suppressed. The Higgs FCNC of $\bar{t}^\prime_Lt_R$ and $\bar{t}_R t'_L$ type is more suppressed by a factor $\mathcal{O}\left(v_L /v_R \right) $ compared to the   $\bar{t}_Lt'_R$ and $\bar{t}'_R t_L$ type. 

\subsubsection{Bottom sector}  
 In the same way, from Eq.(\ref{E2}) we collect the interactions between Higgs boson and the bottom quark $(b)$ and heavy bottom quark $(b')$. By expressing $\mathcal{Z}_{B_L}$ and $\mathcal{Z}_{B_R}$ in terms of their elements, we obtain,
\begin{align}
	\mathcal{L}_{hH} \supset \mathcal{L}_{hb}&=-\left[\frac{\cos\phi}{v_L}\cos^2\phi_{B_L} m_b -\frac{\sin\phi}{v_R}\left( \sin^2\phi_{B_L} \cos^2\beta_{B_R} m_b \right. \right. \nn\\&\hspace{120pt} \left.\left. -\sin\phi_{B_L} \cos\phi_{B_L} \sin\beta_{B_R} \cos\beta_{B_R}m_{b'} \right) \right]\bar{b}bh \nn\\&\quad + \left[\frac{\cos\phi}{v_L}\sin\phi_{B_L}\cos\phi_{B_L} m_{b'} + \frac{\sin\phi}{v_R}\left( \sin\phi_{B_L}\cos\phi_{B_L} \sin^2\beta_{B_R} m_{b'} \right. \right. \nn\\&\hspace{120pt} \left.\left. -\sin^2\phi_{B_L} \sin\beta_{B_R} \cos\beta_{B_R}m_{b} \right) \right](\bar{b}_Lb'_R +h.c.) h \nn\\&\quad + \left[\frac{\cos\phi}{v_L}\sin\phi_{B_L}\cos\phi_{B_L} m_{b} + \frac{\sin\phi}{v_R}\left( \sin\phi_{B_L}\cos\phi_{B_L} \cos^2\beta_{B_R} m_{b} \right. \right. \nn\\&\hspace{120pt} \left.\left. -\cos^2\phi_{B_L} \sin\beta_{B_R} \cos\beta_{B_R}m_{b'} \right) \right](\bar{b}^\prime_Lb_R +h.c.) h \nn\\&\quad  -\left[\frac{\cos\phi}{v_L}\sin^2\phi_{B_L} m_{b'} - \frac{\sin\phi}{v_R}\left( \cos^2\phi_{B_L} \sin^2\beta_{B_R} m_{b'} \right. \right. \nn\\&\hspace{120pt} \left.\left. -\sin\phi_{B_L} \cos\phi_{B_L} \sin\beta_{B_R} \cos\beta_{B_R}m_{b} \right) \right]\bar{b}'b'h. \label{higgs bottom sector}
\end{align}
We use the approximation for the mixing angles in Eqs.(\ref{approximate mixing angle}), (\ref{approximate mixing angle beta}), and (\ref{phi approximate}). In addition, by using the hierarchy in the bottom sector $v_L < v_R \ll M_B$, we obtain the interaction between Higgs and bottom-sector quarks as follows,
\begin{align}
	\mathcal{L}_{hb} &\simeq - \cos\phi \frac{m_b}{v_L} \left( 1 -\frac{\lambda_{LR}}{\lambda_R} \frac{v_L^2}{v_R^2}\right)\bar{b}bh - \cos\phi \frac{m_b m_{b'}}{m_{d_L}m_{d_R}} \left(1+\frac{\lambda_{LR}}{\lambda_R} \frac{v_L^2}{M_B^2}\right) (\bar{b}_Lb'_R + \bar{b}'_R b_L) h \nn\\ &\quad - \frac{v_L}{v_R} \left(\frac{\lambda_{LR}}{\lambda_R} + \frac{v_R^2}{M_B^2}  \right) (\bar{b}^\prime_Lb_R + \bar{b}_R b'_L) h - \cos\phi \frac{m_{d_L}}{m_{b'}} \left(1- \frac{\lambda_{LR}}{\lambda_R} \right)\bar{b}'b'h. \label{bottom sector higgs FCNC} 
\end{align}
Similar to the top sector, the interaction between the Higgs and the bottom quark pairs receives a small correction compared to the SM. The interaction between the Higgs and the heavy bottom quark pairs is suppressed by a factor $\mathcal{O}(v_L/M_B)$. The Higgs FCNC of $\bar{b}^\prime_Lb_R$ and $\bar{b}_R b'_L$ type is suppressed by a factor $\mathcal{O}(v_L/v_R)$. On the other hand, the Higgs FCNC of $\bar{b}_Lb'_R$ and $\bar{b}'_R b_L$ type  is not suppressed. This is because we assume $Y^3_{d_L} \simeq 1 $. 

\subsection{Z FCNC}
In this subsection we discuss about interaction between the $Z$ boson and quarks. We begin by extracting the interaction terms between $Z_L - Z_R$ and quarks from Eq.(\ref{quark Lagrangian5}), which reads as
\begin{align}
	\mathcal{L}_q \supset \mathcal{L}_{Z Z'} &=- \left[\frac{g_L}{2\cos\theta_W}(j_{3L}^\mu) - e \tan \theta_W (j^\mu_{\text{em}})\right] Z_{L\mu} \nn\\ &\quad - \left[\frac{g_R}{2\cos\theta_R}(j_{3R}^\mu) - g' \tan \theta_R \left(j^\mu_{\text{em}} - \frac{1}{2}(j^\mu_{3L}) \right)  \right] Z_{R\mu}  \label{F1}.
\end{align}
Here $j_{3L}^\mu, j_{3R}^\mu,$ and $j^\mu_{\text{em}}$ are defined in Eqs.(\ref{left handed weak isospin final})-(\ref{electromagnetic current final}), respectively. Next, we change the basis from the $Z_L - Z_R$ basis to the $Z - Z'$ basis using Eq.(\ref{Z mixing matrix}), and it leads to,
\begin{align}
	\mathcal{L}_{Z Z'} &=-\left[\frac{1}{2\cos \theta_W}(g_L \cos\theta - e \tan \theta_R \sin\theta)j^{\mu}_{3L} -\frac{g_R \sin\theta}{2\cos\theta_R}j^\mu_{3R}\right.\nn\\ &\hspace{100pt} \left.-\frac{e}{\cos\theta_W} (\sin\theta_W \cos\theta - \tan\theta_R \sin\theta) j^\mu_{\text{em}} \right]Z_\mu \nn\\ &\quad -\left[\frac{1}{2\cos \theta_W}(g_L \sin\theta + e \tan \theta_R \cos\theta)j^{\mu}_{3L} +\frac{g_R \cos\theta}{2\cos\theta_R}j^\mu_{3R}\right.\nn\\ &\hspace{100pt} \left.-\frac{e}{\cos\theta_W} (\sin\theta_W \sin\theta + \tan\theta_R \cos\theta) j^\mu_{\text{em}} \right]Z'_\mu \label{F2}.
\end{align}
In this discussion, we will focus on the interaction between SM $Z$-boson with quarks. We expressed the $Z$-boson interaction in terms of vector and axial-vector couplings as follows,
\begin{align}
	\mathcal{L}_{Z Z'} \supset \mathcal{L}^{Z}_{\bar{q}q} &= -\frac{g_L}{2 \cos\theta_W} \sum_{\alpha,\beta=1}^{4} \overline{(\hat{u}^m)^\alpha} \gamma^\mu  \bigg[(g_V)^{\alpha\beta}_{u} -(g_A)^{\alpha\beta}_{u} \gamma^5 \bigg](\hat{u}^m)^\beta Z_{\mu}\nn\\ &\quad - \frac{g_L}{2 \cos\theta_W} \sum_{\alpha,\beta=1}^{4} \overline{(\hat{d}^m)^\alpha } \gamma^\mu  \bigg[(g_V)^{\alpha\beta}_{d} -(g_A)^{\alpha\beta}_{d} \gamma^5 \bigg](\hat{d}^m)^\beta Z_{\mu}\label{vector and axial vector},
\end{align}
where,
\begin{align}
	(g_V)^{\alpha\beta}_u &= \frac{1}{2} \left( 	(\kappa_{T_L})^{\alpha\beta}  - (\kappa_{T_R})^{\alpha\beta} \right) - 2 \kappa Q_u \delta^{\alpha\beta} \label{vector coupling up type},\\
	(g_A)^{\alpha\beta}_u &= \frac{1}{2} \left( 	(\kappa_{T_L})^{\alpha\beta}  + (\kappa_{T_R})^{\alpha\beta} \right) \label{axial vector coupling up type},\\
	(g_V)^{\alpha\beta}_d &= -\frac{1}{2} \left( (\kappa_{B_L})^{\alpha\beta} - 	(\kappa_{B_R})^{\alpha\beta} \right) - 2 \kappa Q_d \delta^{\alpha\beta} \label{vector coupling down type}, \\
	(g_A)^{\alpha\beta}_d &= -\frac{1}{2} \left( (\kappa_{B_L})^{\alpha\beta} + 	(\kappa_{B_R})^{\alpha\beta} \right) \label{axial vector coupling down type}, \\
	(\kappa_{T_L})^{\alpha\beta} &= (\cos \theta - \sin \theta_W \tan \theta_R \sin \theta) (\mathcal{Z}^{\text{all}}_{T_L})^{\alpha\beta} \label{kappa TL},\\
	(\kappa_{T_R})^{\alpha\beta} &= \frac{\sin \theta_W \sin \theta}{\sin \theta_R \cos \theta_R} (\mathcal{Z}^{\text{all}}_{T_R})^{\alpha\beta} \label{kappa TR},\\
	(\kappa_{B_L})^{\alpha\beta} &= (\cos \theta - \sin \theta_W \tan \theta_R \sin \theta) (\mathcal{Z}^{\text{all}}_{B_L})^{\alpha\beta}\label{kappa BL},\\
	(\kappa_{B_R})^{\alpha\beta} &= \frac{\sin \theta_W \sin \theta}{\sin \theta_R \cos \theta_R} (\mathcal{Z}^{\text{all}}_{B_R})^{\alpha\beta}\label{kappa BR},\\
	\kappa & = \sin^2 \theta_W \cos \theta - \sin \theta_W \tan \theta_R \sin \theta \label{kappa}.
\end{align}
The matrix forms of $4\times 4$ unitary matrices  $\mathcal{Z}^{\text{all}}_{T_L}, \mathcal{Z}^{\text{all}}_{B_L}, \mathcal{Z}^{\text{all}}_{T_R}$, and $\mathcal{Z}^{\text{all}}_{B_R}$ are given as follows,
\begin{align}
	\mathcal{Z}^{\text{all}}_{T_L} = \left(\begin{array}{cc}
		I_2 & 0_2 \\
		0_2 & \mathcal{Z}_{T_L}
	\end{array} \right), 	\mathcal{Z}^{\text{all}}_{T_R} = \left(\begin{array}{cc}
	I_2 & 0_2 \\
	0_2 & \mathcal{Z}_{T_R}
	\end{array} \right),	\mathcal{Z}^{\text{all}}_{B_L} = \left(\begin{array}{cc}
	I_2 & 0_2 \\
	0_2 & \mathcal{Z}_{B_L}
	\end{array} \right),	\mathcal{Z}^{\text{all}}_{B_R} = \left(\begin{array}{cc}
	I_2 & 0_2 \\
	0_2 & \mathcal{Z}_{B_R}
	\end{array} \right),
\end{align}
where $I_2$ and $0_2$ are $2\times 2$ unit matrix and zero matrix respectively. The $2 \times 2$ submatrix $\mathcal{Z}_{T_L}, \mathcal{Z}_{B_L}, \mathcal{Z}_{T_R}$, and $\mathcal{Z}_{B_R}$ are given in Eqs.(\ref{ZTL}),(\ref{ZBL}),(\ref{ZTR}),(\ref{ZBR}) respectively. $Q_u=2/3$, $Q_d=-1/3$ are the electric charge of up-type and down-type quarks.

\subsubsection{Up sector}
In this part, we analyze the interaction between $Z$-boson with the up sector in our model. From Eq.(\ref{vector and axial vector}), it reads as,
\begin{align}
	\mathcal{L}^{Z}_{\bar{q}q}\supset 	\mathcal{L}^{Z}_t &= -\frac{g_L}{2 \cos\theta_W} \left\lbrace  \overline{(\hat{u}^m)^1} \gamma^\mu  \bigg[(g_V)^{11}_{u} -(g_A)^{11}_{u} \gamma^5 \bigg](\hat{u}^m)^1 \right. \nn\\ &\quad \left. + \overline{(\hat{u}^m)^2} \gamma^\mu  \bigg[(g_V)^{22}_{u} -(g_A)^{22}_{u} \gamma^5 \bigg](\hat{u}^m)^2 +  \overline{t} \gamma^\mu  \bigg[(g_V)^{33}_{u} -(g_A)^{33}_{u} \gamma^5 \bigg]t \right. \nn\\ &\quad \left. + \overline{t} \gamma^\mu  \bigg[(g_V)^{34}_{u} -(g_A)^{34}_{u} \gamma^5 \bigg]t'  + \overline{t'} \gamma^\mu  \bigg[(g_V)^{43}_{u} -(g_A)^{43}_{u} \gamma^5 \bigg]t \right. \nn\\ &\quad \left. + \overline{t'} \gamma^\mu  \bigg[(g_V)^{44}_{u} -(g_A)^{44}_{u} \gamma^5 \bigg]t'  \right\rbrace   Z_{\mu},
\end{align}
where the vector coupling $(g_V)_u$ and axial-vector coupling $(g_A)_u$ are defined in Eqs.(\ref{vector coupling up type})-(\ref{axial vector coupling up type}) respectively. By using the definition of $\kappa_{T_L},\kappa_{T_R}$ and $\kappa$ which are written in Eqs.(\ref{kappa TL}),(\ref{kappa TR}), and (\ref{kappa}), we obtain
\begin{align}
	(\kappa_{T_L})^{11} &= 	(\kappa_{T_L})^{22}= \cos \theta\left( 1 - \sin \theta_W \tan \theta_R \mathcal{O}\left(\frac{v_L^2}{v_R^2} \right) \right), \\
	(\kappa_{T_R})^{11} &= (\kappa_{T_R})^{22}=\frac{\sin \theta_W \cos \theta}{\sin \theta_R \cos \theta_R} \mathcal{O}\left(\frac{v_L^2}{v_R^2}\right),  \\
	(\kappa_{T_L})^{33} &= \cos \theta\left( 1 - \sin \theta_W \tan \theta_R \mathcal{O}\left(\frac{v_L^2}{v_R^2} \right) \right), \\
	(\kappa_{T_R})^{33} &= \frac{\sin \theta_W \cos \theta}{\sin \theta_R \cos \theta_R} \mathcal{O}\left(\frac{v_L^2}{v_R^2}\right) \frac{M_T^2}{m^2_{u_R}}, \\
	(\kappa_{T_L})^{34} &=(\kappa_{T_L})^{43} = \cos \theta\left( 1 - \sin \theta_W \tan \theta_R \mathcal{O}\left(\frac{v_L^2}{v_R^2} \right) \right) \frac{m_{u_L} M_T}{m^2_{u_R}}\label{kappa TL 34},\\
	(\kappa_{T_R})^{34} &= (\kappa_{T_R})^{43}=-\frac{\sin \theta_W \cos \theta}{\sin \theta_R \cos \theta_R} \mathcal{O}\left(\frac{v_L^2}{v_R^2}\right) \frac{M_T}{m_{u_R}}\label{kappa TR 34},\\
	(\kappa_{T_L})^{44} &= \cos \theta\left( 1 - \sin \theta_W \tan \theta_R \mathcal{O}\left(\frac{v_L^2}{v_R^2} \right) \right) \frac{m^2_{u_L} M_T^2}{m^4_{u_R}}, \\
	(\kappa_{T_R})^{44} &= \frac{\sin \theta_W \cos \theta}{\sin \theta_R \cos \theta_R} \mathcal{O}\left(\frac{v_L^2}{v_R^2}\right),  \\
	\kappa & = \cos \theta \left(\sin^2 \theta_W - \sin \theta_W \tan \theta_R \mathcal{O}\left(\frac{v_L^2}{v_R^2}\right) \right).
\end{align}
Here we write the suppression coming from the small mixing angle $\theta$ as $\mathcal{O}(v^2_L/v_R^2)$. The exact form of the mixing angle $\theta$ is given in Eq.(\ref{Z mixing angle exact}). From Eqs.(\ref{kappa TL 34}) and (\ref{kappa TR 34}), the $\kappa_{T_L}$ and $\kappa_{T_R}$ terms related to the $Z$-boson FCNC process with the top and heavy-top quarks are suppressed by $\mathcal{O}(v_L M_T / v_R^2)$ and $\mathcal{O}(v^2_L M_T/v_R^3)$, respectively. This indicates that the $Z$-mediated FCNC process in the up sector is suppressed within our model. In addition, the interaction between $Z$-boson and heavy top quark is also suppressed. Moreover, the deviation of the SM-like terms in $(\kappa_{T_L})^{ii}$ and $\kappa$, with $i\in \{1,2,3\}$ are suppressed by a factor $\mathcal{O}(v_L^2/v_R^2)$.  
\subsubsection{Down sector}
In this part, we analyze the interaction between $Z$-boson and the down sector in our model. From Eq.(\ref{vector and axial vector}), we extract,
\begin{align}
	\mathcal{L}^{Z}_{\bar{q}q}\supset 	\mathcal{L}^{Z}_b &= -\frac{g_L}{2 \cos\theta_W} \left\lbrace  \overline{(\hat{d}^m)^1} \gamma^\mu  \bigg[(g_V)^{11}_{d} -(g_A)^{11}_{d} \gamma^5 \bigg](\hat{d}^m)^1 \right. \nn\\ &\quad \left. + \overline{(\hat{d}^m)^2} \gamma^\mu  \bigg[(g_V)^{22}_{d} -(g_A)^{22}_{d} \gamma^5 \bigg](\hat{d}^m)^2 +  \overline{b} \gamma^\mu  \bigg[(g_V)^{33}_{d} -(g_A)^{33}_{d} \gamma^5 \bigg]b \right. \nn\\ &\quad \left. + \overline{b} \gamma^\mu  \bigg[(g_V)^{34}_{d} -(g_A)^{34}_{d} \gamma^5 \bigg]b'  + \overline{b'} \gamma^\mu  \bigg[(g_V)^{43}_{d} -(g_A)^{43}_{d} \gamma^5 \bigg]b \right. \nn\\ &\quad \left. + \overline{b'} \gamma^\mu  \bigg[(g_V)^{44}_{d} -(g_A)^{44}_{d} \gamma^5 \bigg]b'  \right\rbrace   Z_{\mu},
\end{align}
where the vector coupling $(g_V)_d$ and axial-vector coupling $(g_A)_d$ are defined in Eqs.(\ref{vector coupling down type})-(\ref{axial vector coupling down type}) respectively. By using the definition of $\kappa_{B_L},\kappa_{B_R}$ and $\kappa$ written in Eqs.(\ref{kappa BL}),(\ref{kappa BR}), and (\ref{kappa}) respectively, we obtain
\begin{align}
	(\kappa_{B_L})^{11} &= 	(\kappa_{B_L})^{22}= \cos \theta\left( 1 - \sin \theta_W \tan \theta_R \mathcal{O}\left(\frac{v_L^2}{v_R^2} \right) \right), \\
	(\kappa_{B_R})^{11} &= (\kappa_{B_R})^{22}=\frac{\sin \theta_W \cos \theta}{\sin \theta_R \cos \theta_R} \mathcal{O}\left(\frac{v_L^2}{v_R^2}\right),  \\
	(\kappa_{B_L})^{33} &= \cos \theta\left( 1 - \sin \theta_W \tan \theta_R \mathcal{O}\left(\frac{v_L^2}{v_R^2} \right)\right),  \\
	(\kappa_{B_R})^{33} &= \frac{\sin \theta_W \cos \theta}{\sin \theta_R \cos \theta_R} \mathcal{O}\left(\frac{v_L^2}{v_R^2}\right),  \\
	(\kappa_{B_L})^{34} &=(\kappa_{B_L})^{43} = \cos \theta\left( 1 - \sin \theta_W \tan \theta_R \mathcal{O}\left(\frac{v_L^2}{v_R^2} \right) \right) \frac{m_{d_L}}{M_B} \label{kappa BL 34},\\
	(\kappa_{B_R})^{34} &= (\kappa_{B_R})^{43}=-\frac{\sin \theta_W \cos \theta}{\sin \theta_R \cos \theta_R} \mathcal{O}\left(\frac{v_L^2}{v_R^2}\right) \frac{m_{d_R}}{M_B}\label{kappa BR 34}, \\
	(\kappa_{B_L})^{44} &= \cos \theta\left( 1 - \sin \theta_W \tan \theta_R \mathcal{O}\left(\frac{v_L^2}{v_R^2} \right)  \right) \frac{m^2_{d_L}}{M^2_B}, \\
	(\kappa_{B_R})^{44} &= \frac{\sin \theta_W \cos \theta}{\sin \theta_R \cos \theta_R} \mathcal{O}\left(\frac{v_L^2}{v_R^2}\right) \frac{m^2_{d_R}}{M^2_B},  \\
	\kappa & = \cos \theta \left(\sin^2 \theta_W - \sin \theta_W \tan \theta_R \mathcal{O}\left(\frac{v_L^2}{v_R^2}\right) \right).
\end{align}
The FCNC process in the down sector is suppressed, similar to the up sector. As shown in Eqs.(\ref{kappa BL 34}) and (\ref{kappa BR 34}), the $\kappa_{B_L}$ and $\kappa_{B_R}$ are suppressed by factor $\mathcal{O}(v_L / M_B)$ and $\mathcal{O}(v^2_L/v_R M_B)$, respectively. In addition, the interaction between $Z$-boson and heavy bottom quark is also suppressed. Furthermore, the deviation of the SM-like terms in $(\kappa_{B_L})^{ii}$ and $\kappa$, with $i\in \{1,2,3\}$ are suppressed by a factor $\mathcal{O}(v_L^2/v_R^2)$.  

\section{Conclusion}
We have presented a systematic  analysis of the quark sector in the universal seesaw model. We derived the Lagrangian of the model, including the quark sector, Higgs sector, and kinetic terms of the gauge fields. We start by writing the Lagrangian which is invariant under $\mathrm{SU(2)_L\times SU(2)_R \times U(1)_{Y'}}$. After $\mathrm{SU(2)_R}$ Higgs doublet acquires non-zero vev, we obtain the Lagrangian, which is invariant under SM gauge symmetry. Furthermore,  the SM gauge group is broken into $\mathrm{U(1)_{\text{em}}}$ after $\mathrm{SU(2)_L}$ Higgs doublet acquires non-zero vev. In the gauge interactions sector, we classify the terms based on the number of fields, such as linear, quadratic, cubic, and quartic interactions. In addition, we found that the massless Nambu-Goldstone bosons are mixed to become new states $\chi_{Z}$ and $\chi_{Z'}$. We have shown clearly that $\chi_Z$ and $\chi_{Z'}$ become the longitudinal components of the massive $Z$ and $Z'$ bosons, respectively.

Our model focuses on the third family of quark sector. Within this framework we explain the hierarchy between the top and bottom quark masses by mixing with the heavy VLQs. We use the direct measurement of the top quark mass and the running mass of the bottom quark. Additionally, the lower bounds on the heavy top and heavy bottom quark masses also serve as constraints. The lower mass limit of the $Z'$-boson, linked to the $W_R$ boson mass, also imposes a stringent constraint on $v_R$. By setting $g_R$ and the Yukawa couplings equal to 1, the lower limit of $v_R$ is 10 TeV in this model. We obtained that the heavy top quark mass is in the order of $v_R$ ($m_{t'}=7.13$ TeV) and the heavy bottom mass is in the order of $M_B$ ($m_{b'}=293.82$ TeV). We confirmed that the hierarchy of VLQ's mass parameters, $v_L$, and $v_R$ in our model is $v_L<M_T< v_R\ll M_B$. 

Moreover, the presence of VLQs in the model induces the flavor-changing neutral current (FCNC) at the tree level. In the SM, the FCNC process is highly suppressed and only occurs at the loop level due to the GIM mechanism. In our model, we have shown that the $Z$-boson mediated FCNC process is suppressed for both (up and down) sectors. The deviation from the SM values are suppressed by $\mathcal{O}(v_L^2/v_R^2)$, which comes from the small mixture in the lighter mass eigenstate $Z$ from $Z_R$. On the other hand, Higgs mediated FCNC of $\bar{b}_Lb'_R$ and $\bar{b}'_R b_L$ type are not suppressed when $Y^3_{d_L}\simeq 1$.

\section*{Acknowledgement}
We would like to thank Apriadi Salim Adam, Yuta Kawamura, Yusuke Shimizu, Hironori Takei, and Kei Yamamoto for the useful comment and discussion.  

\begin{appendices} \renewcommand{\thesection}{\Alph{section}.}
	\counterwithin*{equation}{section}
	\renewcommand\theequation{\thesection\arabic{equation}}
	\section{Weak-basis of Yukawa interaction}\label{sec:appendix weak basis}
	In this appendix, we show how to obtain the Yukawa interaction which is written in Eq.(\ref{quark Lagrangian}). We start from the general Yukawa interaction terms,
	\begin{align}
		\mathcal{L}_{\text{YM}}&= - \overline{q_L^i}y_{u_L}^i\tilde{\phi}_L T_R - \overline{T_L} y_{u_R}^{i\ast}\tilde{\phi}_R^\dagger q_R^i - \overline{T_L}M_T T_R -h.c. \nn\\&\hspace{10pt} - \overline{q^i_{L}} y^i_{d_L} \phi_L B_R - \overline{B_L} y^{i\ast}_{d_R} \phi_R^\dagger q_R^i- \overline{B_L} M_B B_R -h.c.. \label{general yukawa}
	\end{align}
	The Yukawa couplings are general complex vectors in $\mathbb{C}^3$ with the following parameterization,
	\begin{align}
		y_{u_{L(R)}}^i=\mathbf{y}_{u_{L(R)}}&=\left(\begin{array}{c}
		\sin\theta^u_{L(R)} \sin \phi^u_{L(R)} e^{i \alpha^1_{u_{L(R)}}} \\
		\sin\theta^u_{L(R)} \cos \phi^u_{L(R)} e^{i \alpha^2_{u_{L(R)}}}	\\
			\cos\theta^u_{L(R)} e^{i \alpha^3_{u_{L(R)}}}
		\end{array} \right) Y^3_{u_{L(R)}}, \\
			y_{d_{L(R)}}^i=\mathbf{y}_{d_{L(R)}}&=\left(\begin{array}{c}
			\sin\theta^d_{L(R)} \sin \phi^d_{L(R)} e^{i \alpha^1_{d_{L(R)}}} \\
			\sin\theta^d_{L(R)} \cos \phi^d_{L(R)} e^{i \alpha^2_{d_{L(R)}}}	\\
			\cos\theta^d_{L(R)} e^{i \alpha^3_{d_{L(R)}}}
		\end{array} \right) Y^3_{d_{L(R)}} \label{down yukawa general},
	\end{align}
	where $Y^3_{u_{L(R)}}$ and $Y^3_{d_{L(R)}}$ are real positive numbers. Define the following weak-basis transformations (WBTs) as follows,
	\begin{align}
		(q'_L)^i &= e^{-i \alpha^i_{u_L}} q^i_L,\\
		(q'_R)^i &= e^{-i \alpha^i_{u_R}} q^i_R.
	\end{align}
	Applying this WBT into Eq.(\ref{general yukawa}), we obtain
		\begin{align}
		\mathcal{L}_{\text{YM}}&= - \overline{(q'_L)^i}(y'_{u_L})^i\tilde{\phi}_L T_R - \overline{T_L} (y'_{u_R})^{i\ast}\tilde{\phi}_R^\dagger (q'_R)^i - \overline{T_L}M_T T_R -h.c. \nn\\&\hspace{12pt} - \overline{(q'_{L})^i} y^i_{d_L}  \phi_L B_R - \overline{B_L} y^{i\ast}_{d_R} \phi_R^\dagger (q'_R)^i- \overline{B_L} M_B B_R -h.c. \label{general yukawa2},
	\end{align}
	where
	\begin{align}
		(y'_{u_L})^i &= y^i_{u_L} e^{-i \alpha^i_{u_L}}, \\
		(y'_{u_R})^i &= y^i_{u_R} e^{-i \alpha^i_{u_R}}
	\end{align}
	are real vectors. On the other hand, $y^i_{d_L}$ and $y^i_{d_R}$ remain complex vectors with the redefined phases.
	
	Next we write the $(y'_{u_L})^i$ Yukawa coupling explained above as,
	\begin{align}
		(y'_{u_{L}})^i &=\left(\begin{array}{c}
			\sin\theta^u_{L} \sin \phi^u_{L}  \\
			\sin\theta^u_{L} \cos \phi^u_{L} 	\\
			\cos\theta^u_{L} 
		\end{array} \right) Y^3_{u_{L}} \nn \\
		&=\mathbf{e}^u_{L_3} Y^3_{u_{L}} \label{up left yukawa}
	\end{align}
	and defining another WBT,
	\begin{equation}
		(q'_L)^i = (V_{u_L})^{ij} (q^{\prime\prime}_L)^j,
	\end{equation}
	where in general $V_{u_L}$ is $3\times 3$ unitary matrix formed by three orthonormal vectors with the third column chosen as $\mathbf{e}^u_{L_3}$ in Eq.(\ref{up left yukawa}),
	\begin{equation}
		V_{u_L} = \left(\begin{array}{ccc}
		\mathbf{e}^u_{L_1}	& \mathbf{e}^u_{L_2} & \mathbf{e}^u_{L_3}
		\end{array} \right), \label{matrix Vul} 
	\end{equation} 
	which leads the product $(V^\dagger_{u_L})^{ji}(y'_{u_{L}})^i = \delta^{j3}  Y^3_{u_{L}}  $. 
	
	For the $(y'_{u_R})^i$, Yukawa coupling can be derived similarly by changing $L\rightarrow R$ in Eq.(\ref{up left yukawa}) - (\ref{matrix Vul}). For the down-sector, the product of Eq.(\ref{matrix Vul}) and the down-type Yukawa coupling yields down-type Yukawa coupling on another basis. For example, $(V^\dagger_{u_L})^{ji}(y_{d_{L}})^i = (y^{\prime\prime}_{d_{L}})^j$. Therefore, the Lagrangian in Eq.(\ref{general yukawa2}) becomes,
	\begin{align}
		\mathcal{L}_{\text{YM}}&= - Y^3_{u_{L}} \overline{(q^{\prime\prime}_L)^3}\tilde{\phi}_L T_R -Y^3_{u_{R}} \overline{T_L} \tilde{\phi}_R^\dagger (q^{\prime\prime}_R)^3 - \overline{T_L}M_T T_R -h.c. \nn\\&\hspace{12pt} - \overline{(q^{\prime\prime}_{L})^i} (y^{\prime\prime}_{d_{L}})^i  \phi_L B_R - \overline{B_L} (y^{\prime\prime}_{d_{R}})^{i\ast} \phi_R^\dagger (q^{\prime\prime}_R)^i- \overline{B_L} M_B B_R -h.c., \label{general yukawa3}
	\end{align}
	and it has the form that the Yukawa couplings of up-type quark doublet ($Y^3_{u_L}$ and $Y^3_{u_R}$) are given by the real positive numbers while the Yukawa couplings of down-type quark are general complex vectors as written in Eq.(\ref{quark Lagrangian}).
	
	\section{Parameterization of $V_{d_R}$ and $V_{d_L}$}\label{sec:appendix removing phase}
	In this appendix, we explain in more detail how to parameterize and remove the unphysical phases of $V_{d_R} $ and $V_{d_L}$. Both $V_{d_R} $ and $V_{d_L}$ have the following form,
\begin{equation}
	V = \left(\begin{array}{ccc}
		\mathbf{v}_1	& \mathbf{v}_2 & \mathbf{v}_3
	\end{array} \right)\label{V1},
\end{equation} 
where the third column is related to either $y_{d_R}$ or $y_{d_L}$ and is parameterized by,
\begin{equation}
	\mathbf{v}_3 = \left(\begin{array}{c}
		\sin\theta \sin \phi e^{i \alpha_1} \\
		\sin\theta \cos \phi e^{i \alpha_2}	\\
		\cos\theta e^{i \alpha_3}
	\end{array} \right). 
\end{equation}
	Since $V$ is a unitary matrix, the column vectors satisfy $\mathbf{v^\dagger_i}\cdot \mathbf{v_j} = \delta_{i j}$ and $V$ has matrix form as follow,
	\begin{equation}
		V=(\alpha_1,\alpha_2,\alpha_3)R_{12}(\phi)R_{23}(\theta)(0,\delta,0)R_{12}(\alpha)(\rho,\sigma,0) \label{V},
	\end{equation}
	where $(\alpha_1,\alpha_2,\alpha_3) = \mathrm{diag}(e^{i\alpha_1},e^{i\alpha_2},e^{i\alpha_3})$; $(0,\delta,0) = \mathrm{diag}(1,e^{i\delta},1)$;$(\rho,\sigma,0) =\mathrm{diag} (e^{i\rho},e^{i\sigma},1)$  and
	
	\begin{align}
		R_{12}(\phi) =\left( \begin{array}{ccc}
			\cos\phi& \sin\phi &0 \\
			-\sin\phi& \cos\phi &0  \\
			0& 0 & 1
		\end{array}\right)&,\quad 	R_{23}(\theta) =\left( \begin{array}{ccc}
			1& 0 &0 \\
			0& \cos\theta &\sin\theta  \\
			0& -\sin\theta & \cos\theta
		\end{array}\right), \nn\\	R_{12}(\alpha) =&\left( \begin{array}{ccc}
			\cos\alpha& \sin\alpha &0 \\
			-\sin\alpha& \cos\alpha &0  \\
			0& 0 & 1
		\end{array}\right).
	\end{align}
	We have the freedom to rotate $V$ by $\mathrm{U(2)}$ transformations from both sides. As shown in Eqs.(\ref{removing phase Vdr}) and (\ref{removing phase Vdl}), we can remove the unphysical phases and angles in Eq.(\ref{V}) by following,
\begin{equation}
	\widetilde{V} = \widetilde{U}^\dagger V \widetilde{W},
\end{equation}
where $\widetilde{U}$ and $\widetilde{W}$ are $3\times 3$ unitary matrices which have the following expressions,
\begin{align}
	\widetilde{U}^\dagger &= (0,\frac{\alpha_3}{2},0)R^{-1}_{12}(\phi)(-\alpha_1,-\alpha_2,0),\nn\\
	\widetilde{W} &=(-\rho,-\sigma,0)R^{-1}_{12}(\alpha)(0,-\delta,0)(0,-\frac{\alpha_3}{2},0)\label{B9}.
\end{align} 
Thus, we obtain,
\begin{equation}
	\widetilde{V} = \left( \begin{array}{ccc}
		1 & 0 & 0 \\
		0 & \cos\theta & \sin\theta e^{i \frac{\alpha_3}{2} }  \\
		0 & -\sin\theta e^{i \frac{\alpha_3}{2} }  & \cos\theta e^{i\alpha_3}
	\end{array}\right). \label{B10}
\end{equation} 
	
   \section{Diagonalization of quark mass matrix}\label{appendix quark mass}
   In this appendix, we derive the exact mass eigenvalues of the top-bottom SM quarks and the heavy VLQ partners, as well as the matrices used for the diagonalization procedure. We will show the diagonalization procedure for the top sector. The bottom sector can be done similarly because the form of $\mathbb{M}_b$ is the same as $\mathbb{M}_t$. We start from Eq.(\ref{top mass matrix}), explicitly writing the $(W_{T_R})^{43}$ and $(W_{T_R})^{44}$ values,
   \begin{equation}
   		\mathbb{M}_t \equiv \left(\begin{array}{cc}
   		-\frac{Y_{u_L}^3Y_{u_R}^3 v_L v_R}{2m_{u_4}}	&  Y_{u_L}^3\frac{v_L}{\sqrt{2}}\frac{M_T}{m_{u_4}} \\
   		0 & m_{u_4}
   	\end{array} \right) = \left(\begin{array}{cc}
   	-m_{t_1}	&  m_{t_2} \\
   	0 & m_{u_4}
   	\end{array} \right), \label{top mass matrix 2}
   \end{equation}
   where $m_{t_1}$ and $m_{t_2}$ in Eq.(\ref{top mass matrix 2}) are defined as follows,
   \begin{equation}
   	m_{t_1} = \frac{Y_{u_L}^3Y_{u_R}^3 v_L v_R}{2m_{u_4}},\qquad m_{t_2} = Y_{u_L}^3\frac{v_L}{\sqrt{2}}\frac{M_T}{m_{u_4}}.
   \end{equation}
   The top quark mass matrix in Eq.(\ref{top mass matrix 2}) can be diagonalized by bi-unitary transformation, which gives,
   \begin{equation}
   		K^\dagger_{T_L} \mathbb{M}_t K_{T_R} = (m^{\text{diag}}_t) = \text{diag}(m_t,m_{t'}) \label{diagonalizing top mass matrix 2}. \\
   \end{equation}
   Initially, we transform $\mathbb{M}_t$ into a real symmetric matrix by multiplying it on the left side with an orthogonal matrix $S_t$, which yields
   \begin{equation}
   	\mathbb{M}^\prime_t = S_t\mathbb{M}_t,
   \end{equation}
   where,
   \begin{equation}
   	S_t = \left( \begin{array}{cc}
   		\cos \phi_{T_l} &- \sin \phi_{T_l} \\
   		\sin \phi_{T_l} & \cos \phi_{T_l}
   	\end{array} \right).
   \end{equation}
   $\mathbb{M}_t'$ becomes a real symmetric matrix with the following expression
   \begin{equation}
   	\mathbb{M}_t' = \left( \begin{array}{cc}
   		-m_{t_1}\cos \phi_{T_l}  & 	-m_{t_1}\sin \phi_{T_l}   \\
   		-m_{t_1}\sin \phi_{T_l}  & m_{t_2}\sin \phi_{T_l} + m_{u_4} \cos \phi_{T_l}
   	\end{array}\right)	
   \end{equation}
   if the mixing angle satisfies the following condition:
   \begin{equation}
   	\tan \phi_{T_l} = \frac{m_{t_2}}{m_{u_4}-m_{t_1}} \label{theta l}.
   \end{equation}
   Then, a real symmetric matrix can be diagonalized by multiplying from both sides another $2 \times 2$ orthogonal matrix $R_t$ and its transpose,
   \begin{equation} \label{diagonalization 2}
   R_t\hspace{2pt} \mathbb{M}_t'R_t^T = \text{diag}(-m_t , m_{t'}), 
   \end{equation}
   where,
   \begin{equation}
   	R_t = \left( \begin{array}{cc}
   		\cos \phi_{T_R} & \sin \phi_{T_R} \\
   		-\sin \phi_{T_R} & \cos \phi_{T_R}
   	\end{array} \right).
   \end{equation}
   The minus sign inside the diagonal matrix on the right-hand side of Eq.(\ref{diagonalization 2}) arises because the determinant of the top quark mass matrix $\mathbb{M}_t$ is negative. Since $m_t$ is lighter than $m_{t'}$, we assign the minus sign to $m_t$. However, we could eliminate the minus sign by multiplying Eq.(\ref{diagonalization 2}) by $-\tau_3$ on the right side, where $\tau_3$ is the third component of the Pauli matrices. The mixing angle can then be obtained as:
   \begin{equation}
   	\tan 2\phi_{T_R}=\frac{2m_{t_1} m_{t_2}}{m_{u_4}^2+m_{t_2}^2 - m_{t_1}^2}\label{phi TR}.
   \end{equation}
  The eigenvalues of Eq.(\ref{diagonalization 2}) can be computed using the following equation,
  \begin{equation}
  	\lambda^2 - (\text{tr} \mathbb{M}_t')\lambda + \text{det} \mathbb{M}_t' = 0.
  \end{equation}
  After performing the calculations, we obtain
  \begin{align}
  	\lambda_1 &= -m_t =  \frac{\sqrt{m_{t_2}^2 + (m_{u_4} - m_{t_1})^2}}{2} - \frac{\sqrt{m_{t_2}^2 + (m_{u_4} + m_{t_1})^2}}{2},\\
  	\lambda_2 &= m_{t'} =  \frac{\sqrt{m_{t_2}^2 + (m_{u_4} - m_{t_1})^2}}{2} + \frac{\sqrt{m_{t_2}^2 + (m_{u_4} + m_{t_1})^2}}{2}.
  \end{align}
  We can also equivalently express them with the parameters of the mass matrix as follows,
   \begin{align}
  m_t &= - \frac{\sqrt{M_T^2 + (m_{u_R} - m_{u_L})^2}}{2} +  \frac{\sqrt{M_T^2 + (m_{u_R} + m_{u_L})^2}}{2},\\
  m_{t'} &=  \frac{\sqrt{M_T^2 + (m_{u_R} - m_{u_L})^2}}{2} +  \frac{\sqrt{M_T^2 + (m_{u_R} + m_{u_L})^2}}{2}, \label{top sector exact mass eigenvalues}
  \end{align}
  where,
  \begin{equation}
  	m_{u_R} = Y^3_{u_R}\frac{v_R}{\sqrt{2}}, \qquad m_{u_L} = Y^3_{u_L}\frac{v_L}{\sqrt{2}}. \label{mur and mul}
  \end{equation}
  Finally, we can summarize all the matrix transformations explained above as,
   \begin{equation}
   	R_t\hspace{1pt} S_t \hspace{1pt} \mathbb{M}_t\hspace{1pt} R_t^T (-\tau_3) = \text{diag}(m_t,m_{t'}). \label{summarize diagonalization}
   \end{equation}
   Additionally, the product of two orthogonal matrices is also an orthogonal matrix. Then we can define $O_t$ as,
   \begin{equation}
   	O_t=R_t S_t = \left( \begin{array}{cc}
   		\cos \phi_{T_L} & \sin \phi_{T_L} \\
   		-\sin \phi_{T_L} & \cos \phi_{T_L}
   	\end{array}\right)
   \end{equation} 
   with $\phi_{T_L} =\phi_{T_R}-\phi_{T_l}$. Hence, by comparing Eq.(\ref{summarize diagonalization}) and Eq.(\ref{diagonalizing top mass matrix 2}) we obtain the expression for the mixing matrices as follows.
   \begin{align}
   	K^{\dagger}_{T_L} & =
   	\left( \begin{array}{cc}
   		\cos \phi_{T_L} & \sin \phi_{T_L} \\
   		-\sin \phi_{T_L} & \cos \phi_{T_L}
   	\end{array}\right),  \label{KTL explicit}\\
   	K_{T_R} &=  \left( \begin{array}{cc}
   		\cos \phi_{T_R} & -\sin \phi_{T_R} \\
   		\sin \phi_{T_R} & \cos \phi_{T_R}
   	\end{array}\right) \left( \begin{array}{cc}
   		-1 &0 \\
   		0 & 1
   	\end{array}\right) = \left( \begin{array}{cc}
   		-\cos \phi_{T_R} & -\sin \phi_{T_R} \\
   		-\sin \phi_{T_R} & \cos \phi_{T_R}
   	\end{array}\right). \label{KTR explicit}
   \end{align}
   For the bottom sector, we can derive the results similarly by replacing $t$ with $b$, $T$ with $B$, and $u$ with $d$. Thus, we write the mass eigenvalues and the mixing matrices for the bottom sector as follows,
   \begin{align}
   	m_b &= - \frac{\sqrt{M_B^2 + (m_{d_R} - m_{d_L})^2}}{2} +  \frac{\sqrt{M_B^2 + (m_{d_R} + m_{d_L})^2}}{2},\\
   	m_{b'} &=  \frac{\sqrt{M_B^2 + (m_{d_R} - m_{d_L})^2}}{2} +  \frac{\sqrt{M_B^2 + (m_{d_R} + m_{d_L})^2}}{2} \label{bottom sector exact mass eigenvalues},
   \end{align}
   where,
   \begin{equation}
   	m_{d_R} = Y^3_{d_R}\frac{v_R}{\sqrt{2}}, \qquad m_{d_L} = Y^3_{d_L}\frac{v_L}{\sqrt{2}} \label{mdr and mdl},
   \end{equation}
   \begin{align}
   	K^{\dagger}_{B_L} & =
   	\left( \begin{array}{cc}
   		\cos \phi_{B_L} & \sin \phi_{B_L} \\
   		-\sin \phi_{B_L} & \cos \phi_{B_L}
   	\end{array}\right),  \label{KBL explicit}\\
   	K_{B_R} &=  \left( \begin{array}{cc}
   		\cos \phi_{B_R} & -\sin \phi_{B_R} \\
   		\sin \phi_{B_R} & \cos \phi_{B_R}
   	\end{array}\right) \left( \begin{array}{cc}
   		-1 &0 \\
   		0 & 1
   	\end{array}\right) = \left( \begin{array}{cc}
   		-\cos \phi_{B_R} & -\sin \phi_{B_R} \\
   		-\sin \phi_{B_R} & \cos \phi_{B_R}
   	\end{array}\right) \label{KBR explicit}.
   \end{align}
   For the approximate masses already written in Eq.(\ref{eq:top mass})-(\ref{eq:bottom VLQ mass}) and the approximate mixing angle yields,
   \begin{align}
   	&\sin\phi_{T_L} \simeq - \frac{m_{u_L} M_T}{M_T^2 + m_{u_R}^2}, \quad \cos\phi_{T_L} \simeq 1,\quad \sin\phi_{T_R} \simeq \frac{m^2_{u_L} m_{u_R} M_T }{(M_T^2 + m_{u_R}^2)^2},\quad \cos\phi_{T_R} \simeq 1 \nn\\ 
   	&\sin\phi_{B_L} \simeq - \frac{m_{d_L}}{M_B}, \quad \cos\phi_{B_L} \simeq 1,\quad \sin\phi_{B_R} \simeq \frac{m^2_{d_L} m_{d_R}}{M_B^3},\quad \cos\phi_{B_R} \simeq 1 \label{approximate mixing angle}
   \end{align}
   or in the approximate matrix form as follows,
   \begin{align}
   	K^{\dagger}_{T_L}  &\simeq
   	\left( \begin{array}{cc}
   		1 & - \frac{m_{u_L} M_T}{M_T^2 + m_{u_R}^2} \\
   	\frac{m_{u_L} M_T}{M_T^2 + m_{u_R}^2} & 1
   	\end{array}\right), \qquad  	K_{T_R} \simeq  \left( \begin{array}{cc}
   	1 & -\frac{m^2_{u_L} m_{u_R} M_T }{(M_T^2 + m_{u_R}^2)^2} \\
   	-\frac{m^2_{u_L} m_{u_R} M_T }{(M_T^2 + m_{u_R}^2)^2} & 1
   	\end{array}\right)  \label{KT approx} \\ 
   	K^{\dagger}_{B_L}  &\simeq
   	\left( \begin{array}{cc}
   		1 & - \frac{m_{d_L}}{M_B} \\
   		\frac{m_{d_L}}{M_B} & 1
   	\end{array}\right), \qquad \qquad \quad   	K_{B_R} \simeq  \left( \begin{array}{cc}
   		-1 & -\frac{m^2_{d_L} m_{d_R}}{M_B^3} \\
   		-\frac{m^2_{d_L} m_{d_R}}{M_B^3} & 1
   	\end{array}\right) \label{KB approx}. 
   \end{align}

   \section{CKM-like matrices}\label{appendix CKM matrix}
   In this appendix, we will discuss CKM-like matrices in this model and the rephasing of the CKM-like matrices. CKM-like matrix, which appears for the first time in Section \ref{sec:quark doublet and yukawa interaction}, is an ``intermediate" right-handed CKM-like matrix which has explicit form as follows, 
    \begin{equation}
   	V_R^{\text{CKM}} = \left(\begin{array}{cccc}
   		1& 0 &0  & 0 \\
   		0	& c_{\theta^d_{R}} & s_{\theta^d_{R}} c_{\theta_{B_R}} e^{i \frac{\alpha^3_{d_R}}{2}}& s_{\theta^d_{R}}s_{\theta_{B_R}} e^{i \frac{\alpha^3_{d_R}}{2}} \\
   		0	& -c_{\theta_{T_{R}}}s_{\theta^d_{R}} e^{i \frac{\alpha^3_{d_R}}{2}} & c_{\theta_{T_{R}}}c_{\theta^d_{R}}c_{\theta_{B_R}} e^{i \alpha^3_{d_R}} &c_{\theta_{T_R}}c_{\theta^d_{R}}s_{\theta_{B_R}} e^{i \alpha^3_{d_R}} \\
   		0&-s_{\theta_{T_R}}s_{\theta^d_{R}}e^{i \frac{\alpha^3_{d_R}}{2}}&s_{\theta_{T_R}}c_{\theta^d_{R}}c_{\theta_{B_R}}e^{i \alpha^3_{d_R}}&s_{\theta_{T_R}}c_{\theta^d_{R}}s_{\theta_{B_R}}e^{i \alpha^3_{d_R}}
   	\end{array} \right)  \label{VRCKM},
   \end{equation}
   where,
   \begin{align}
   	c_{\theta^d_{R}}&=\cos\theta^d_{R},\quad s_{\theta^d_{R}}=\sin\theta^d_{R},\quad c_{\theta_{T_{R}}}=\cos\theta_{T_{R}},\nn\\ s_{\theta_{T_{R}}}&=\sin\theta_{T_{R}},\quad c_{\theta_{B_{R}}}=\cos\theta_{B_{R}},\quad s_{\theta_{B_{R}}}=\sin\theta_{B_{R}}.
   \end{align}
   After Step 6 is done, we have the expressions of the left-handed CKM-like matrix and right-handed CKM-like matrix, which are defined in Eq.(\ref{left CKM final}) and Eq.(\ref{right CKM final}), respectively. The matrix forms of the left-handed CKM-like matrix and right-handed CKM-like matrix are as follows,
   \begin{equation}
   	\mathcal{V}_L^{\text{CKM}} = \left(\begin{array}{cccc}
   		1& 0 &0  & 0 \\
   		0	& c_{\theta^d_{L}} & s_{\theta^d_{L}} c_{\phi_{B_L}} e^{i \frac{\alpha^3_{d_L}}{2}}& -s_{\theta^d_{L}}s_{\phi_{B_L}} e^{i \frac{\alpha^3_{d_L}}{2}} \\
   		0	& -c_{\phi_{T_{L}}}s_{\theta^d_{L}} e^{i \frac{\alpha^3_{d_L}}{2}} & c_{\phi_{T_{L}}}c_{\theta^d_{L}}c_{\phi_{B_L}} e^{i \alpha^3_{d_L}} &-c_{\phi_{T_L}}c_{\theta^d_{L}}s_{\phi_{B_L}} e^{i \alpha^3_{d_L}} \\
   		0&s_{\phi_{T_L}}s_{\theta^d_{L}}e^{i \frac{\alpha^3_{d_L}}{2}}&-s_{\phi_{T_L}}c_{\theta^d_{L}}c_{\phi_{B_L}}e^{i \alpha^3_{d_L}}&s_{\phi_{T_L}}c_{\theta^d_{L}}s_{\phi_{B_L}}e^{i \alpha^3_{d_L}}
   	\end{array} \right)  \label{VL CKM final},
   \end{equation}
   where
   \begin{align}
   	c_{\theta^d_{L}}&=\cos\theta^d_{L},\quad s_{\theta^d_{L}}=\sin\theta^d_{L},\quad c_{\phi_{T_{L}}}=\cos\phi_{T_{L}},\nn\\ s_{\phi_{T_{L}}}&=\sin\phi_{T_{L}},\quad c_{\phi_{B_{L}}}=\cos\phi_{B_{L}},\quad s_{\phi_{B_{L}}}=\sin\phi_{B_{L}}.
   \end{align}
   and
    \begin{equation}
   	\mathcal{V}_R^{\text{CKM}} = \left(\begin{array}{cccc}
   		1& 0 &0  & 0 \\
   		0	& c_{\theta^d_{R}} & -s_{\theta^d_{R}} c_{\beta_{B_R}} e^{i \frac{\alpha^3_{d_R}}{2}}& s_{\theta^d_{R}}s_{\beta_{B_R}} e^{i \frac{\alpha^3_{d_R}}{2}} \\
   		0	& c_{\beta_{T_{R}}}s_{\theta^d_{R}} e^{i \frac{\alpha^3_{d_R}}{2}} & c_{\beta_{T_{R}}}c_{\theta^d_{R}}c_{\beta_{B_R}} e^{i \alpha^3_{d_R}} &-c_{\beta_{T_R}}c_{\theta^d_{R}}s_{\beta_{B_R}} e^{i \alpha^3_{d_R}} \\
   		0&-s_{\beta_{T_R}}s_{\theta^d_{R}}e^{i \frac{\alpha^3_{d_R}}{2}}&-s_{\beta_{T_R}}c_{\theta^d_{R}}c_{\beta_{B_R}}e^{i \alpha^3_{d_R}}&s_{\beta_{T_R}}c_{\theta^d_{R}}s_{\beta_{B_R}}e^{i \alpha^3_{d_R}}
   	\end{array} \right)  \label{VR CKM final},
   \end{equation}
   where
   \begin{align}
   	c_{\theta^d_{R}}&=\cos\theta^d_{R},\quad s_{\theta^d_{R}}=\sin\theta^d_{R},\quad c_{\beta_{T_{R}}}=\cos\beta_{T_{R}},\nn\\ s_{\beta_{T_{R}}}&=\sin\beta_{T_{R}},\quad c_{\beta_{B_{R}}}=\cos\beta_{B_{R}},\quad s_{\beta_{B_{R}}}=\sin\beta_{B_{R}}, \nn\\  \beta_{T_R} & = \theta_{T_R} -\phi_{T_R}, \quad \beta_{B_R} = \theta_{B_R} -\phi_{B_R}.
   \end{align}
   Recall the mass terms in the diagonal mass basis (including the massless two lightest quark fields) as follows,
   \begin{align}
   	\mathcal{L}_q \supset \mathcal{L}_{\text{mass}}& = - \overline{(u_L^m)^\alpha}(m_t^{\text{diag}})^\alpha (u_R^m)^\alpha -h.c. \nn\\ & \hspace{12pt} -\overline{(d_L^m)^\alpha}(m_b^{\text{diag}})^\alpha (d_R^m)^\alpha -h.c.. \label{mass term before rephasing}
   \end{align}
   We have the freedom to rephase the quark fields with the following transformations,
   \begin{align}
   	(u_{L(R)}^m)^\alpha &= (\theta_{u_{L(R)}})^\alpha \delta^{\alpha \beta} (\hat{u}_{L(R)}^m)^\beta \label{rephase up type},\\
   	(d_{L(R)}^m)^\alpha &= (\theta_{d_{L(R)}})^\alpha \delta^{\alpha \beta} (\hat{d}_{L(R)}^m)^\beta \label{rephase down type},
   \end{align}
   where $\theta_{u_{L(R)}} = \mathrm{diag}(e^{i\theta_{u_{L(R)1}}},e^{i\theta_{u_{L(R)2}}},e^{i\theta_{u_3}},e^{i\theta_{u_4}})$ and $\theta_{d_{L(R)}}=\mathrm{diag}(e^{i\theta_{d_{L(R)1}}},e^{i\theta_{d_{L(R)2}}},e^{i\theta_{d_3}},e^{i\theta_{d_4}})$. One can show that Eq.(\ref{mass term before rephasing}) is invariant under transformation in Eq.(\ref{rephase up type})-(\ref{rephase down type}).
   
   We apply this rephasing transformation into the $\mathcal{L}_q$. The left-handed and right-handed CKM-like matrices are rephased and become,
   \begin{align}
   	\mathcal{\hat{V}}_L^{\text{CKM}} = \theta^\dagger_{u_L} \mathcal{V}_L^{\text{CKM}} \theta_{d_L}, \qquad 	\mathcal{\hat{V}}_R^{\text{CKM}} = \theta^\dagger_{u_R} \mathcal{V}_R^{\text{CKM}} \theta_{d_R},
   \end{align}
   By choosing proper phase and phase difference, we could rephase the left-handed and right-handed CKM-like matrices and they become the following matrix forms,
   \begin{align}
   		\mathcal{\hat{V}}_L^{\text{CKM}}& =  \left(\begin{array}{cccc}
   			1& 0 &0  & 0 \\
   			0	& c_{\theta^d_{L}} & s_{\theta^d_{L}} c_{\phi_{B_L}} & -s_{\theta^d_{L}}s_{\phi_{B_L}}  \\
   			0	& -c_{\phi_{T_{L}}}s_{\theta^d_{L}}  & c_{\phi_{T_{L}}}c_{\theta^d_{L}}c_{\phi_{B_L}}  &-c_{\phi_{T_L}}c_{\theta^d_{L}}s_{\phi_{B_L}} \\
   			0&s_{\phi_{T_L}}s_{\theta^d_{L}}&-s_{\phi_{T_L}}c_{\theta^d_{L}}c_{\phi_{B_L}}&s_{\phi_{T_L}}c_{\theta^d_{L}}s_{\phi_{B_L}}
   		\end{array} \right)  \label{VL CKM rephased app}, \\ \nn\\
   		\mathcal{\hat{V}}_R^{\text{CKM}}& =  \left(\begin{array}{cccc}
   			1& 0 &0  & 0 \\
   			0	& c_{\theta^d_{R}} & -s_{\theta^d_{R}} c_{\beta_{B_R}} e^{i \frac{\delta}{2}}& s_{\theta^d_{R}}s_{\beta_{B_R}} e^{i \frac{\delta}{2}} \\
   			0	& c_{\beta_{T_{R}}}s_{\theta^d_{R}} e^{i \frac{\delta}{2}} & c_{\beta_{T_{R}}}c_{\theta^d_{R}}c_{\beta_{B_R}} e^{i \delta} &-c_{\beta_{T_R}}c_{\theta^d_{R}}s_{\beta_{B_R}} e^{i \delta} \\
   			0&-s_{\beta_{T_R}}s_{\theta^d_{R}}e^{i \frac{\delta}{2}}&-s_{\beta_{T_R}}c_{\theta^d_{R}}c_{\beta_{B_R}}e^{i \delta}&s_{\beta_{T_R}}c_{\theta^d_{R}}s_{\beta_{B_R}}e^{i \delta}
   		\end{array} \right)  \label{VR CKM rephased app},
   \end{align}
   where we redefine the phase difference as $\delta = \alpha^3_{d_R}-\alpha^3_{d_L}$. Therefore, in this model, we have one CP violating phase $\delta$ and in our choice, it is included in the right-handed CKM-like as shown in Eq.(\ref{VR CKM rephased app}).
   
   Moreover, the mixing angle $\beta_{T_R}$ and $\beta_{B_R}$ can be expressed in the approximate form as,
   \begin{align}
   	\sin\beta_{T_R} \simeq \frac{m_{u_R}}{\sqrt{M_T^2 + m^2_{u_R}}}, \quad \cos\beta_{T_R} \simeq \frac{M_T}{\sqrt{M_T^2 + m^2_{u_R}}}, \quad \sin\beta_{B_R} \simeq \frac{m_{d_R}}{M_B},\quad \cos\beta_{B_R} \simeq 1 \label{approximate mixing angle beta}.
   \end{align}

\end{appendices}


\end{document}